%% file: main.tex
\documentclass[manuscript,acmsmall]{acmart}

\input{dlab_macros}

\usepackage{tabu}
\usepackage{arydshln}
\usepackage{subcaption}
\usepackage{siunitx}
\usepackage{balance}
\usepackage{float}
\AtBeginDocument{%
  \providecommand\BibTeX{{%
    \normalfont B\kern-0.5em{\scshape i\kern-0.25em b}\kern-0.8em\TeX}}}

\begin{document}

\title[A Large-Scale Characterization of How Readers Browse Wikipedia]{A Large-Scale Characterization of How Readers Browse Wikipedia}

\author{Tiziano Piccardi}
\affiliation{%
  \institution{EPFL}
  \city{Lausanne}
  \country{Switzerland}
  }
\email{tiziano.piccardi@epfl.ch}

\author{Martin Gerlach}
\affiliation{%
  \institution{Wikimedia Foundation}
  \country{USA}
  }
\email{mgerlach@wikimedia.org}

\author{Akhil Arora}
\affiliation{%
  \institution{EPFL}
  \city{Lausanne}
  \country{Switzerland}
  }
\email{akhil.arora@epfl.ch}

\author{Robert West}
\authornote{Robert West is a Wikimedia Foundation Research Fellow.}
\affiliation{%
  \institution{EPFL}
  \city{Lausanne}
  \country{Switzerland}
  }
\email{robert.west@epfl.ch}


\begin{abstract}
Despite the importance and pervasiveness of Wikipedia as one of the largest platforms for open knowledge, surprisingly little is known about how people navigate its content when seeking information.
To bridge this gap, we present the first systematic large-scale analysis of how readers browse Wikipedia. 
Using billions of page requests from Wikipedia's server logs, we measure how readers reach articles, how they transition between articles, and how these patterns combine into more complex navigation paths.
We find that navigation behavior is characterized by highly diverse structures. Although most navigation paths are shallow, comprising a single pageload, there is much variety, and the depth and shape of paths vary systematically with topic, device type, and time of day.
We show that Wikipedia navigation paths commonly mesh with external pages as part of a larger online ecosystem, and we describe how naturally occurring navigation paths are distinct from targeted navigation in lab-based settings.
Our results further suggest that navigation is abandoned when readers reach low-quality pages.
Taken together, these insights contribute to a more systematic understanding of readers' information needs and allow for improving their experience on Wikipedia and the Web in general.
\end{abstract}

\begin{CCSXML}
<ccs2012>
   <concept>
       <concept_id>10002951.10003260</concept_id>
       <concept_desc>Information systems~World Wide Web</concept_desc>
       <concept_significance>500</concept_significance>
       </concept>
   <concept>
       <concept_id>10010405.10010476.10003392</concept_id>
       <concept_desc>Applied computing~Digital libraries and archives</concept_desc>
       <concept_significance>500</concept_significance>
       </concept>
   <concept>
       <concept_id>10002951.10003317</concept_id>
       <concept_desc>Information systems~Information retrieval</concept_desc>
       <concept_significance>500</concept_significance>
       </concept>
 </ccs2012>
\end{CCSXML}

\ccsdesc[500]{Information systems~World Wide Web}
\ccsdesc[500]{Applied computing~Digital libraries and archives}
\ccsdesc[500]{Information systems~Information retrieval}

\keywords{wikipedia, web navigation, server logs, information needs}

\setcopyright{acmlicensed}
\acmJournal{TWEB}
\acmYear{2023} \acmVolume{1} \acmNumber{1} \acmArticle{1} \acmMonth{1} \acmPrice{15.00}\acmDOI{10.1145/3580318}


\maketitle

\section{Introduction}
% \noindent
Evolution has optimized humans for information seeking, and humans have in turn optimized the world around them to facilitate access to information. Many of the most consequential evolutionary, cultural, and technological advances in humans---from the development of language and writing systems to modern telecommunication---have enhanced their ability to find, ingest, process, and transfer information.
Given the central importance of information seeking to human nature---epitomized by the view of humans as informavores \cite{machlup1983study}---, understanding the dynamics of how humans seek information and engage with knowledge is of key significance across disciplines, both in the basic and applied sciences. In the basic sciences, biologists, psychologists, anthropologists, among others, stand to gain fundamental insights into how humans function, whereas in the applied sciences, such insights can enable the design of more effective tools and information environments, such that humans can more readily find relevant knowledge in an ever-surging flood of information.

% 3. Why is it hard? (E.g., why do naive approaches fail?)

However, closely observing humans as they seek information is challenging, since it requires measuring predominantly cognitive behaviors at a great level of detail.
As a consequence, although much work has been dedicated to shedding light on human information seeking behavior (see \Secref{sec:related_work}), it has faced important limitations:
surveys \cite{younger2010internet} and
thinking-out-loud studies \cite{muramatsu2001transparent} are prone to cognitive biases, as humans generally perform poorly at introspection \cite{nisbett1977telling}.
Lab-based experiments \cite{Lydon-Staley2021} typically involve small samples consisting of biased populations (\eg, university students) and are thus frequently not representative and might lack statistical power.
Studies based on surrogate tasks (\eg, navigation games \cite{Wikispeedia}), although measuring navigation\hyp related skills, do not capture real-world, self\hyp motivated information seeking and may thus lack external validity \cite{olteanu2019social}.
Finally, studies based on aggregated versions of real-world information seeking traces (specifying page-to-page transition counts instead of full traces \cite{LinkSuccessfulWikipedia,Gildersleve2018Inspiration}), although capturing local, page-level choices accurately, may lack relevant trace-level information (\eg, relating the start of a trace to its end).

% 4. Why hasn't it been solved before? (Or, what's wrong with previous proposed solutions? How does mine differ?)

In this work, we provide a complementary perspective in the context of encyclopedic information seeking---an important special case of human information seeking---by leveraging a large-scale dataset of digital traces compiled from one month's worth of English Wikipedia's complete server logs, which offer unprecedented opportunities for observing humans interacting with knowledge in great detail.
% \edited{%
% We define knowledge-seeking as a specialization of online information-seeking focused on encyclopedic content.
% Regardless of the underlying reasons for visiting Wikipedia \cite{singer_why_2017}, the content consumed by the readers is composed of thematic articles, and the navigation space is centered around the concepts covered by the encyclopedia.}

Wikipedia is a primary source of encyclopedic knowledge and plays a unique role in the global knowledge ecosystem, fulfilling a wide range of information needs \cite{singer_why_2017,lemmerich_why_2019}. It is the largest encyclopedia ever built, with almost 60M articles in more than 300 languages. It is freely accessible across the globe and attracts more than 1.5B unique devices generating billions of pageloads every month, and it is the most popular website (except for search engines, Facebook, and YouTube) in 43 countries (more than any other website) \cite{hostinger_tutorials_2022}. Wikipedia thus reaches an audience whose representativeness far surpasses that of lab-based studies.
Since Wikipedia's server logs contain a record of all pageloads, the logs are uniquely suited for providing a geographically and temporally complete mirror of real-world, self-motivated encyclopedic information seeking.

In contrast to prior work, which has leveraged Wikipedia's server logs to shed light on specific aspects of reader behavior (including
reasons for visiting Wikipedia \cite{singer_why_2017,lemmerich_why_2019},
engagement with citations and external links \cite{piccardi2020quantifying,piccardi2021gateway, maggio2020meta}, studying variation in dwell time \cite{DwellingTime}, and measuring geo-localized collective behavior \cite{Tizzoni2020impact}),
this paper is the first to employ the logs in a principled, broad analysis with the goal of systematically elucidating the nature and structure of encyclopedic information seeking pathways.
% Whereas previous, complementary research has mostly focused on \textit{why} people use Wikipedia, in this paper we describe the mechanisms of content consumption in order to investigate the question of \textit{how} people use Wikipedia.
%
By analyzing billions of navigation traces extracted from the logs (\Secref{sec:data}) at various levels of aggregation, we consider three research questions:

\begin{enumerate}
    \item[\textbf{RQ1}] How do readers reach Wikipedia articles? (\hide{Unigram level; }\Secref{sec:unigrams})
    \item[\textbf{RQ2}] How do readers transition from one article to the next? (\hide{Bigram and trigram level; } \Secref{sec:bi_trigrams})
    \item[\textbf{RQ3}] What are the properties of entire reading sessions? (\hide{Session level; }\Secref{sec:session_level})
\end{enumerate}

% \edited{To answer the first two questions, we represent the sequence of articles loaded by the same reader with ``$n$-grams''. We progressively analyze unigrams, bigrams, and trigrams and then move to the last research question, where we investigate the full sessions.}
We find that Wikipedia navigation traces expose a wide variety of structures. While shallow sessions consisting of single pageloads dominate, we observe a long tail of long, complex traces, whose depth and shape vary systematically with topic, device type, and time of day.
Although it is known that search engines play a key role in driving readers to Wikipedia, we further highlight their importance for navigation between pages, showing that browsing Wikipedia  does not happen in isolation, but is embedded in sessions where users transition fluidly to and from the external Web, frequently via search engines.
We describe the interaction between article content and reader navigation, finding strong evidence that users stop navigating when reaching articles of low quality or the periphery of the network.
Finally, we show important differences between in-the-wild Wikipedia usage on the one hand and targeted navigation behavior captured by lab-based studies on the other hand.

%This aspect, as well as other differences that emerge, distinguishes real-world, in-the-wild Wikipedia usage from the targeted navigation behavior captured by lab-based studies.
%We highlight that Wikipedia navigation does not happen in isolation, but is embedded in sessions where users transition fluidly to and from the external Web.
%Finally, we find strong evidence that users stop navigating when reaching low-quality articles.
These findings are complemented with a description of best practices for analyzing readers' navigation using Wikipedia's server logs. We examine different ways of aggregating user sessions as well as their impact on the conclusions drawn.
 
Our results have important implications for Wikipedia and beyond.
Understanding how readers explore content on Wikipedia is critical for framing its role in fulfilling information needs and for making design decisions regarding its structure, format, accessibility, and supportive tools such as recommender systems.
Going beyond Wikipedia, these findings may help deepen our understanding of how humans navigate information when seeking knowledge. 
% We summarize the findings and their implications in \Secref{sec:discussion}.

%%%%%%%%%%%%%%%%%%%%%%%%%%%%%%%%%%%%%%%%%%%%%%%%%%%%%%%%%

\section{Related work}
\label{sec:related_work}

% This paper is related to research on navigation on information networks and readers' behavior on Wikipedia.

\xhdr{Information-seeking behavior}
Over time, information-seeking behavior has received attention from sociologists, cognitive psychologists, and, more recently, computer scientists, thanks to the availability of digital trace data. The study of information seeking investigates the strategies used by humans to find a piece of information to satisfy an information need \cite{wilson1981user}. Whereas the definition of ``need'' is unclear and relatively hard to formalize, seeking behavior is observable and easier to model, especially in information systems \cite{wilson1997information,wilson1999models}.
A complementary hypothesis from cognitive psychology argues that humans are \textit{informavores} and seek information with the same dynamics used by animals in searching for food \cite{machlup1983study}. This idea inspired the formulation of information foraging theory \cite{pirolli1999information}, which describes humans as behaving akin to predators in the information space, relying on ``information scent'' \cite{chi2001using} to find the paths that maximize the chances of leading them to the desired piece of information \cite{suh2010want,rotman2011slacktivism,mangel2013invasion}. Similarly, complementary models applied to the Web describe users' navigation as driven by the relatedness of a link or image with the desired goal \cite{kitajima2000comprehension}. Finally, additional cognitive models include ``berrypicking'' \cite{bates1989design}, which describes the search for information as a dynamic process where users collect small portions of information bit by bit, and ``exploratory search'' \cite{qu2008model}, which describes the information-seeking behavior of users unfamiliar with the topic of their search or with an unclear goal.

\xhdr{Navigation on the Web and log analysis}
Characterizing user navigation on the Web is a challenging task because of the limited availability of data. Previous work focused on modeling navigation patterns based on server logs of large websites or by using modified browser versions. A common finding is that people frequently revisit the same content multiple times \cite{RevisitationWWWNavigation, RepeatConsumption}. This repeat consumption behavior, which is abandoned when the person becomes bored of the content \cite{UserConsumption}, makes human mobility on the Web predictable \cite{kulshrestha2020web}. 
Although researchers found that Web users are not strictly Markovian (the page visited next does not depend exclusively on the current page) \cite{UsersReallyMarkovian}, many prediction models approximate the navigation of users on a network with Markov chains \cite{Pirolli1999,mabroukeh2009semantic,deshpande2004selective} and hybrid models \cite{narvekar2015predicting,khalil2009integrated,WebNavigationPrediction,awad2007web}.

A significant effort in investigating Web navigation has focused on search engines and how people find content from relevant keywords \cite{jiang2013mining}. Log-based analyses of the navigation following a Web search show that people's behavior exhibits a high level of variability \cite{white2007investigating} and that different search queries and origins are associated with different navigation patterns \cite{bilenko2008mining,ibanez2022comparison}. Beyond characterizing users' information needs, digital trails can be exploited to improve search engine results \cite{white2010assessing,fox2005evaluating,eickhoff2014lessons,singla2010studying}, \eg, by using the collective interest of a destination page as a metric of relevance \cite{white2007studying}.
%
% \xhdr{\edited{Methods and applications of log analysis}}
Similarly, navigation traces %are also investigated as a tool to improve the website.
have proven useful as a tool to improve website navigability by identifying missing links \cite{ParanjapeImproving,EvaluatingImprovingNavigability,West2015Mining} and other usability issues that normally require the work of domain experts \cite{RuiliImproving}.
Finally, navigation logs can be used to compute the semantic relatedness of pages by studying what content is typically accessed together
\cite{SemanticRelatednessHumanNavigational,Dallmann2016Extracting}.

\xhdr{Reader behavior on Wikipedia}
Researchers have also studied how readers behave when reading Wikipedia. Recent work focuses on the interaction with external links \cite{piccardi2021gateway} and references \cite{piccardi2020quantifying,maggio2020meta}, and on the reading time of articles \cite{DwellingTime}. Researchers have concluded that Wikipedia users have reading patterns that fall in different categories, such as exploration, focus, trending, and passing \cite{ReaderPreferences}, and that readers prefer links that lead to the periphery of the network, about semantically similar content and located at the top of the article \cite{StructureArticlesNavigation,LinkSuccessfulWikipedia}. 
Other studies have investigated the inter-event time in the navigation logs of Wikipedia and found strong regularities in the temporal rhythms, which suggest a reasonable rule of thumb for segmenting sessions after inactivity periods of one hour \cite{InterActivityTime}.

These studies are complemented by investigations of the motivations for visiting Wikipedia \cite{lemmerich_why_2019,singer_why_2017}, which describe a variety of factors such as current events, media coverage of a topic, personal curiosity, work or school assignments, or boredom.

% \xhdr{Navigation on Wikipedia}
Closest in spirit to the present work, multiple approaches have been used to study human navigation on Wikipedia. The public clickstream~\cite{Wulczyn2015clickstream} 
%\footnote{\url{https://meta.wikimedia.org/wiki/Research:Wikipedia_clickstream}}
contains transition counts for pairs of articles. Although the clickstream constitutes an aggregated and filtered version of the server logs, it has been shown that it can serve as a useful approximation in many practical applications  \cite{arora2022wikipedia}.
It has been used to study how different topics relay more traffic than others \cite{SearchNavigationWikipedia,Gildersleve2018Inspiration}, and how readers' navigation paths tend to start general and become incrementally more focused at every step \cite{SearchStrategies}.

Other approaches to understanding readers' navigation have identified different types of curiosity during Wikipedia exploration by relying on data shared by volunteers \cite{Lydon-Staley2021},
%where it has been observed how different types of curiosity drive the exploration of Wikipedia. 
while yet others have characterized human navigation as manifested in digital traces obtained via Wikipedia navigation games such as Wikispeedia \cite{Wikispeedia}, where players start from a random article and are tasked to reach a target page in as few clicks as possible by following links only. 
These trajectories, denoted as \textit{targeted navigation} here, show how efficient people are at finding short paths \cite{HumanWayfinding,Helic,West_Leskovec_2012}.
% , but with performance comparable to automatic agents that rely only on a local view \cite{West_Leskovec_2012}.
In contrast to natural navigation, targeted navigation posits an unambiguous definition of success (\ie, reaching the target article), which allows researchers to study how users drift away from the best path and when they abandon their search \cite{PredictingNavigationSuccess,LastClick}.
%These trajectories differ from natural navigation because the researchers working with this data know exactly where the user is headed. 
Targeted navigation behavior as observed in navigation games may, however, differ from natural navigation behavior, which limits the utility of such traces for studying the real-world usage of Wikipedia.

% \xhdr{Using navigation traces}

% Information needs, cognition and curiosity (library) \cite{Lydon-Staley2021}  \cite{SearchingBehaviorDifficultyInformationTasks} \cite{MURDOCK2017117} \cite{Clickbaits} \cite{InformationSeekingLibrary} Information scent \cite{InformationScent} \cite{InformationScentUnderstandSearch} interplay with search (SERPs) \cite{McMahon2017Substantial,Vincent2021Deeper} 

% Modeling topics on Wikipedia \cite{ORES} \cite{LanguageAgnosticORES} \cite{WikiPDA}

% Wikipedia and Learning: Wikipedia plays an important role in online learning ecosystem \cite{Kross2021Characterizing}, prerequisite relations for learning \cite{Sayyadiharikandeh2019Finding}

% measuring collective attention \cite{Garcia-Gavilanes2017Memory,Miz2020Trending}, e.g. diseases~\cite{Tizzoni2020impact,De_Toni2021general}

\section{Materials and methods}
\label{sec:data}

The data sources exploited in this study include user traces mined from Wikipedia's server logs and features extracted from articles.

\subsection{Pageloads}
To study how readers navigate Wikipedia, we analyze the server logs of the English language edition
% (``enwiki'')
collected for four weeks between 1 and 28 March 2021. This data contains an entry for each time a Wikipedia page is loaded. It is continuously and automatically collected for analytic purposes on Wikimedia's infrastructure and deleted after 90 days. 

We limit our analysis to the pageload requests for articles (MediaWiki namespace 0), filtering out requests from bots.
To protect readers' privacy, we remove sensitive information in several steps: discarding pageloads from readers who edited or were logged in during the time of data collection; discarding all requests from countries with at least one day with fewer than 300 pageloads; generating (pseudo) user identifiers by hashing IPs and user agent strings, as done in previous work \cite{ParanjapeImproving}; and dropping IP, user agent, and fine-grained geo information. In total, these anonymization steps lead to the removal of around 3\% of the data.
In addition, we perform the following filtering steps. 
First, we drop pageloads of the \textit{Main\_Page} article, as it does not represent any specific entity.
% (removing 37.8M users who loaded only the home page).
These requests may, \eg, come from users who set Wikipedia as the browser's default page.
Second, we remove traffic from massively common IPs, which would make it hard to study individual users' activities, by dropping all user identifiers with more than 2,800 pageloads, or on average 100 per day, thus removing 28k (0.0019\%) user identifiers. After the above steps, each request entry includes the anonymous user identifier, the page title, the timezone-corrected timestamp, the access method (mobile or desktop), and the referrer URL.
The final dataset contains 6.52B pageloads associated with 1.47B user identifiers. 

%\mg{not crucial to have this data}
% We obtained 7.1B pageloads distributed across 6.7M articles associated with 1.5B different unique ids. 
% Mobile devices are the most common way to access Wikipedia, with around 4.2B pageloads (59.5\%), while readers requested 2.9B (40.5\%) pages from desktop devices. The five most common locations are countries with a large English-speaking population, such as the United States (40\%), United Kingdom (11\%), India (8\%), Canada (4.8\%), and Australia (2.8\%). The most visited article is the landing page named \textit{Main\_Page} that was loaded 353M times (4.9\% of requests). 

%\mg{user identifier is described above already}
%To study how the readers move across the articles of Wikipedia, we group pageloads by the user identifier and sort by their timestamp. Since the user identifier is obtained by combining IP address and user-agent, this only constitutes a proxy for unique users (see \cite{singer_why_2017} for a detailed discussion on the limitations of this approach). 

\subsection{Article features} 
\label{sec:features}
To characterize the content viewed by readers, we collect a set of article features. To ensure alignment between the server logs and the articles' content, we compute the features for the revisions of the public snapshot released at the end of March 2021.

We obtain article features such as the number of outgoing links, the PageRank, article quality score, and topic. We assign the quality of the articles using the \textit{articlequality} model of ORES\footnote{\url{https://www.mediawiki.org/wiki/ORES}} \cite{ORES}, Wikipedia's official scoring platform. This model offers a way to obtain a score~\cite{halfaker2017interpolating} that summarizes the structural properties of the article, such as the number of sections, references, and the presence of infoboxes. To represent articles semantically, we use two approaches:
(1)~the probabilities for 64 manually curated topics obtained from the ORES \cite{ORES} \textit{articletopic} model let us assign topical labels to articles;
(2)~the crosslingual WikiPDA~\cite{WikiPDA} topic model lets us place articles in a 300-dimensional topic space.

\begin{table}[t]
\centering
% \small
\begin{tabu}{l|l|l||r}
\textbf{Origin} & \textbf{Desktop} & \textbf{Mobile} & \textbf{Total}\\
\hline
Search engines & 45.97\% & 48.77\% & 47.71\% \\
Wikipedia & & & \\
\hspace{5mm}{Articles} & 35.64\% & 35.75\% & 35.72\% \\
\hspace{5mm}{Main page} & 1.65\% & 0.70\% & 1.06\%  \\
\hspace{5mm}{Lang.\ switching} & 1.62\% & 0.50\% & 0.92\%  \\
\hspace{5mm}{Categories} & 0.59\% & 0.25\% & 0.39\%  \\
\hspace{5mm}{Search page} & 0.38\% & 0.22\% & 0.29\% \\
\hspace{5mm}{Special pages} & 0.07\% & 0.01\% & 0.03\%  \\
\hspace{5mm}{Portals} & 0.03\% & 0.01\% & 0.02\%  \\
\hspace{5mm}{Others} & 0.07\% & 0.01\% & 0.03\%  \\
Unspecified origin & 12.64\% & 13.03\% & 12.88\% \\
External websites & 1.36\% & 0.70\% & 0.95\%   \\
% Others & 0\% & 0.01\% & 0.01\%  & \\
% \hdashline
\end{tabu}
 \vspace{10mm}
\caption{
  Statistics of referrers of single pageloads.
%   Internal transitions marked with *
  }
  \label{table:origin_stats}
%   \vspace{-5mm}
\end{table}

% \section{Results: unigram level}
\section{RQ1: How do readers reach Wikipedia articles?}
\label{sec:unigrams}

In this work, we use the term ``$n$-gram'' to designate a sequence of $n$ subsequent Wikipedia pageloads from the same user, where the ``vocabulary'' consists of all articles available on Wikipedia.
We start our analysis with unigrams ($n=1$) to investigate individual pageloads and enumerate how readers can reach Wikipedia articles. 
We classify Web traffic according to HTTP referrers and quantify the frequency of each referrer type (\Tabref{table:origin_stats}). In total, 4B (61.5\%) pageloads have external or empty referrers and are thus entry points to Wikipedia.

\xhdr{Search engines} The most common way to reach the content of Wikipedia is through external search engines, at 3.1B pageloads (45.9\% of all recorded traffic, or 77.5\% of external traffic). This volume reflects the significant value offered by Wikipedia in fulfilling the information needs of search engine users
% More fine-grained analysis on the special relationship between Wikipedia and search engines can be found in a previous study on Wikipedia in SERPs~
\cite{Vincent2021Deeper,techblogsearch}.
% and in the publicly available search referrer dataset~\cite{techblogsearch}.

\xhdr{Wikipedia}
Clicks from other articles account for 35.7\% of all traffic. 
Interestingly, as observed in previous work \cite{mitrevski2020wikihist}, 6.6\% of these pageloads happen through links that do not exist in the link network itself, but likely through other interactions such as Wikipedia's search drop down menu. 
Content can also be reached from other pages on the Wikipedia platform: (1) the main page, (2) category pages, (3) Wikipedia's internal search, (4) portals, or (5) other Wikipedia pages, including talk pages or pages in other languages (language switching). 

\xhdr{Unspecified origin} In 12.9\% of all traffic, we observe an empty referrer field (20.9\% of external traffic). Multiple reasons can produce a request without an explicit origin, including direct access via the browser history, redirects from apps, bookmarks, search toolbars, or when the link source has explicitly turned on the \textit{noreferrer} property.

\xhdr{External websites} In total, only 0.95\% of the requests originated from external websites that are not search engines nor Wikipedia domains (1.55\% of the external traffic). Among those, the most common sources are Facebook (15.6\%), Reddit (9.6\%), YouTube (8.0\%), and Twitter (4.3\%). 
% Previous work explored the effect of traffic from social media on the risk of vandalism, finding no evidence of negative impact~\cite{Morgan2021social}.
% \tp{add \% of external websites relative to external traffic}\mg{Not needed imo.}

\xhdr{Others}
Other external visits (0.015\% of external traffic) come from Android Web views and custom embedded visualizations, with the most common being the Telegram and Reddit sync apps, and Facebook on Android devices.
% (\textit{android-app://}). 

\section{RQ2: How do readers transition from one article to the next?}
\label{sec:bi_trigrams}

Next, we move from unigrams ($n=1$) to bigrams ($n=2$), in order to understand how readers transition between Wikipedia articles. We study events aggregated by user identifier and sorted by time to investigate the properties of consecutive pageloads and their inter-event time.
We consider two subsequent pageloads from the same user identifier as a bigram if they are separated by less than one hour~\cite{InterActivityTime}.

Since here we are not interested in the exact article visited, we instead represent each pageload in a bigram with an alias indicating if the reader loaded the same page or different pages. The pattern ``AA'' means that the user revisited sequentially the same article, whereas ``AB'' indicates a load of two different pages.
Here it is important to note that the Wikipedia server instructs the browser to disable the cache,
% \footnote{HTTP header: \textit{cache-control: private, s-maxage=0, max-age=0, must-revalidate}}.
such that the server logs contain essentially all pageload events, including cases when the readers reloaded an article, \eg, by using the back button.

\begin{figure}[t]
    \centering
    \includegraphics[width=0.8\linewidth]{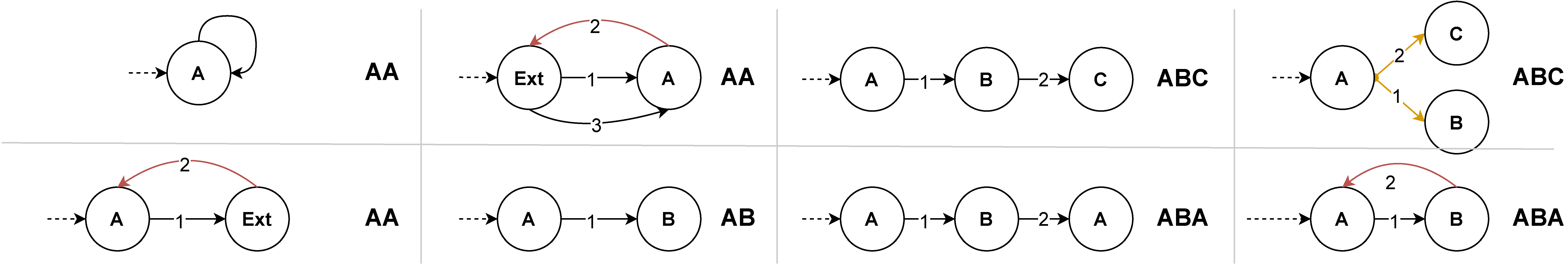}
    \caption{Examples of patterns in the logs and the multitude of client-side behaviors that can generate these patterns. Black arrows represent forward link clicks, red arrows represent back-button clicks, yellow arrows represent clicks that open multiple tabs from the same source page. ``Ext'' represents external (non-Wikipedia) pages. Numbers represent the order of clicks.}
    \label{fig:patterns-combined}
\end{figure}

\begin{table}[t]
\centering
% \small

%   \vspace{-3mm}
\begin{tabular}{l|ll||lllll}
\textbf{Device} & \textbf{AB} & \textbf{AA}& \textbf{ABC} & \textbf{ABA} & \textbf{ABB} & \textbf{AAB} & \textbf{AAA} \\
\hline
Desktop & 0.900 & 0.099 & 0.749 & 0.121 & 0.047 & 0.049 & 0.031 \\
Mobile & 0.880 & 0.119 & 0.719 & 0.143 & 0.055 & 0.053 & 0.027\\
\hline
Total & 0.888 & 0.111 & 0.732 & 0.134 & 0.052 & 0.052 & 0.029\\
\end{tabular}
\vspace{3mm}
  \caption{
  Frequencies of bigram and trigram patterns.
  }
  \label{table:ngrams_patterns}
\end{table}

\xhdr{Bigrams} The logs contain 3.95B instances of bigrams.
The emerging patterns, described next, are summarized in \Tabref{table:ngrams_patterns}.
The most frequent bigram pattern (``AB'' in \Tabref{table:ngrams_patterns})
corresponds to transitions between two different articles. It can happen both through internal and external navigation (\cf\ \Figref{fig:patterns-combined}). This pattern represents around 89\% of all bigrams.
The other possible bigram pattern (``AA'' in \Tabref{table:ngrams_patterns}), corresponds to the consecutive reload of the same article. Representing 11\% of all bigrams, it is rather common (84\% share the same referrer). This pattern appears at least once in 37\% of the navigation histories of readers with at least two pageloads in the month of data collection. The pattern can be generated by different client behaviors (\cf\ \Figref{fig:patterns-combined}), including repeated consumption as described in previous work~\cite{RevisitationWWWNavigation, UserConsumption}, user activities involving external navigation, or artificial reloads by the browser when a tab unloaded from memory is restored.

\xhdr{Trigrams}
Finally, we also briefly consider the 2.98B trigrams present in the logs.
The most common trigram pattern (73\%, ``ABC'' in \Tabref{table:ngrams_patterns}) represents transitions between three different articles. A variety of behaviors can generate this pattern, including sequential clicks or multitab behavior (\cf\ \Figref{fig:patterns-combined}).
The second most common trigram pattern (13\%, ``ABA'' in \Tabref{table:ngrams_patterns}) can be generated by intentionally revisiting the same page in a forward manner or by clicking the back button (\cf\ \Figref{fig:patterns-combined}). In 89\% of ABA instances, the first and last event also share the same referrer.
The remaining trigram patterns (ABB, AAB, AAA) are combinations of the bigrams described above.

\begin{figure*}[t]
% \hfill
    \begin{minipage}[t]{.24\columnwidth}
        \centering
        \includegraphics[height=3.1cm]{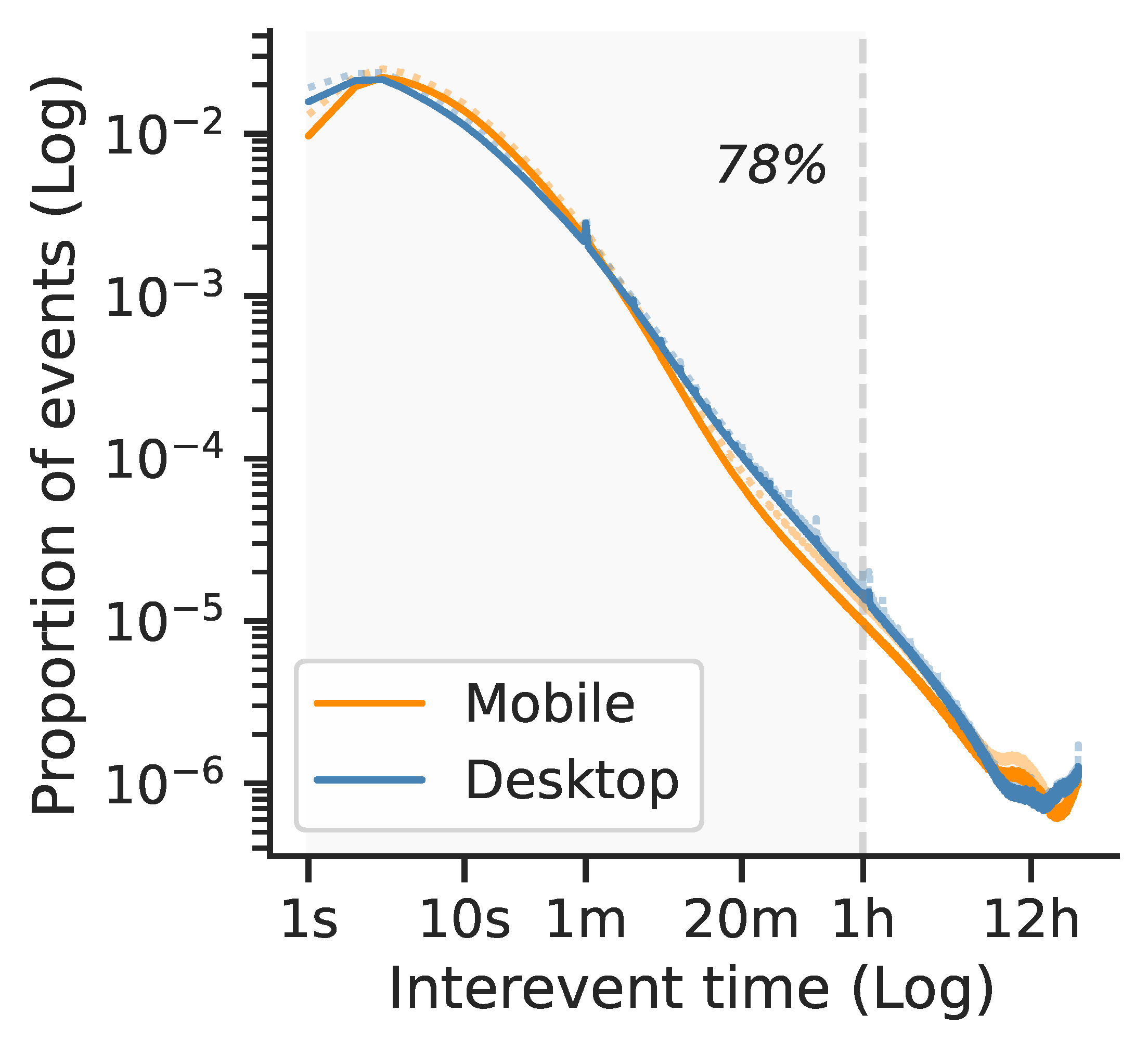}
        \subcaption{Distribution of inter-event times}
        \label{fig:interevent_time}
    \end{minipage}
    \hfill
    \begin{minipage}[t]{.24\columnwidth}
        \centering
        \includegraphics[height=3.1cm]{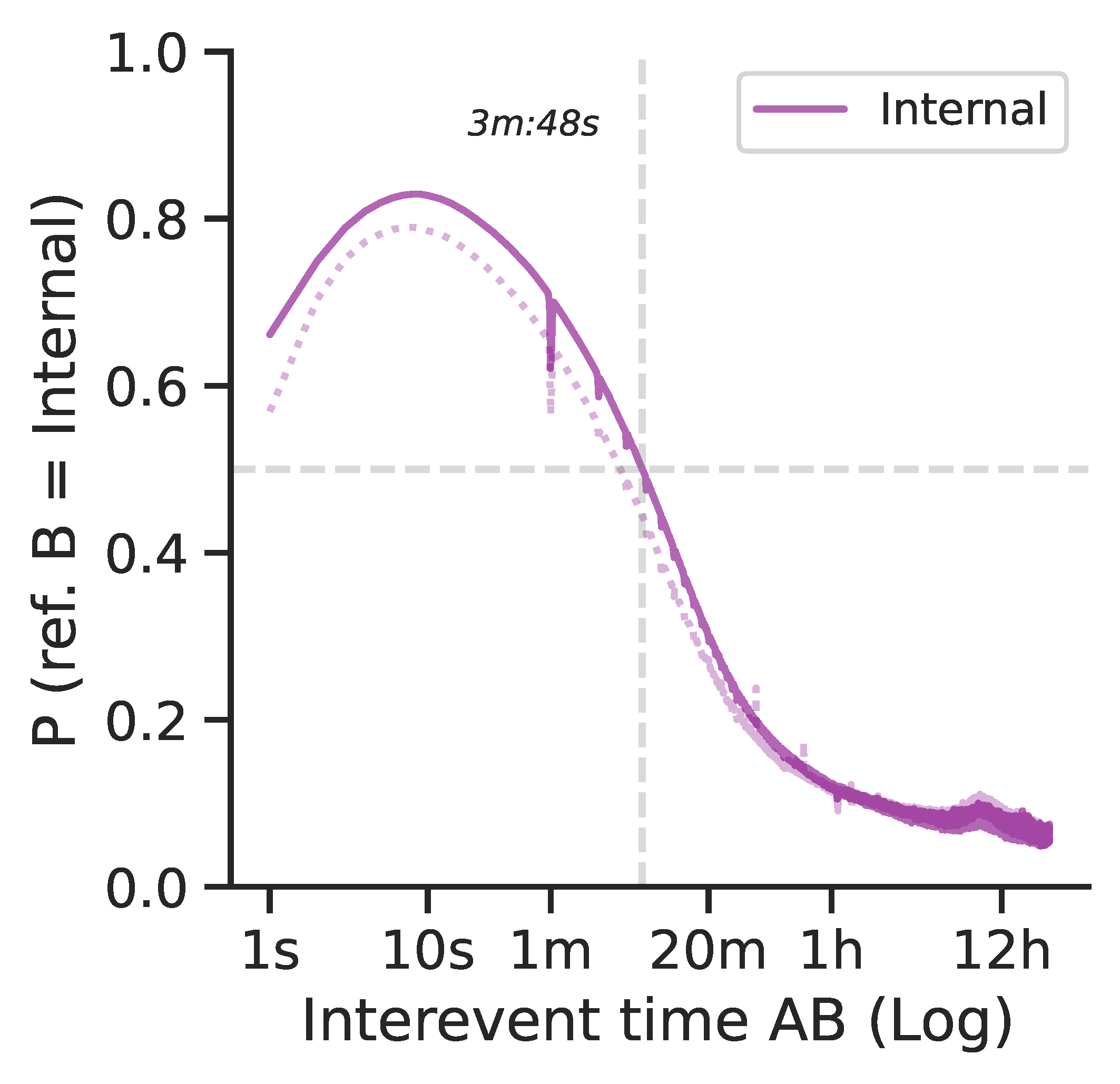}
        \subcaption{Probability of internal transition}
        \label{fig:probability_referer_by_time}
    \end{minipage}
    \hfill
    \begin{minipage}[t]{.24\columnwidth}
        \centering
        \includegraphics[height=3.1cm]{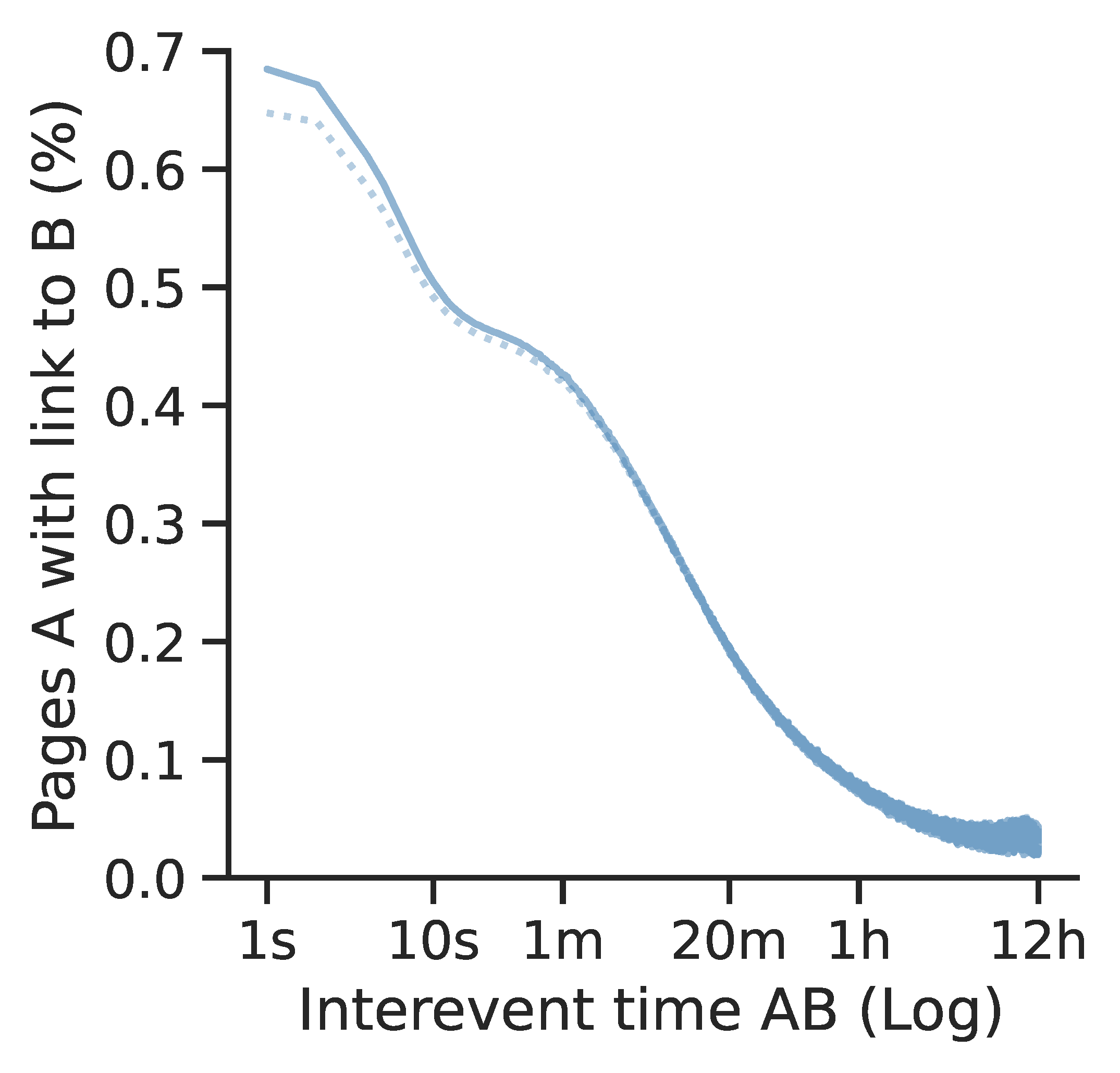}
        \subcaption{External transitions with link}
        \label{fig:out_in_from_search}
    \end{minipage}
    \hfill
    \begin{minipage}[t]{.24\columnwidth}
        \centering
        \includegraphics[height=3.1cm]{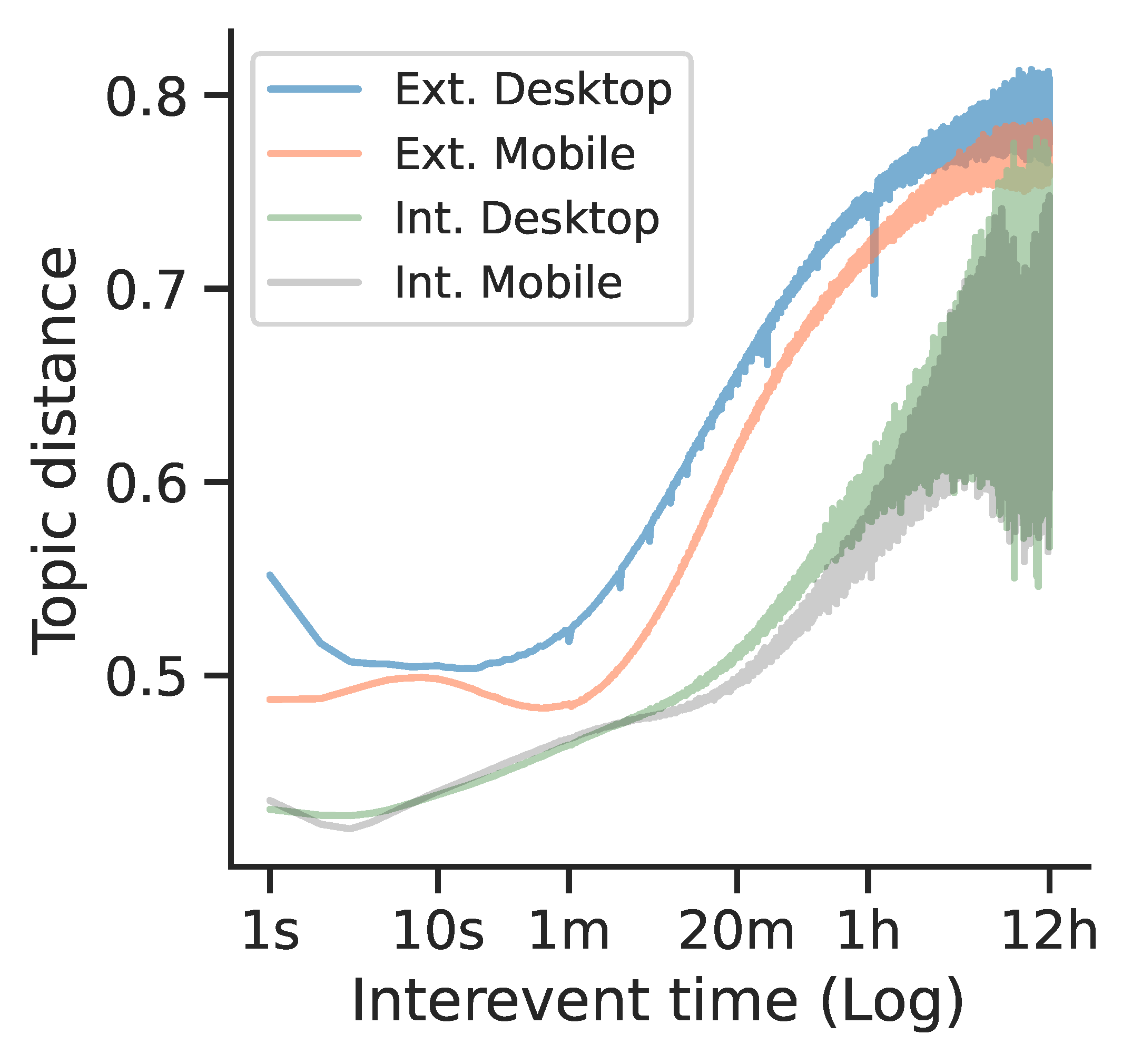}
        \subcaption{Distance between pageloads}\label{fig:topic_distance_interevent}
    \end{minipage}
    % \hfill
    % \vspace{-5mm}
\caption{
Statistics of bigrams as a function of the inter-event time between two pageloads. Dotted curves represent the distributions with AA patterns included.
}
\label{fig:mobile_desktop_time}
\end{figure*}

\xhdr{Dynamics of transitions} 
In order to understand the dynamics of these transitions, we investigate the inter-event time between the two pageloads in each bigram. The interval between two consecutive pageloads peaks at very short times, with a median of 74 seconds (63 and 93 seconds for mobile and desktop devices, respectively). However, as \Figref{fig:interevent_time} shows, the distribution is long-tailed, with 22\% of pairs separated by more than one hour. 

Investigating the referrer of the second page of the bigrams reveals that readers frequently do not use internal links to transition between two articles, but external pages by leaving and re-entering Wikipedia. 
These external transitions are not rare: in 35.2\% (or 40.1\% when including AA patterns) of the bigrams with less than one hour between the two events, the second page was reached through external navigation. This observation is corroborated by  \Figref{fig:probability_referer_by_time}, which shows that for pairs with an inter-event time greater than 3 minutes and 48 seconds, transitions via internal links are even less common than transitions via external navigation. 
External transitions tend to be semantically coherent: considering all 1.4B AB-type bigrams where the second page is reached via search, in 18\% of the cases, the first page explicitly contained the link. This proportion increases to 30\% [56\%] when considering pairs with an inter-event
time of less than one hour [less than 10 seconds] (\Figref{fig:out_in_from_search}).
The topical coherence of these transitions is also visible
% by observing the jump size in topic space.
in \Figref{fig:topic_distance_interevent}, which plots the average topical distance (measured by the cosine of WikiPDA vectors, \cf\ \Secref{sec:features}) as a function of inter-event time, showing that external navigation recorded within a few minutes from the previous pageload shows topical distance comparable to internal navigation.

\section{RQ3: What are the properties of entire reading sessions?}
% \section{Results: session level}
\label{sec:session_level}
Using our insights about navigation at the unigram, bigram, and trigram levels, we can now characterize entire navigation sessions. 
We start by introducing two different approaches to conceptualizing navigation sessions (\Secref{sec:session_definitions}) and discuss how each captures different aspects of reader navigation.
We then describe the properties of reader navigation by focusing on three aspects of the resulting sessions: contextual features defining when and how sessions start (\Secref{sec:session_context}); static properties, such as the structural features of sessions (\Secref{sec:session_structure}); and finally, the dynamic properties of the sessions, such as
the evolution in the content consumed over the course of navigation (\Secref{sec:session_evolution}).

%To understand the evolution of the navigation across articles, we combine the former patterns by using the two aggregation methods introduced previously. 
%After this step, from the original 6.52B events, we obtain 3.7B navigation trees and 2.51B reading sequences.

\subsection{Conceptualizing reader sessions}
\label{sec:session_definitions}
%The previous findings demonstrate that defining a clear concept of sessions is a challenging task. Readers exhibit complex behavior on the platform and consume content by combining a mix of internal and external navigation. 
%This section introduces two strategies to combine these events into consistent user sessions.
Grouping all pageloads of the same user, there is no unique way to operationalize the notion of a reading session. 
Based on different previously employed approaches, we identify two distinct notions of a session:
%We introduce two notions of a user session, each capturing different aspects of navigation pathways (details below):
(1)~\textit{navigation trees} connect pageloads hierarchically based on referrer information, whereas
(2)~\textit{reading sequences} order pageloads linearly based on temporal information.
These capture different aspects of how readers navigate, and which approach is better suited depends on the context and the phenomenon one aims to observe.
From the original 6.52B pageloads, we obtain 3.7B navigation trees and 2.51B reading sequences.

\xhdrNoPeriod{Navigation trees} \cite{ParanjapeImproving} describe how readers traverse Wikipedia by following internal links. 
We generate a tree by connecting pageloads via the referrer contained in HTTP headers.
% When a browser requests a web resource, it typically shares the previous location with the server using an HTTP header field called \textit{referrer}. 
% We use this field to combine the pageload events in structured trees. 
Pages reached through internal transitions (\ie, using internal links) are added as children of the most recent load of the article in the referrer, while pageloads with external or \textit{Main\_Page} referrers generate a new tree. 
If a page is loaded multiple times from the same referrer, the parent node retains only the first instance as a child.
This method has the advantage of representing coherent sessions created through clicks on internal links---regardles of the time spent on one article---and of capturing multitab behavior \cite{huang2010parallel}.
The downside is the difficulty of capturing content consumption over time for subsequent pages not reached through internal clicks, even if close in time (a common pattern, \cf\ \Secref{sec:bi_trigrams}).
Since this aggregation method does not model temporally linear consumption,  loading articles by opening multiple tabs or backtracking to select a different path leads to the same navigation tree.

\xhdrNoPeriod{Reading sequences}
describe how readers consume content in temporal order. 
They are defined as linear sequences of all pageloads by the same user ordered by time.
Sequences are split if the inter-event time between two consecutive pageloads separated by external navigation exceeds a threshold value of one hour, following recommendations from previous studies~\cite{InterActivityTime} and common practice~\cite{singer_why_2017,lemmerich_why_2019}. 
Within such sessions, we keep only the first pageload of each article, in order to only capture the first exposure of the respective content.
This method generates topically less coherent sessions, capturing the temporal and linear sequence of pageloads of a reader within a defined period of time, both via internal and external transitions (\eg, multiple external searches).
This method has the disadvantage of being a simplification of how readers explore the link network, and a fixed threshold of one hour may not be ideal in every context. 

\begin{minipage}[t]{0.48\linewidth}
    \begin{figure}[H]
    % \hfill
        \begin{minipage}[t]{.49\columnwidth}
            \centering
            \includegraphics[height=3.1cm]{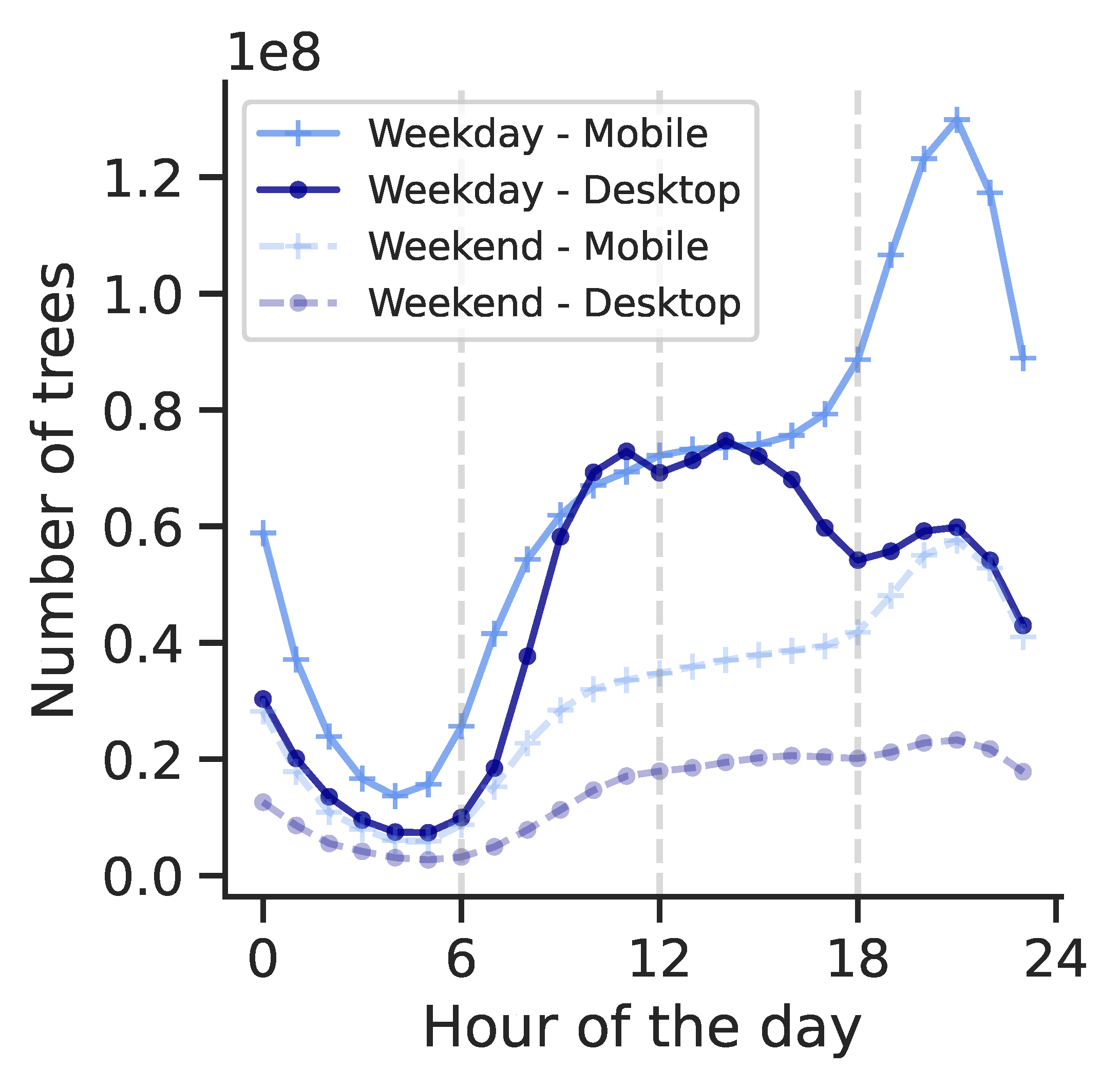}
            \subcaption{Number of trees}\label{fig:total_sessions_by_time_trees}
        \end{minipage}
        \hfill
        \begin{minipage}[t]{.49\columnwidth}
            \centering
            \includegraphics[height=3.1cm]{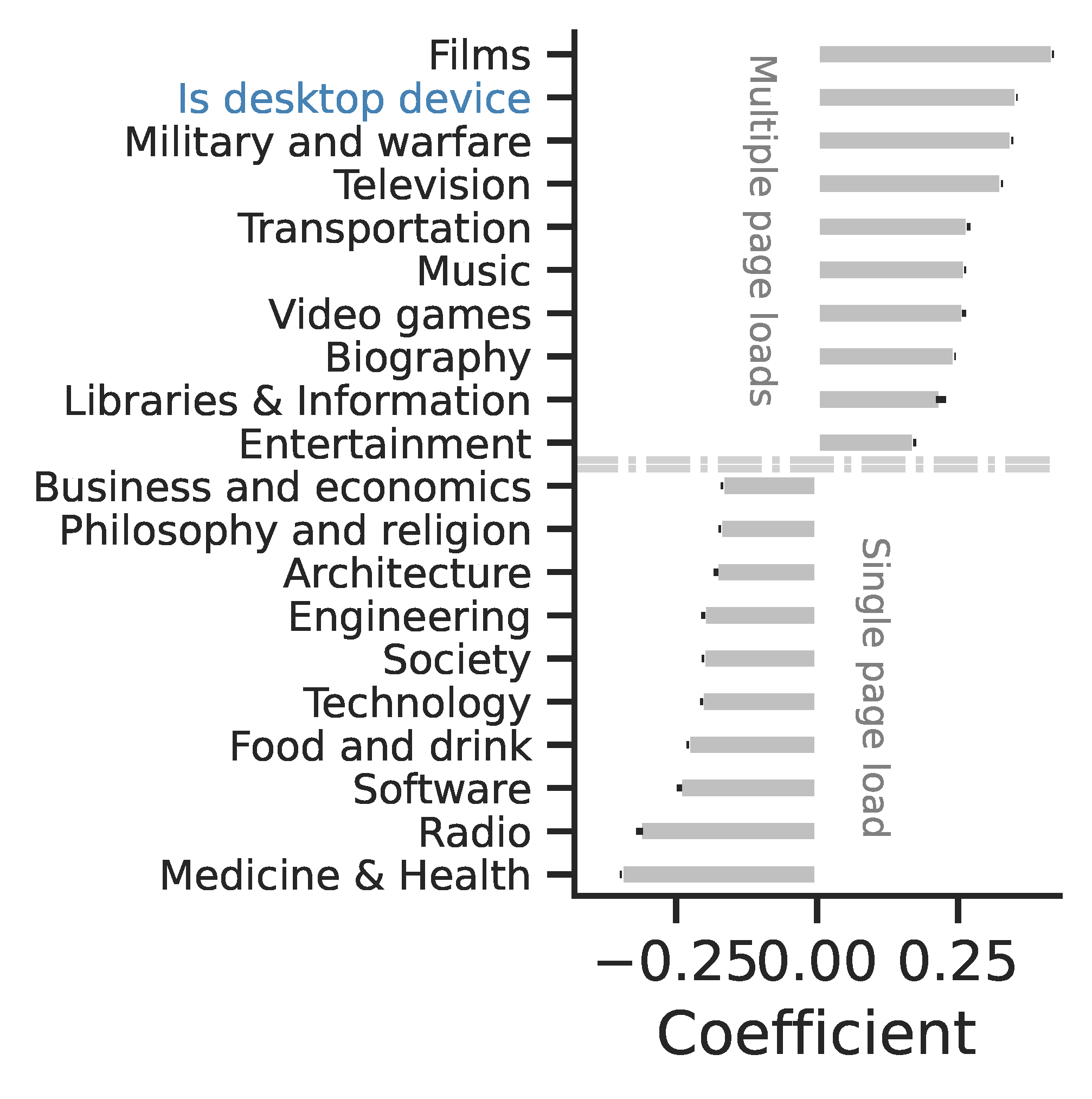}
            \subcaption{Multiple-pageloads}\label{fig:landing_regression_length}
        \end{minipage}
       
        \hfill
        % \vspace{-3mm}
    \caption{
The total number of trees started at different times of day (\Figref{fig:total_sessions_by_time_trees}) and feature contributions to the logistic model predicting if the reading sequence is composed of more than one pageload (\Figref{fig:landing_regression_length}).  }
    \label{fig:time_context}
    \end{figure}
\end{minipage}
\hfill
\begin{minipage}[t]{0.48\linewidth}
\hfill
    \begin{figure}[H]
    % \hfill
 \begin{minipage}[t]{.49\columnwidth}
            \centering
            \includegraphics[height=3.1cm]{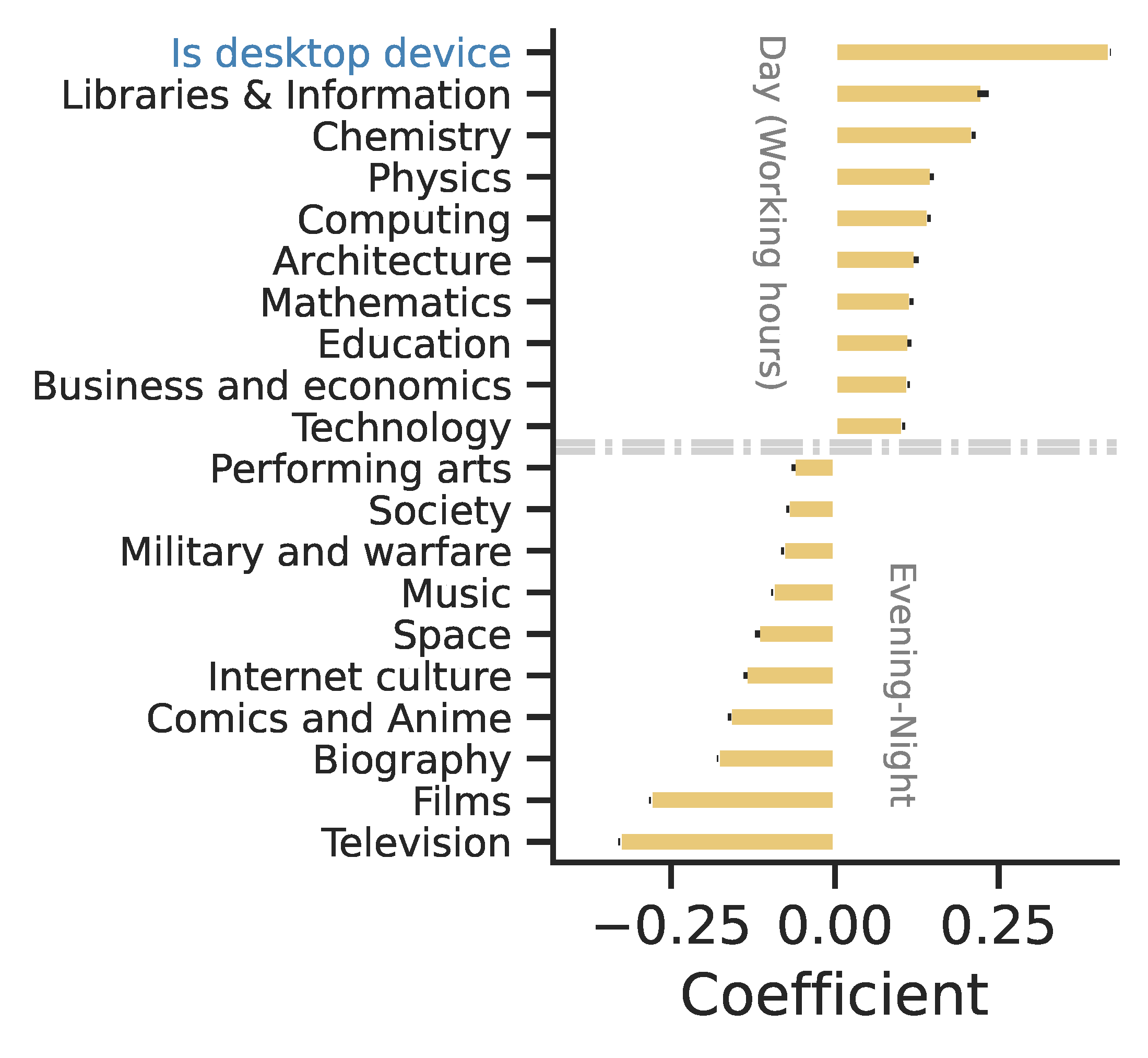}
            \subcaption{Navigation trees}\label{fig:landing_regression_day}
        \end{minipage}
        \hfill
        \begin{minipage}[t]{.49\columnwidth}
            \centering
            \includegraphics[height=3.1cm]{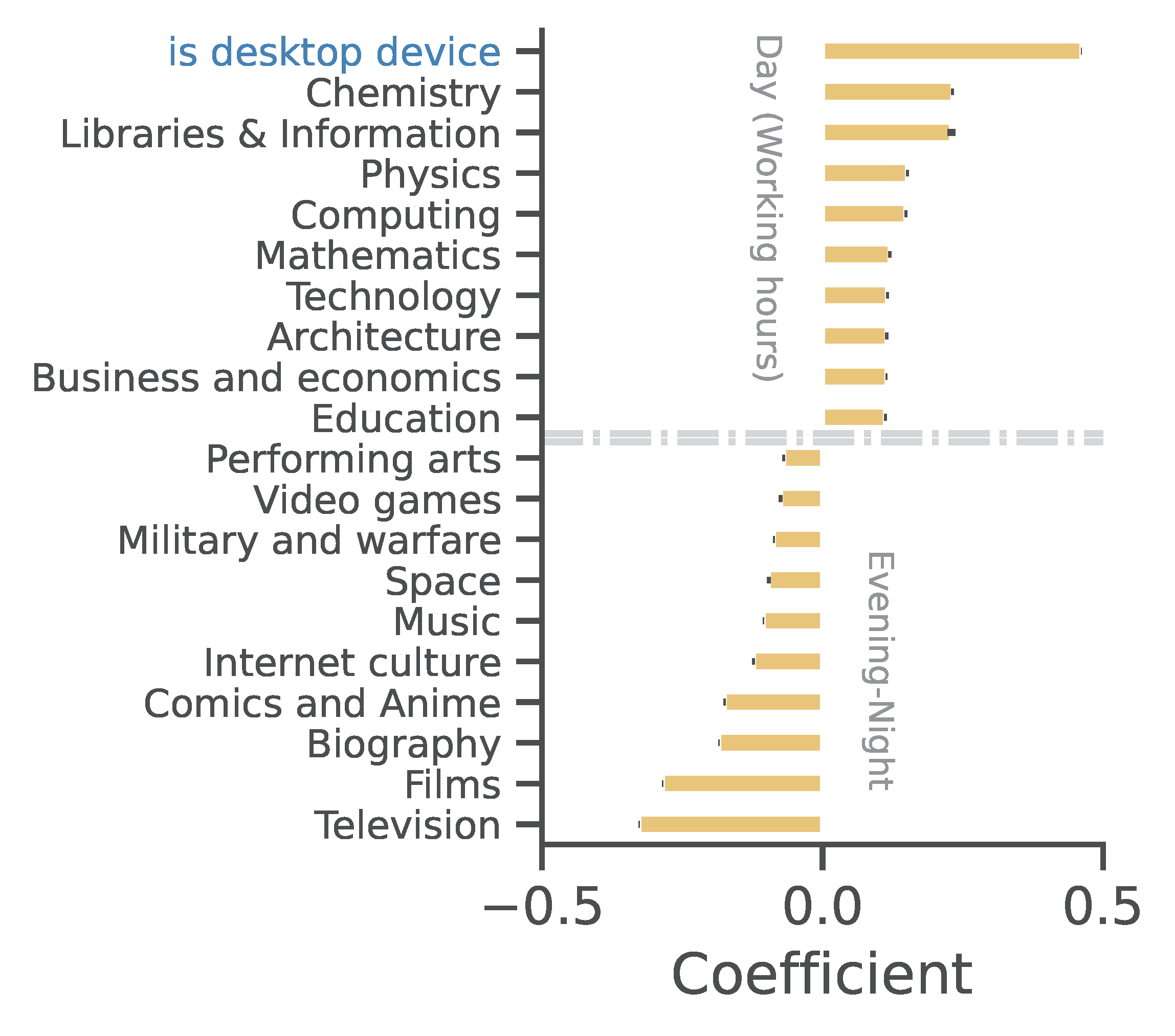}
            \subcaption{Reading sequences}\label{fig:daytime_reading}
        \end{minipage}
        % \hfill
        % \vspace{-3mm}
    \caption{Feature contributions to the logistic model predicting if the reading session started during daytime, for navigation trees (\Figref{fig:landing_regression_day}) and reading sequences (\Figref{fig:daytime_reading}).}
    \label{fig:sessions_by_time}
    \end{figure}
    
\end{minipage}

\begin{minipage}[t]{0.48\linewidth}
    \begin{figure}[H]
    % \hfill
        \begin{minipage}[t]{.49\columnwidth}
            \centering
            \includegraphics[height=3.1cm]{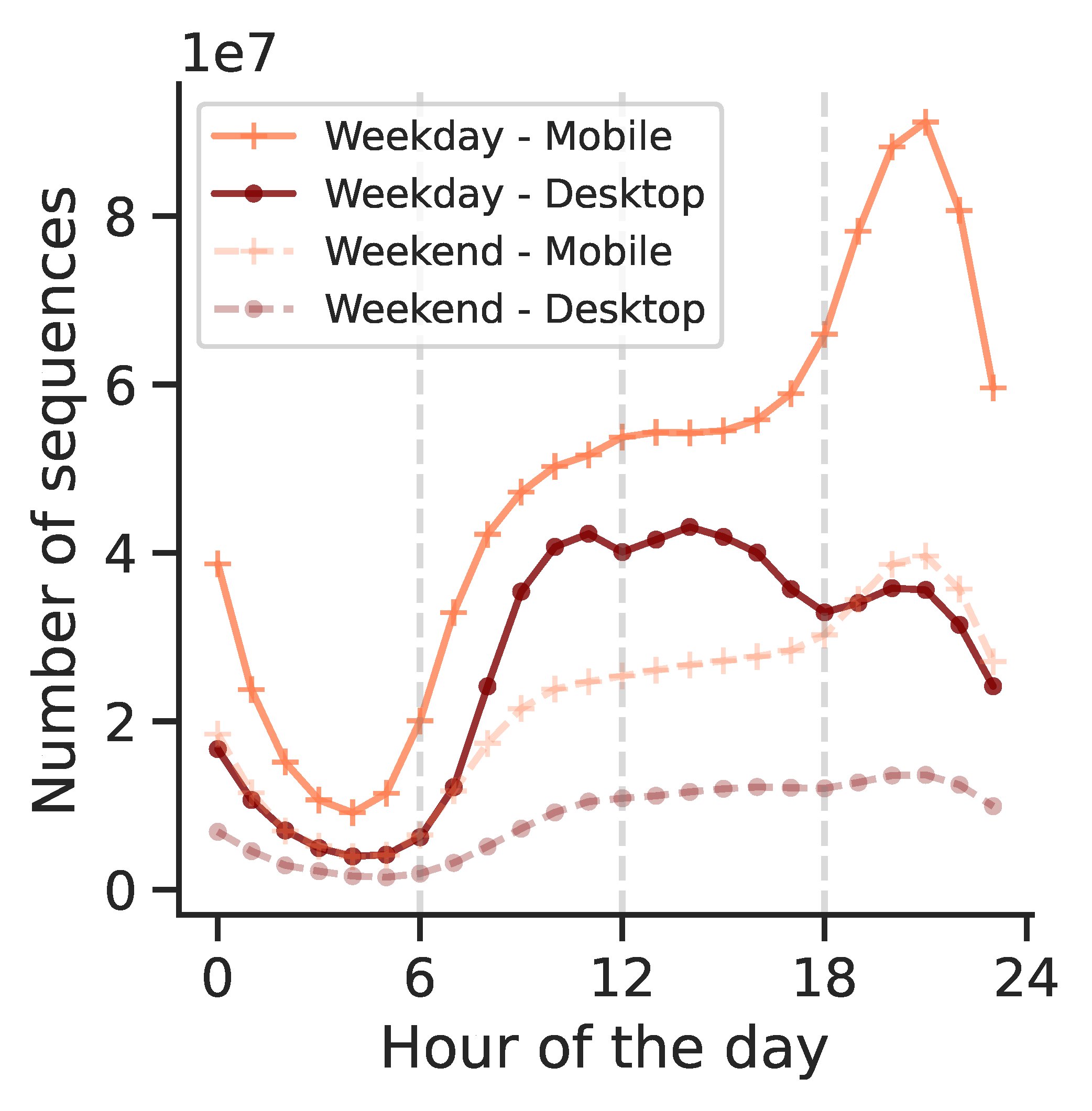}
            \subcaption{Number of reading sequences}\label{fig:total_reading_sessions_by_time}
        \end{minipage}
        \hfill
        \begin{minipage}[t]{.49\columnwidth}
            \centering
            \includegraphics[height=3.1cm]{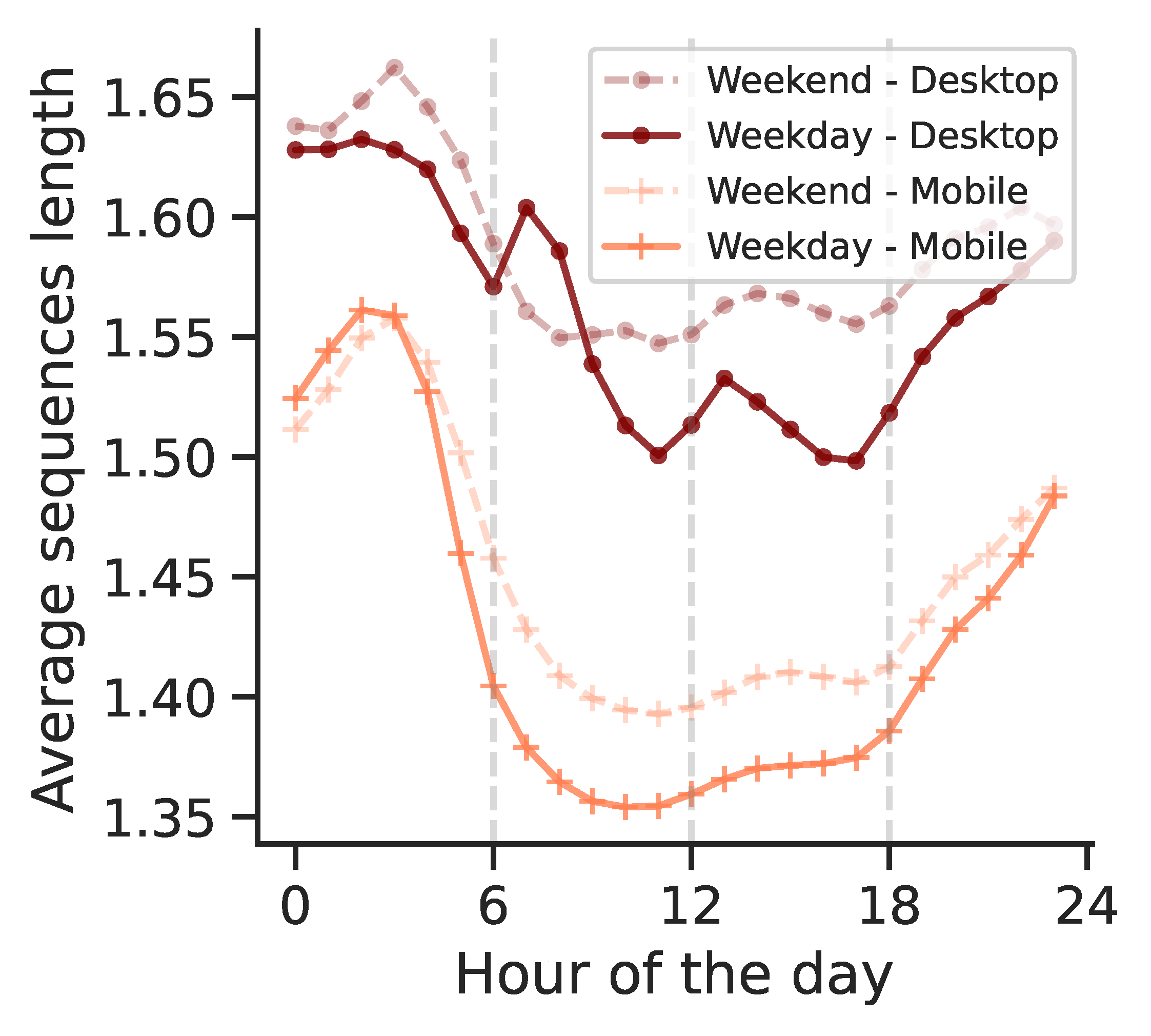}
            \subcaption{Session length by time (for reading sequences)}\label{fig:reading_sessions_len_by_time}
        \end{minipage}
        \hfill
        % \vspace{-3mm}
    \caption{Total count (\Figref{fig:total_reading_sessions_by_time}) and average length (\Figref{fig:reading_sessions_len_by_time}) of reading sequences started at different times of day.
    }
    
    \label{fig:reading_sequences_properties}
    \end{figure}
\end{minipage}
\hfill
\begin{minipage}[t]{0.48\linewidth}

    \begin{figure}[H]
        \begin{minipage}[t]{.49\columnwidth}
            \centering
            \includegraphics[height=3.1cm]{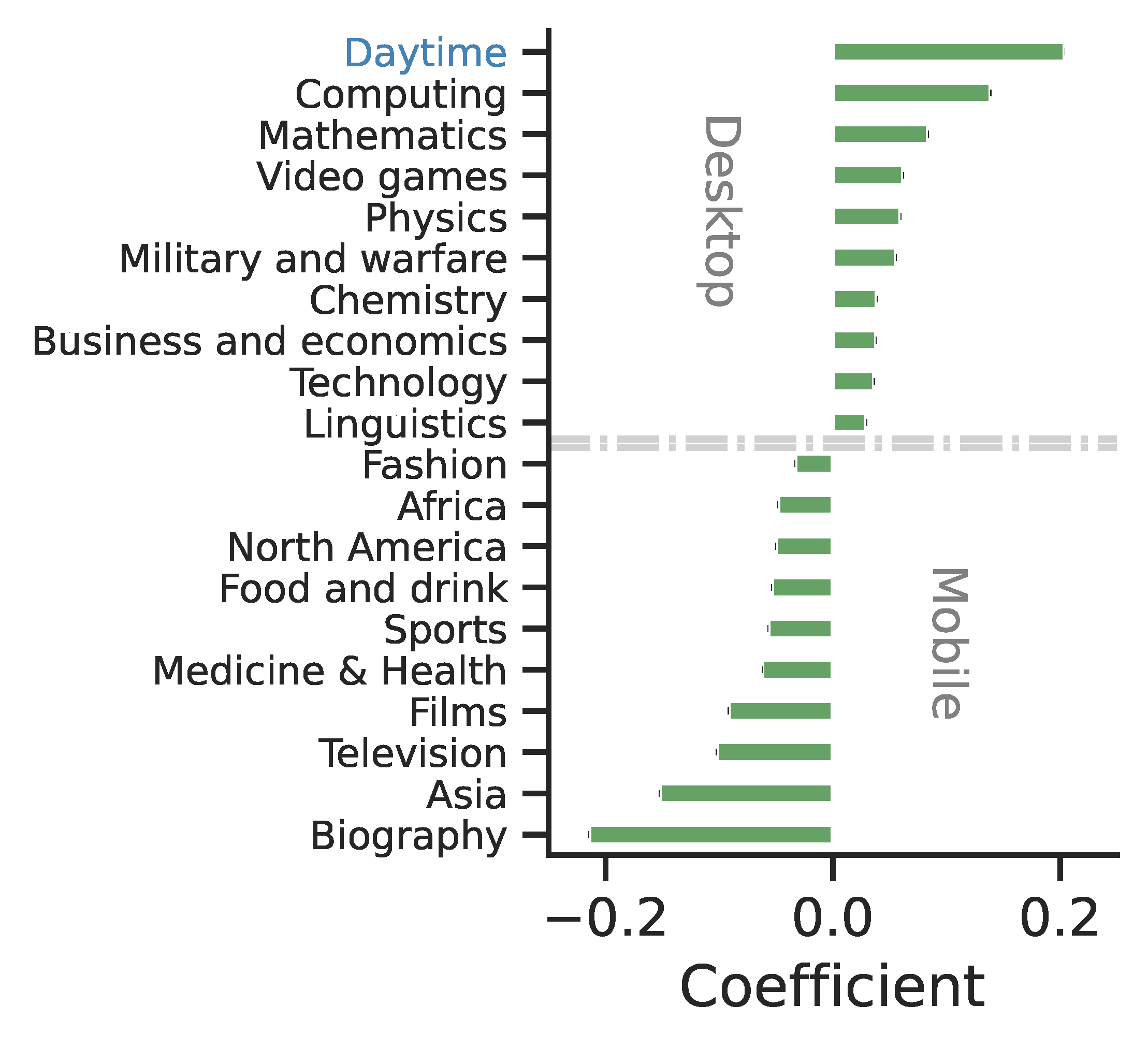}
            \subcaption{Navigation trees}\label{fig:landing_regression_device_reading}
        \end{minipage}
        \hfill
        \begin{minipage}[t]{.49\columnwidth}
            \centering
            \includegraphics[height=3.1cm]{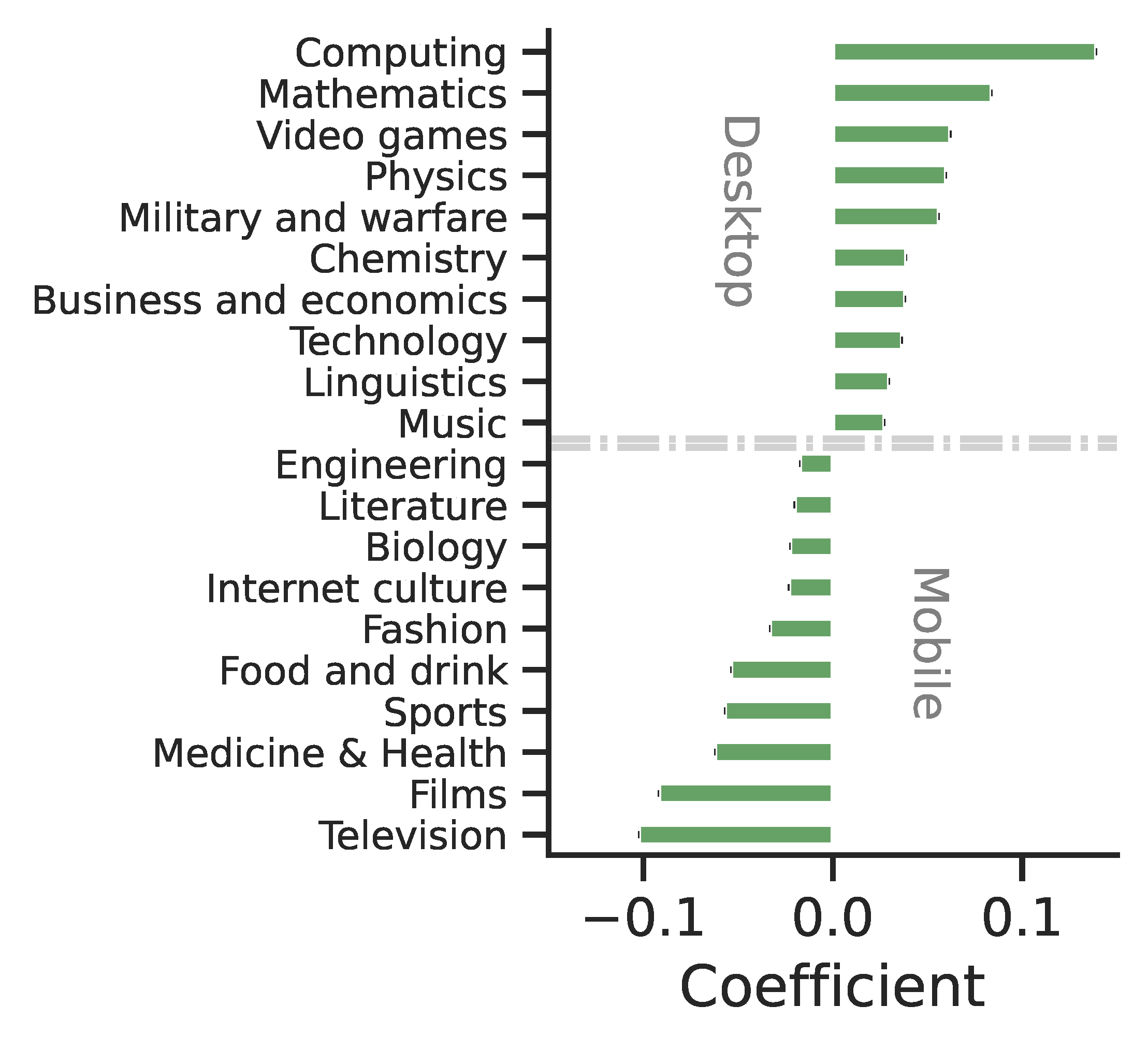}
            \subcaption{Reading sequences}\label{fig:landing_regression_device_trees}
        \end{minipage}

    \caption{Feature contributions to a logistic model predicting if the session is started from a mobile or desktop device.}
    \label{fig:landing_regression_device}
    \end{figure}
    
\end{minipage}

\subsection{Contextual properties: time and device}
\label{sec:session_context}

We study the context of a session by focusing on the time of the first pageload and the device used to access Wikipedia. This section focuses on navigation trees, but reading sequences give qualitatively similar results (\cf\ \Figref{fig:daytime_reading},  \Figref{fig:landing_regression_device_reading}).

\xhdr{Time} 
To remove confounding via different timezones, we use geolocation information to normalize the time of all pageloads to local time. The distribution of session starting times follows a regular circadian rhythm (\Figref{fig:total_sessions_by_time_trees} and \Figref{fig:total_reading_sessions_by_time}). 
%shows the distribution of the total number of navigation trees started at different hours of the day. 
Both access methods (desktop and mobile) show a similar pattern during the day, with a substantial increase of mobile sessions in the evening. 
Wikipedia has fewer sessions during weekends, but with similar temporal distributions as working days. 
The desktop distribution shows dents at 12:00 and 18:00, mirroring work rhythms with a lunch break around noon and the end of work in the evening (and possibly commuting).

In order to understand which features are associated with requests at different times of day, we fitted a 
%We deepened our investigation to understand what type of content is accessed at a different time of the day by training a 
logistic regression model to predict if a pageload was observed during the day or evening\slash night. 
We represent each pageload by its topic probabilities (obtained from ORES, \cf\ \Secref{sec:features}) and the type of device (desktop or mobile). 
Binarizing the target variable by representing daytime (9:00--18:00) as the positive class, we obtain an AUC/ROC of 0.586 on a held-out test set.
Inspecting the fitted feature weights (\Figref{fig:landing_regression_day}) shows that desktop devices and articles associated with STEM and education are associated with sessions starting during the day, whereas topics about entertainment are predictors of sessions starting during the evening or night.
%We binarize the target variable by representing the daytime (roughly representing the working hours between 9AM and 18PM) as the positive class.
%The regression has a AUC of 0.586 and \Figref{fig:landing_regression_day} shows the coefficients. Desktop devices and articles associated with STEM and Education are associated with sessions started during the day, while topics about Entertainment are predictors of sessions started during the evening and night.

\xhdr{Device}
\Figref{fig:total_sessions_by_time_trees} indicates that people prefer different devices at different times of day.
% to access different types of content, especially during night people prefer mobile devices.
Next, we study whether specific topics are associated with device types by representing each pageload with the vector of topic probabilities (obtained from ORES) and a feature indicating if the page was loaded during the daytime. We again fit a logistic regression to predict the device used, with an AUC of 0.639. Inspecting feature importance shows that people tend to access STEM and business content from desktop devices, and biographies, entertainment, and medicine from mobile devices (\Figref{fig:landing_regression_device}).

\begin{figure*}[t]
\hfill
    \begin{minipage}[t]{0.24\textwidth}
        \centering
        \includegraphics[height=3.1cm]{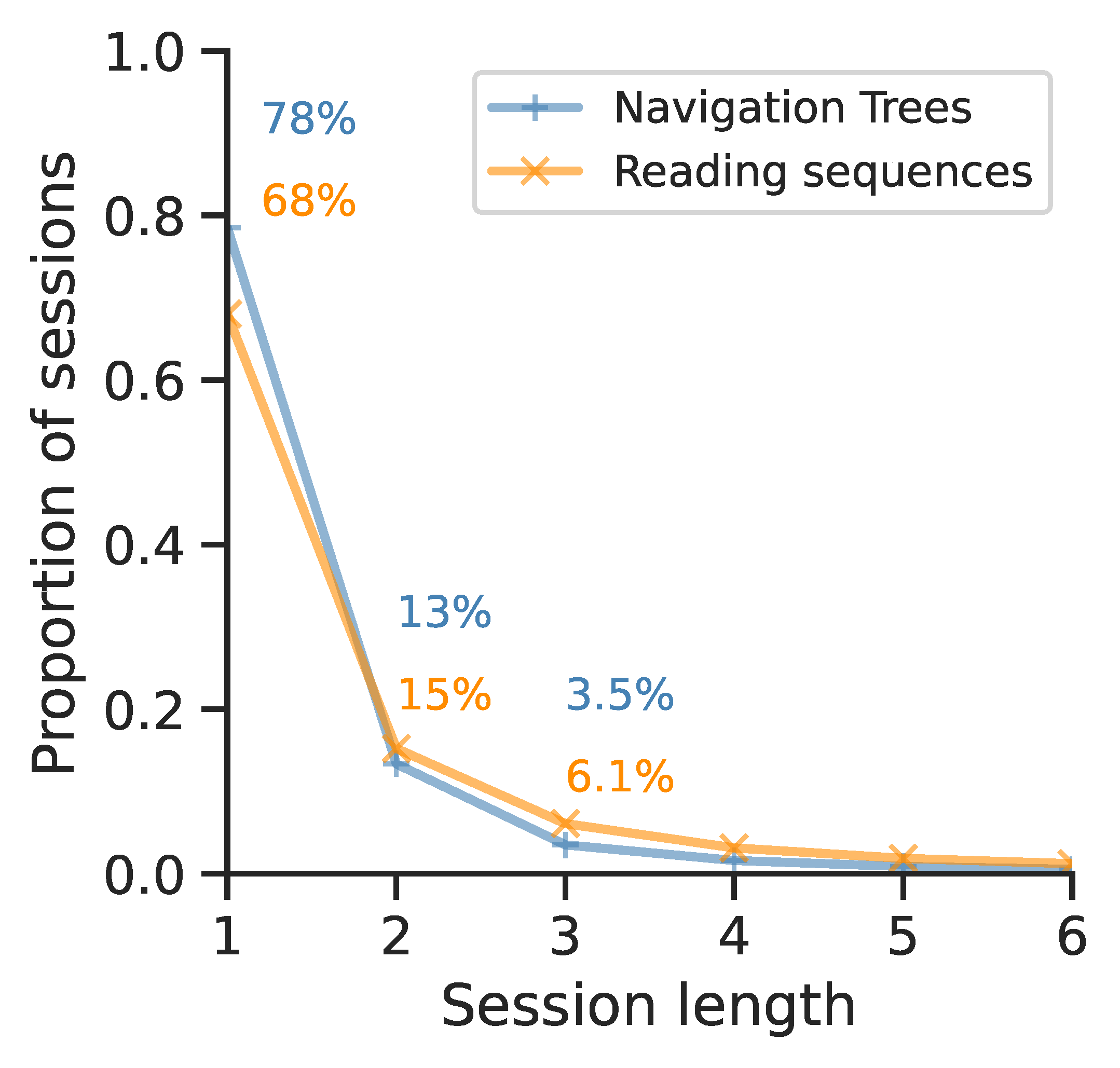}
        \subcaption{Session length histogram}\label{fig:all_sessions_len_perc}
    \end{minipage}
    \hfill
    \begin{minipage}[t]{0.24\textwidth}
        \centering
        \includegraphics[height=3.1cm]{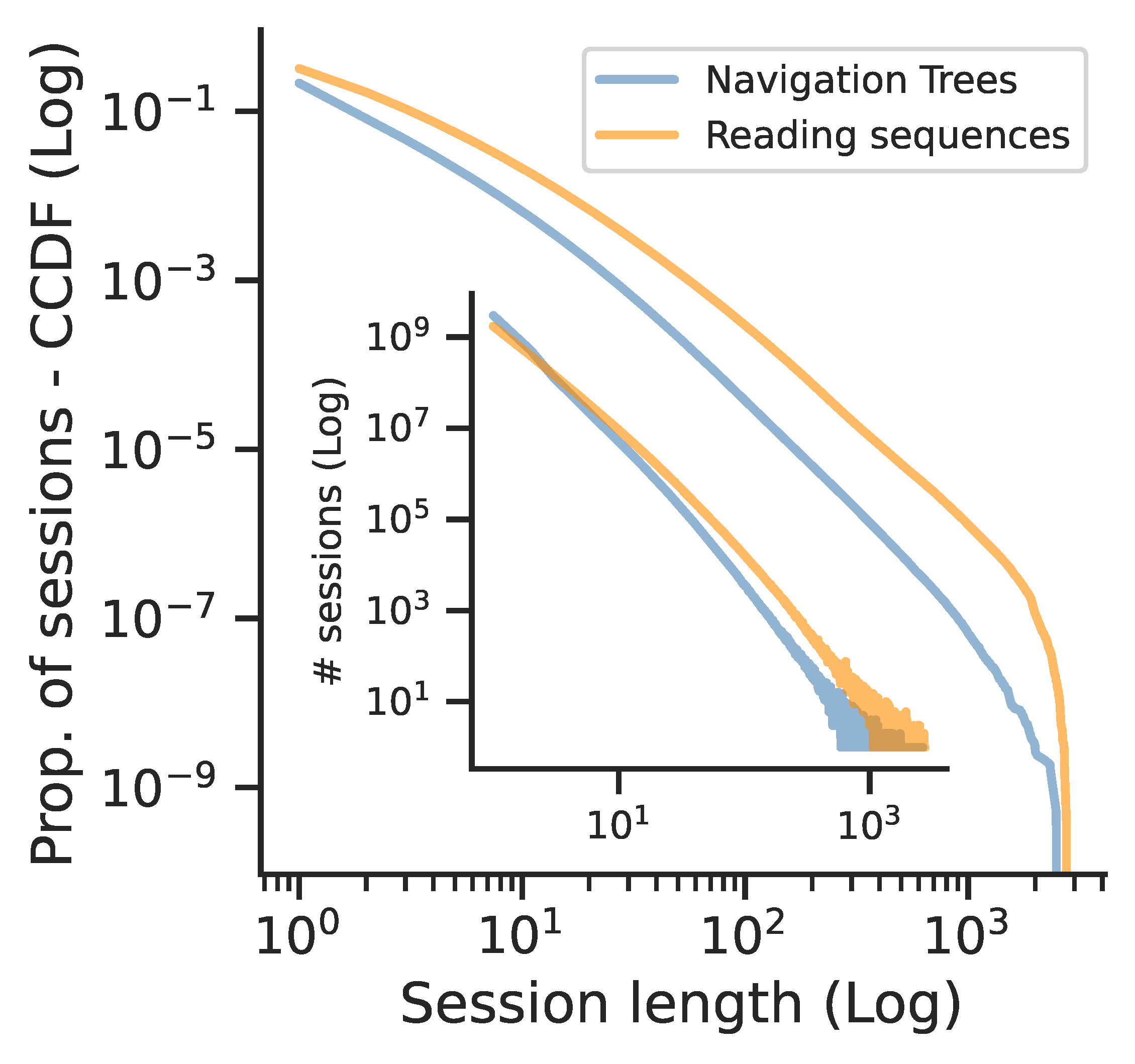}
        \subcaption{Session length CCDF}\label{fig:all_sessions_len}
    \end{minipage}
        \hfill
        % \\
    \begin{minipage}[t]{0.24\textwidth}
        \centering
        \includegraphics[height=3.1cm]{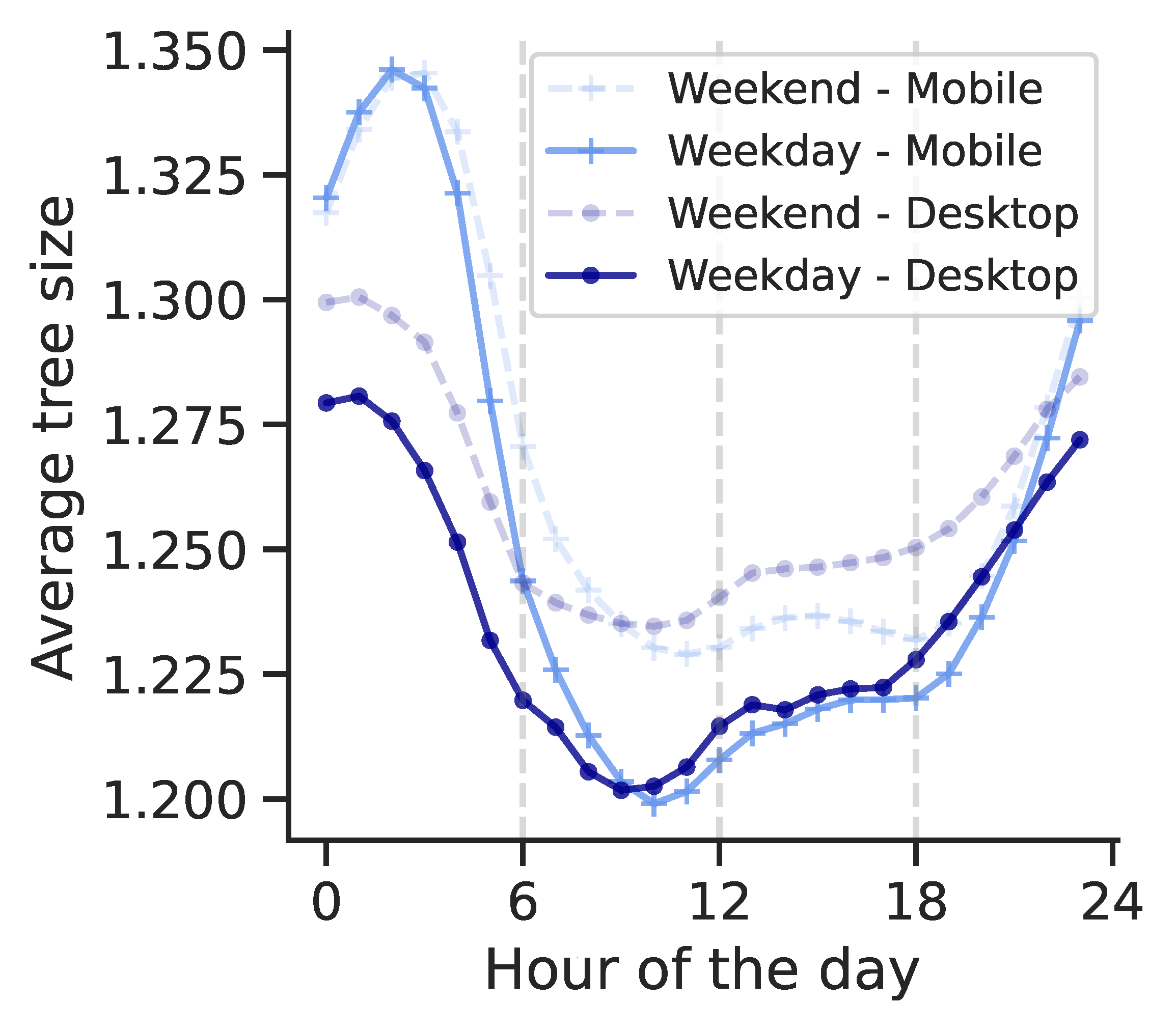}
        \subcaption{Session length by time (for navigation trees)}\label{fig:trees_size_by_time}
    \end{minipage}
    \hfill
    \begin{minipage}[t]{0.24\textwidth}
        \centering
        \includegraphics[height=3.1cm]{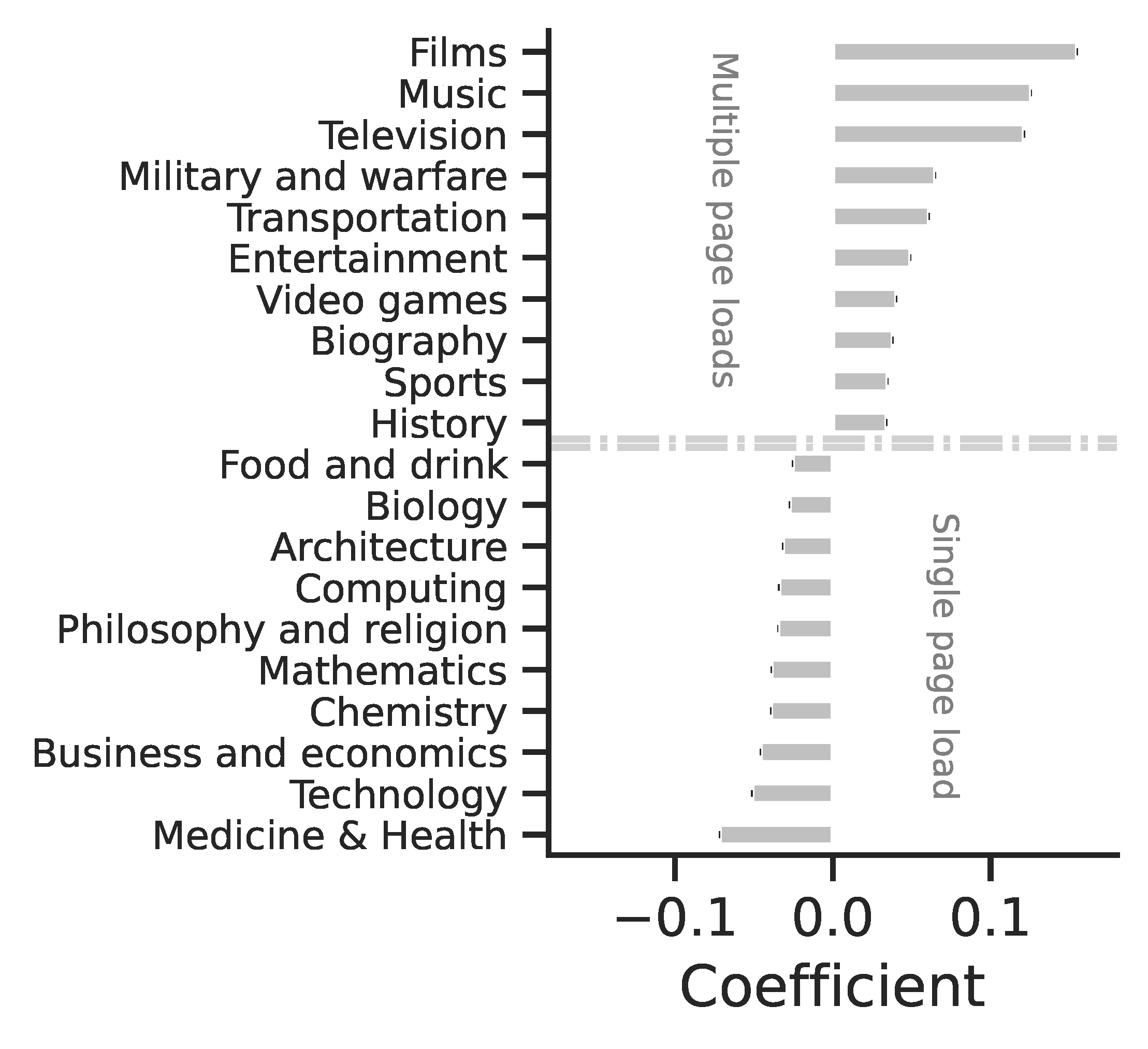}
        \subcaption{Regression coefficients}\label{fig:landing_regression_length}
    \end{minipage}
    \hfill

    % \vspace{-3mm}
\caption{
    Session-length statistics.
   %Distribution of the session lengths for navigation trees and reading sequence (\Figref{fig:all_sessions_len_perc}), complementary CDF for the session lengths with original frequency distribution as insert (\Figref{fig:all_sessions_len}), geometric mean of the size of the trees by the time of the day (\Figref{fig:trees_by_time}), and the features associated with the logistic model predicting if the session has more than one pageload (\Figref{fig:landing_regression_length}).
}
% \vspace*{-3mm}
\label{fig:session_stats}
\end{figure*}

%\subsection{Session size}
\subsection{Static properties: structure of sessions}
\label{sec:session_structure}
\xhdr{Session length}
We measure session length as the number of pageloads in the navigation tree or the reading sequence, respectively. 
Most sessions consist of a single pageload (\Figref{fig:all_sessions_len_perc}), but the length distribution also exposes a long tail (\Figref{fig:all_sessions_len}).
%Given the long-tail nature of the distribution, 
Therefore, we summarize session lengths via the geometric mean (arithmetic mean in parentheses).
% The distribution of the length of the session, i.e. the number of pageloads, shows that 78\% of the navigation trees and 68\% of the reading sequences are composed of a single pageload (\Figref{fig:all_sessions_len_perc})
% As visible in \Figref{fig:all_sessions_len}, the distribution of the length of the sessions is long-tailed. \Figref{fig:all_sessions_len_perc} shows that 78\% of the navigation trees and 68\% of the reading sequences are composed of a single pageload. 
By construction, reading sequences tend to be longer because, unlike navigation trees, they merge both external and internal transitions.

In the case of reading sequences, the average session length shows differences with respect to the access method, with an average length of 1.41 (1.99) for mobile, and 1.54 (2.40) for desktop. This difference is less pronounced for navigation trees, where mobile sessions contain on average 1.23 (1.5) articles, \vs\ 1.24 (1.5) for desktop. 
The average session length varies during the day, with readers engaging in longer sessions during the evening and night, for both navigation trees and reading sequences (\Figref{fig:trees_size_by_time} and \Figref{fig:reading_sessions_len_by_time}).

%Given these substantial differences based on time and access method, we reduce the effect of these confounding with a matched study. For each session started from a desktop device, we selected a random session originated from a mobile device started from the same article, the same country, the same hour of the day, and if it was weekend or not.
%Using a sample of 32M random pairs, the difference between desktop and mobile for the reading sequences is maintained with 1.43 (2.03) pageloads for mobile and 1.58 (2.47) for desktop. In the case of navigation trees, the gap remains minimal, having 1.26 (1.56) on articles on average for mobile and 1.27 (1.58) for desktop. All differences are statistically significant with Wilcoxon signed-rank test ($p < \num{1e-29}$).

\begin{table}[t]
\centering

% \small
\begin{tabular}{lll|ll}
 & \multicolumn{4}{c}{\textbf{Tree size}} \\
 & \multicolumn{2}{c}{\textbf{Top 10 (larger trees)}} & \multicolumn{2}{c}{\textbf{Bottom 10 (smaller trees)}} \\
 \hline
 & \textit{1.377} & Films & \textit{1.152} & Earth and environment \\
 & \textit{1.373} & Entertainment & \textit{1.148} & Food and drink \\
 & \textit{1.340} & Television & \textit{1.145} & Biology \\
 & \textit{1.327} & Military and warfare & \textit{1.138} & Technology \\
 & \textit{1.324} & Music & \textit{1.128} & Physics \\
 & \textit{1.295} & Comics and Anime & \textit{1.122} & Software \\
 & \textit{1.284} & History & \textit{1.114} & Medicine \& Health \\
 & \textit{1.272} & Biography & \textit{1.112} & Computing \\
 & \textit{1.269} & Sports & \textit{1.104} & Mathematics \\
 & \textit{1.264} & Transportation & \textit{1.100} & Chemistry
\end{tabular}
    \caption{Top and bottom 10 topics with respect to (geometric) average tree size (geographical topics excluded).}
    \label{table:tree_size}
\end{table}

To understand what properties are associated with short sessions
% navigation
consisting of a single pageload, we fitted a logistic regression to predict if the reader will continue after loading the first page in a navigation tree (results are qualitatively identical for reading sequences%; \Figref{fig:session_len_reading}
), representing each first pageload with its topic probabilities (obtained from ORES), device type, and time of day, and obtaining a model with an AUC/ROC of 0.606 on a held-out test set. 
Inspecting the coefficients of the regression (\Figref{fig:landing_regression_length}), we find that longer [shorter] sessions are associated with topical content around entertainment [STEM and medicine].
This observation is corroborated by the substantial difference in average navigation tree size across topics (\Tabref{table:tree_size}).
%content associated with navigation beyond the first page are relative to entertainment, while STEM and medicine articles are associated with single-page sessions . 

\begin{figure}[t]
\hfill
\centering
    \begin{minipage}[t]{.85\columnwidth}
        % \raggedleft
        \includegraphics[height=3.1cm]{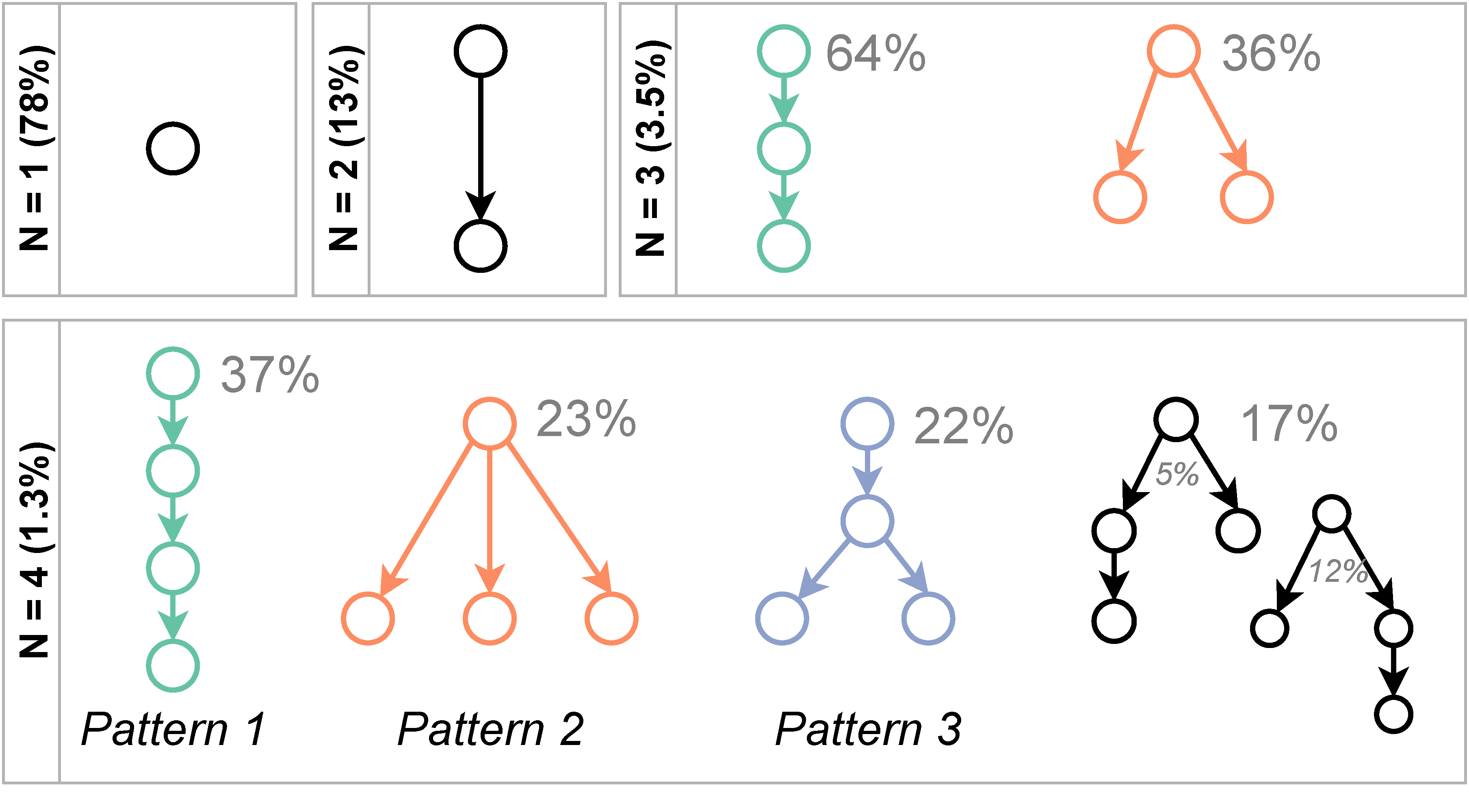}
    % \end{minipage}
    % % \hfill
    \hspace{5mm}
    % \begin{minipage}[t]{.3\columnwidth}
    %     \raggedright
        \includegraphics[height=3.1cm]{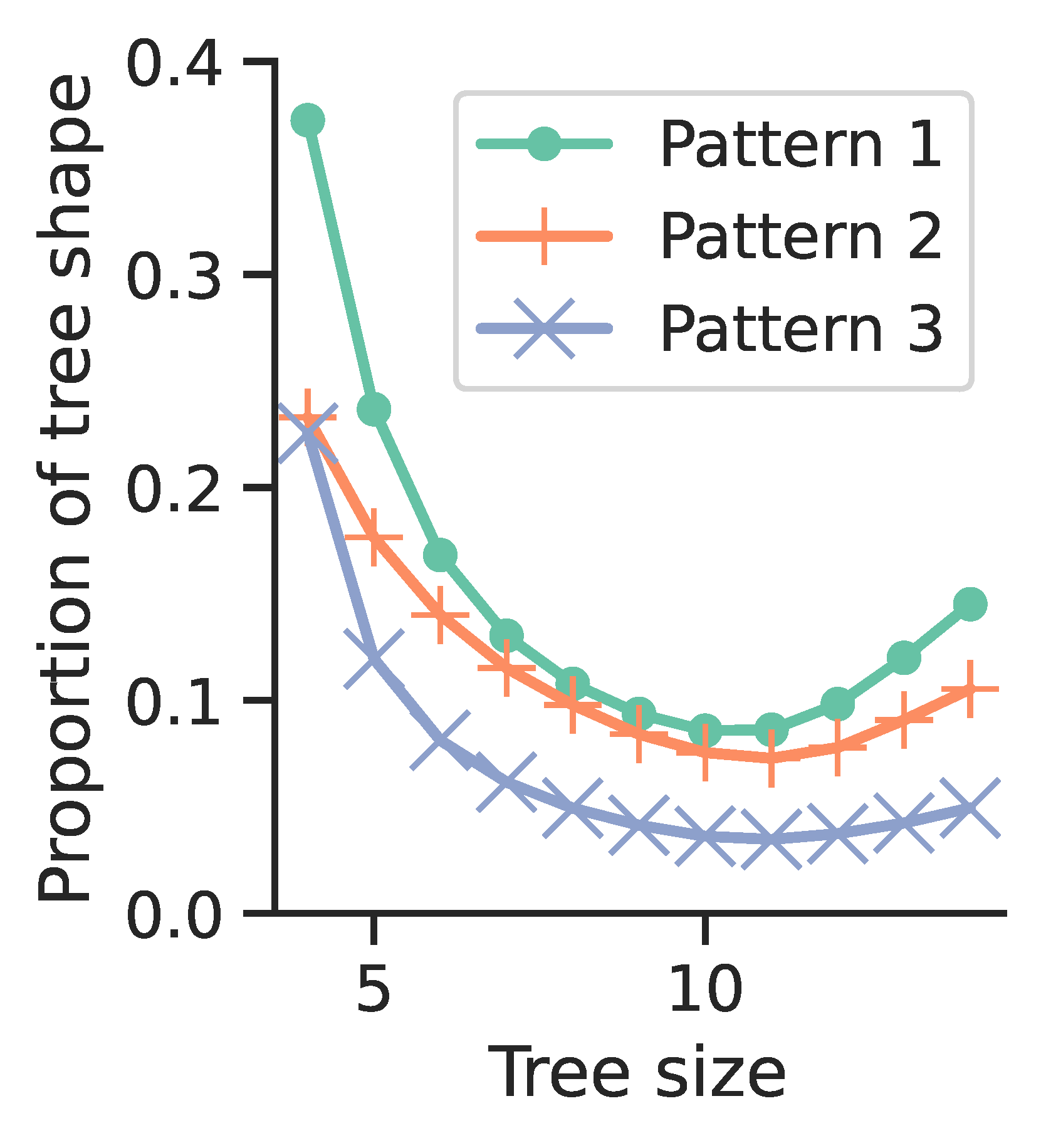}
        % \subcaption{Reading sequences}\label{fig:total_sessions_by_time}
    \end{minipage}
    \hfill
% \vspace{-3mm}
\caption{
Shape of navigation trees. Frequency of patterns for trees size $N\leq 4$ (left panel). Dominance of top three patterns (see main text) for larger trees (right panel). 
%Shape of the trees. Upper figure: shape of trees up to size 4. Lower figure: Distribution of the top 3 shapes in function of the number of pageloads.
}
\label{fig:tree_shapes}
    % \vspace{-5mm}
\end{figure}

%\subsection{Structure of navigation trees}
\xhdr{Shape of navigation trees}
% \label{sec:session_structure}
In order to better understand how readers navigate the link network, 
%To get insights into the type of reading and understanding the navigation dynamics on the links network, 
we analyze the shape of navigation trees (in contrast, the shape of reading sequences is, by construction, always a linear chain).
%\Figref{fig:tree_shapes} summarises the patterns and the frequency of the trees up to size 4.
The three most common patterns (\Figref{fig:tree_shapes}, left) are described as follows, in order of decreasing frequency:
(1)~a linear chain of pageloads;
(2)~fanning out from one page to several different pages, e.g., by opening multiple tabs or rolling back and selecting a different path;
(3)~a combination of the two (one-step chain followed by fanning out).
These three patterns remain the most frequent for all tree sizes (\Figref{fig:tree_shapes}, right).
%and as visible in the figure, it remains the most common for all tree sizes.

\begin{figure*}[t]
    \centering
    \includegraphics[width=0.98\textwidth]{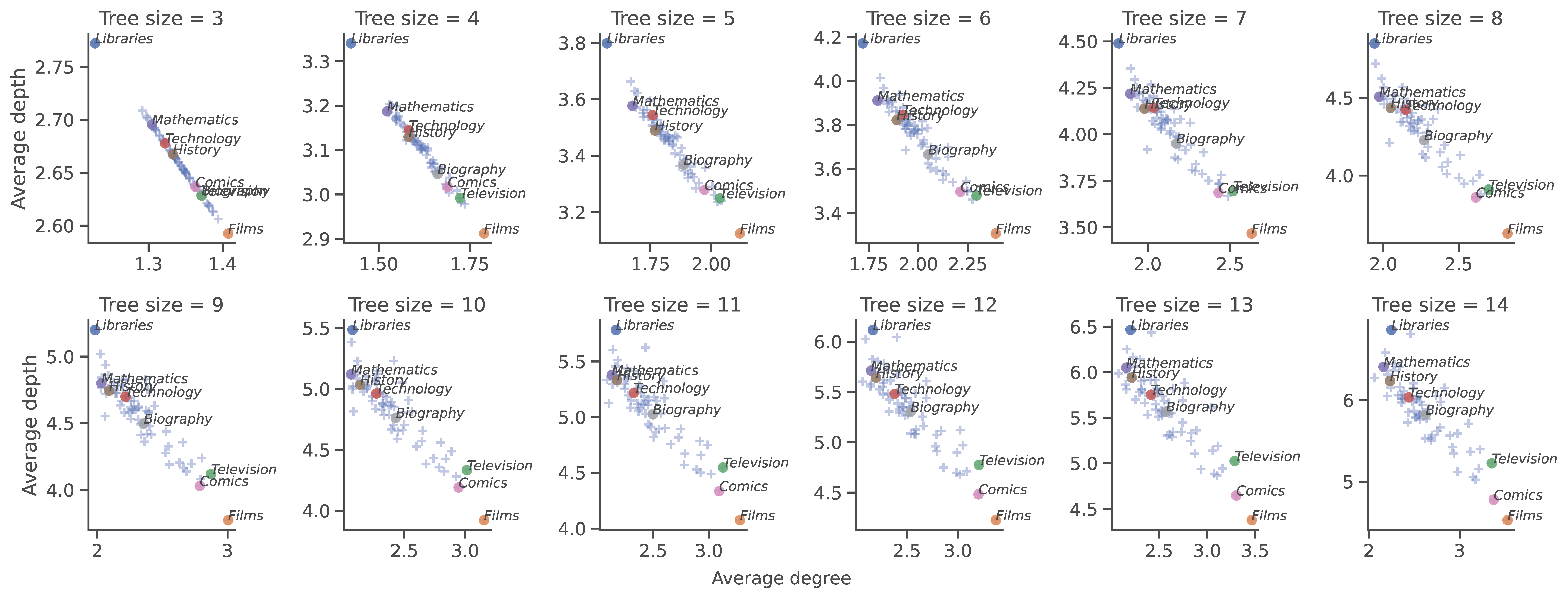}
    \caption{Relation between the average depth and average degree for navigation trees of different sizes.}
    \label{fig:tree_shapes_by_topic}
\end{figure*}

\begin{table}[t]
% \centering
\small
\begin{tabular}{lll|lll}
\multicolumn{3}{c}{\textbf{Top 10 (wider trees)}} & \multicolumn{3}{c}{\textbf{Bottom 10 (deeper trees)}} \\
\textbf{Rank (mean)} & \textbf{SD} & \textbf{Root topic} & \textbf{Rank (mean)} & \textbf{SD} & \textbf{Root topic} \\
1.00 & 0.00 & Films & 27.42 & 2.72 & Linguistics \\
2.50 & 0.87 & Television & 29.42 & 0.95 & Earth and environment \\
3.58 & 0.76 & Entertainment & 29.50 & 1.19 & Space \\
4.50 & 1.85 & Comics and Anime & 30.08 & 2.78 & History \\
4.67 & 1.31 & Education & 31.92 & 1.11 & Computing \\
6.58 & 1.98 & Video games & 32.92 & 1.55 & Software \\
7.92 & 2.43 & Literature & 34.67 & 1.75 & Chemistry \\
8.50 & 2.36 & Fashion & 34.75 & 1.30 & Physics \\
8.83 & 1.07 & Performing arts & 35.50 & 1.26 & Mathematics \\
10.42 & 2.29 & Internet culture & 35.67 & 1.65 & Libraries \& Information
\end{tabular}
    \caption{Rank with respect to average degree of navigation trees, by topic (geographical topics excluded). A separate rank was computed per tree size (3--15), and arithmetic means over tree sizes are reported, alongside standard deviations (SD).}
    \label{table:average_topic_rank}
\end{table}

We further characterize the different strategies associated with navigation trees in terms of tree depth (i.e., average length of paths from the root to the leaves) and breadth (i.e., average out-degree of non-leaves in the tree) for trees starting with different topics.
Noting that the two metrics are almost perfectly anti\hyp correlated and that the relative ordering of topics is stable across all tree sizes (\Figref{fig:tree_shapes_by_topic}), we define an aggregate tree-breadth ranking for each topic based on the average rank across tree sizes (\Tabref{table:average_topic_rank}). This shows that entertainment topics are associated with wider trees with higher branching, and STEM topics are characterized by deeper trees with a more chain-like structure.

%Additionally, to understand the shapes of these trees, we obtained the average degree of the trees by limiting the size to a fixed range between 3 and 15 page views. Branching in navigation can be caused by opening multiple tabs or rolling back and selecting a different path. To make the average degree comparable for different tree sizes, we sorted the topics by average degree and averaged the positions across all the trees sizes. The standard deviation of the individual topic position shows that the ranking is stable, and STEM topics tend to have deep sequences (Pattern 1 in \Figref{fig:tree_shapes}) with less branching than in entertainment content. These results suggest that entertainment-related topics are associated with a broad exploitative pattern, while in STEM articles, readers have sequential navigation patterns. A similar distinction was reported in previous work \cite{singer_why_2017} as navigation guided by boredom or focused on scientific content.

% Additionally, to understand the shapes of these trees, we obtained the average degree of the trees by limiting the size to a fixed range between 3 and 15 page views. Branching can be caused by multi-tab or rollback and selection of a different path. Table \ref{table:average_degree} summarises the results and it shows that when the navigation from STEM topic is long, it tends to be deep as a sequence with less branching than in entertainment. This results suggest that entertainment have a more exploitative pattern, while STEM readers have a more focused navigation.

\begin{figure}[t]
    \centering
    \includegraphics[width=0.6\linewidth]{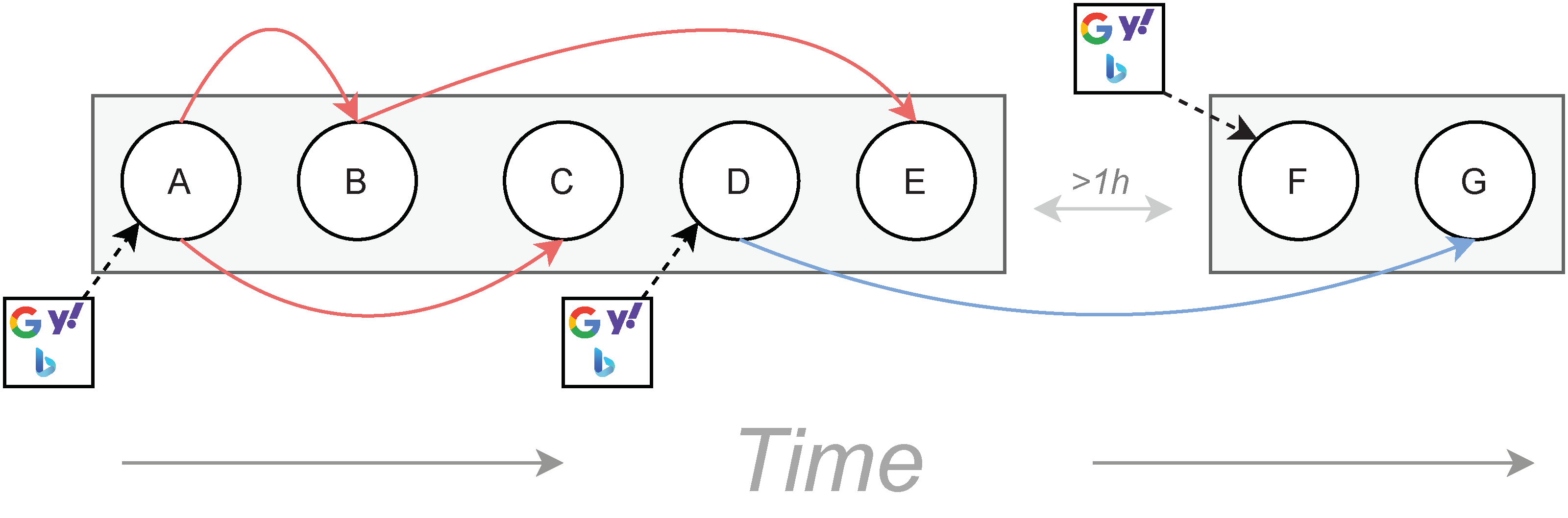}
    \caption{This set of log events yields three navigation trees, represented by arrows and composed of ABCE, DG, and F. The reading sequences method creates two sessions represented as gray boxes: ABCDE and FG. Square boxes are clicks from external origins.}
    \label{fig:session_types}
\end{figure}

\subsection{Dynamic properties: within-session article-property evolution}
\label{sec:session_evolution}

%To understand navigation dynamics, we trace a range of quantities as a function of the position in the session.
To shed light on navigation dynamics, we track the evolution of different article properties within sessions.  
Our evolution analysis revolves around three domains: topic space (distance from the first and previous articles), quality, and network centrality (out-degree and PageRank).
% We describe how the session evolved and how these properties change bases on the position in the session and the total session length.
%We compare sessions obtained with reading sequences and trees. 
Here, reading sequences are represented as defined above,
whereas a navigation tree is represented by the linear path from the root to the temporally last leaf, from where the reader ceased to click further via internal links.
% via internal links until the reader did not click any more links. %-- this corresponds a linear This sequence represents the path composed by internal clicks that leads to terminating the navigation -- where the reader did not click any link. 
% We compare sessions obtained with reading sequences and trees. Navigation trees are represented as the path that lead to the (temporally) last node. This sequence represents the path that lead to conclude the navigation -- where the reader did not click any link. Using a random path to a leaf of the tree shows qualitatively the same findings described in this section.

% It is important to note that the 2 different aggregation methods result in different ordering: \eg, a pageload in position 1 for the navigation trees can be in position 5 in reading sequences. The last node of the sequences can be interpreted as: the reader stop the navigation using the links in the page (navigation trees) or the reader stop to use Wikipedia for at least 1 hour (reading sequences).

It is important to note that these two approaches can produce different sequences of pageloads: \eg, a pageload in position 1 of a navigation tree could be in position 4 of a reading sequence (as in \Figref{fig:session_types}). 
Also, the last pageload of each sequence can have different interpretations: for navigation trees, the reader stopped link-based navigation on that page, whereas for reading sequences, the reader did not load a Wikipedia page for at least one hour.

\begin{figure*}[t]
% \hfill
    \begin{minipage}[t]{0.19\textwidth}
        \centering
        \includegraphics[height=2.7cm]{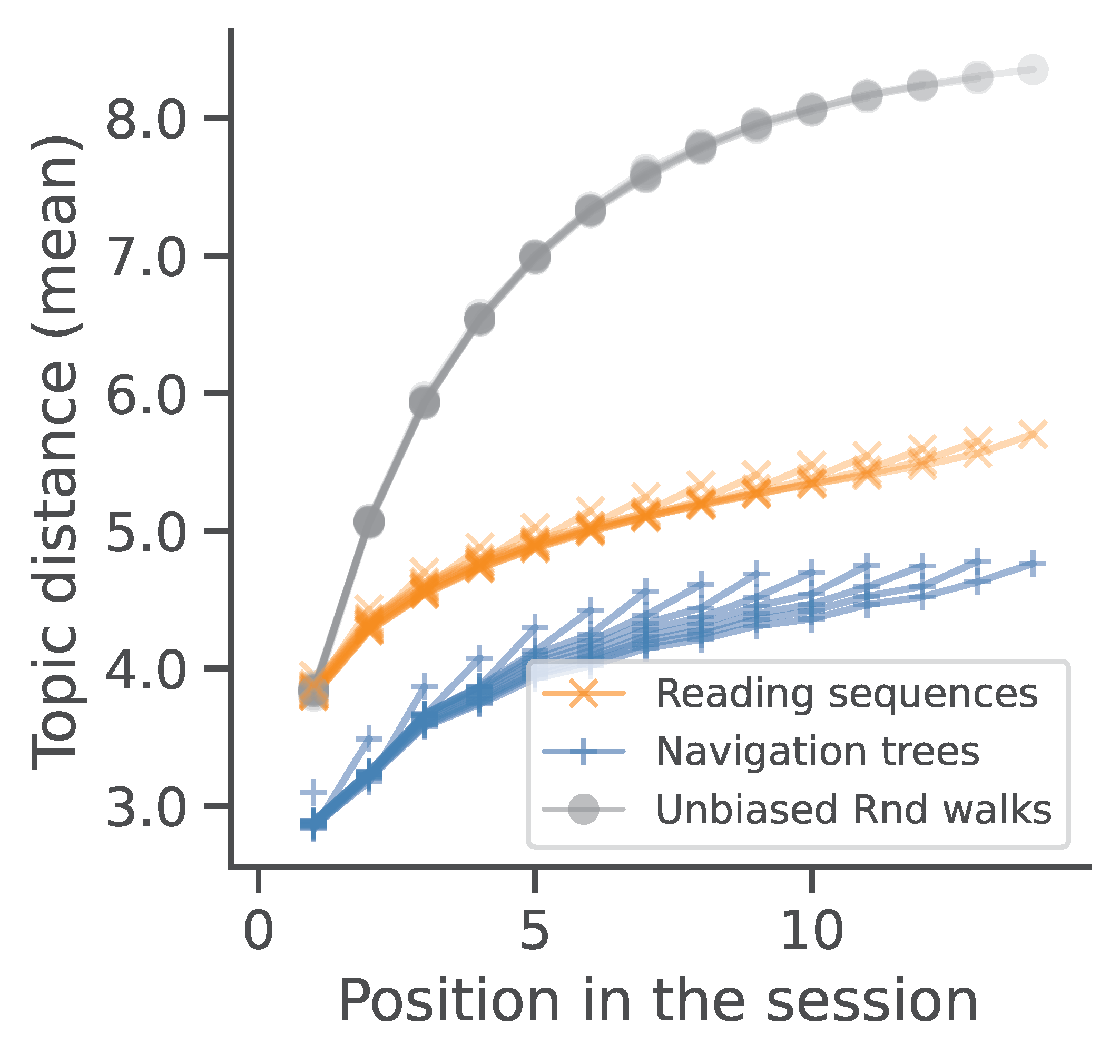}
       \subcaption{From first article}\label{fig:topic_distance_first}
    \end{minipage}
    \hfill
    \begin{minipage}[t]{0.19\textwidth}
        \centering
        \includegraphics[height=2.7cm]{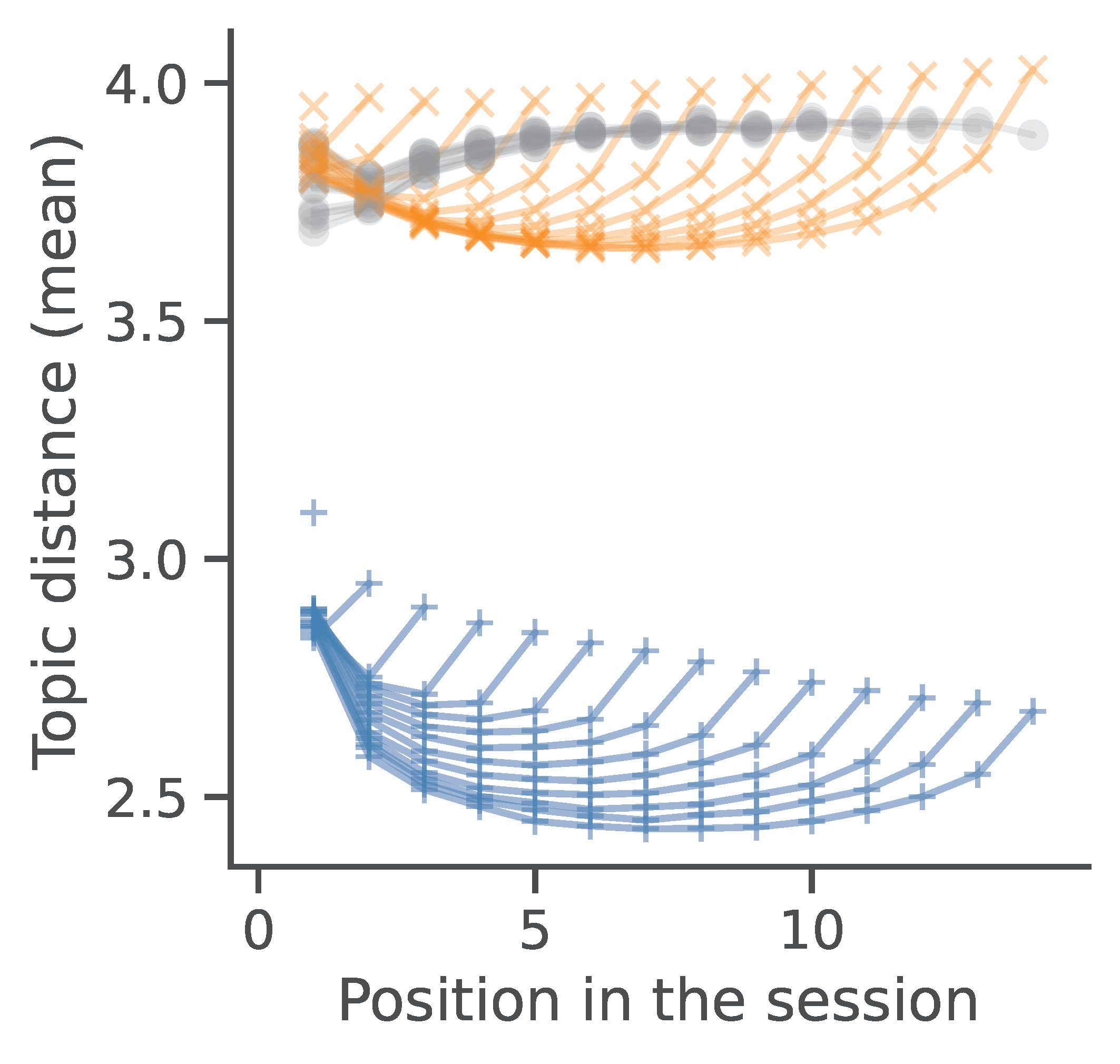}
        \subcaption{From previous article}\label{fig:topic_distance_prev}
    \end{minipage}
    \hfill
    \begin{minipage}[t]{0.19\textwidth}
        \centering
        \includegraphics[height=2.7cm]{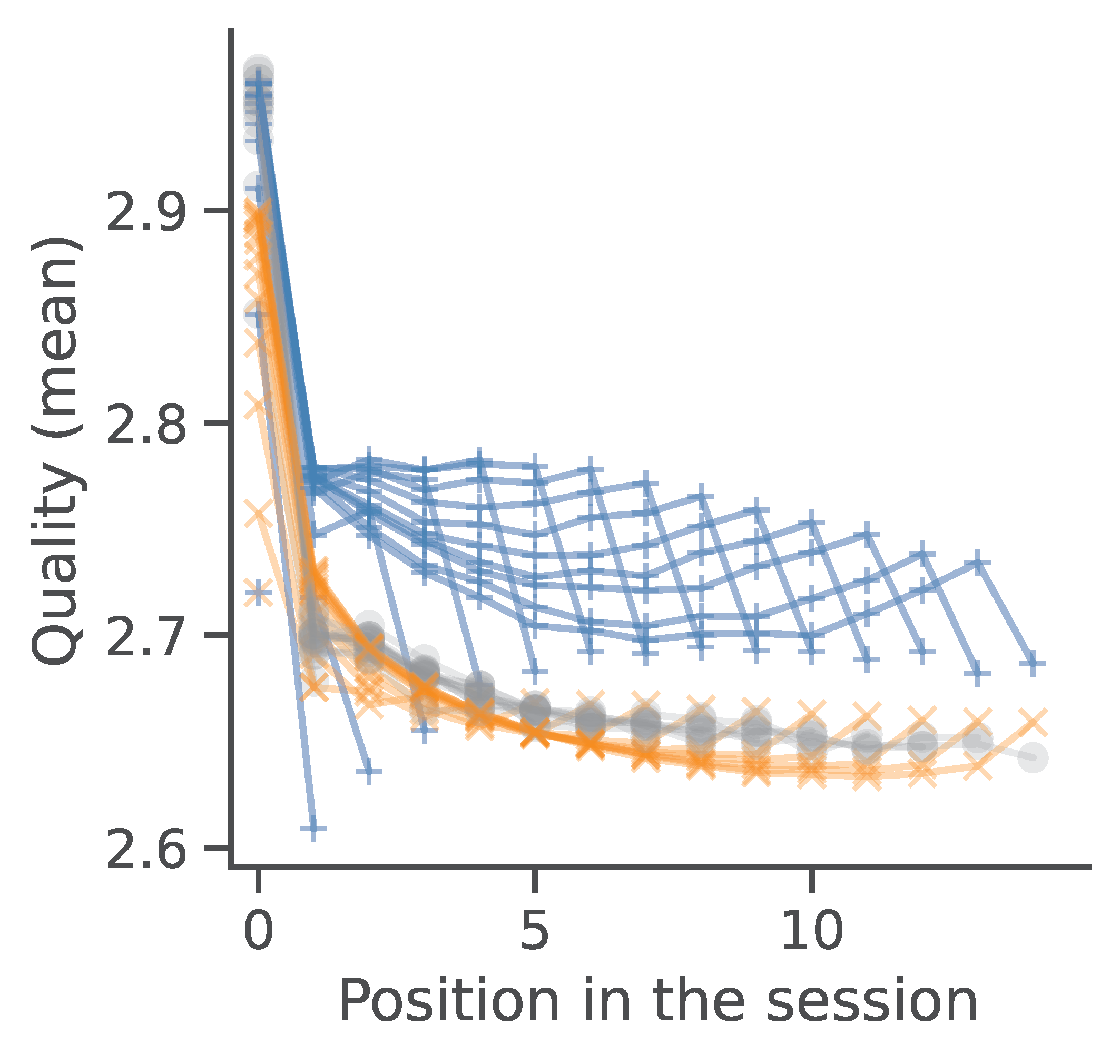}
        \subcaption{Quality}\label{fig:quality}
    \end{minipage}
    \hfill
    \begin{minipage}[t]{0.19\textwidth}
        \centering
        \includegraphics[height=2.7cm]{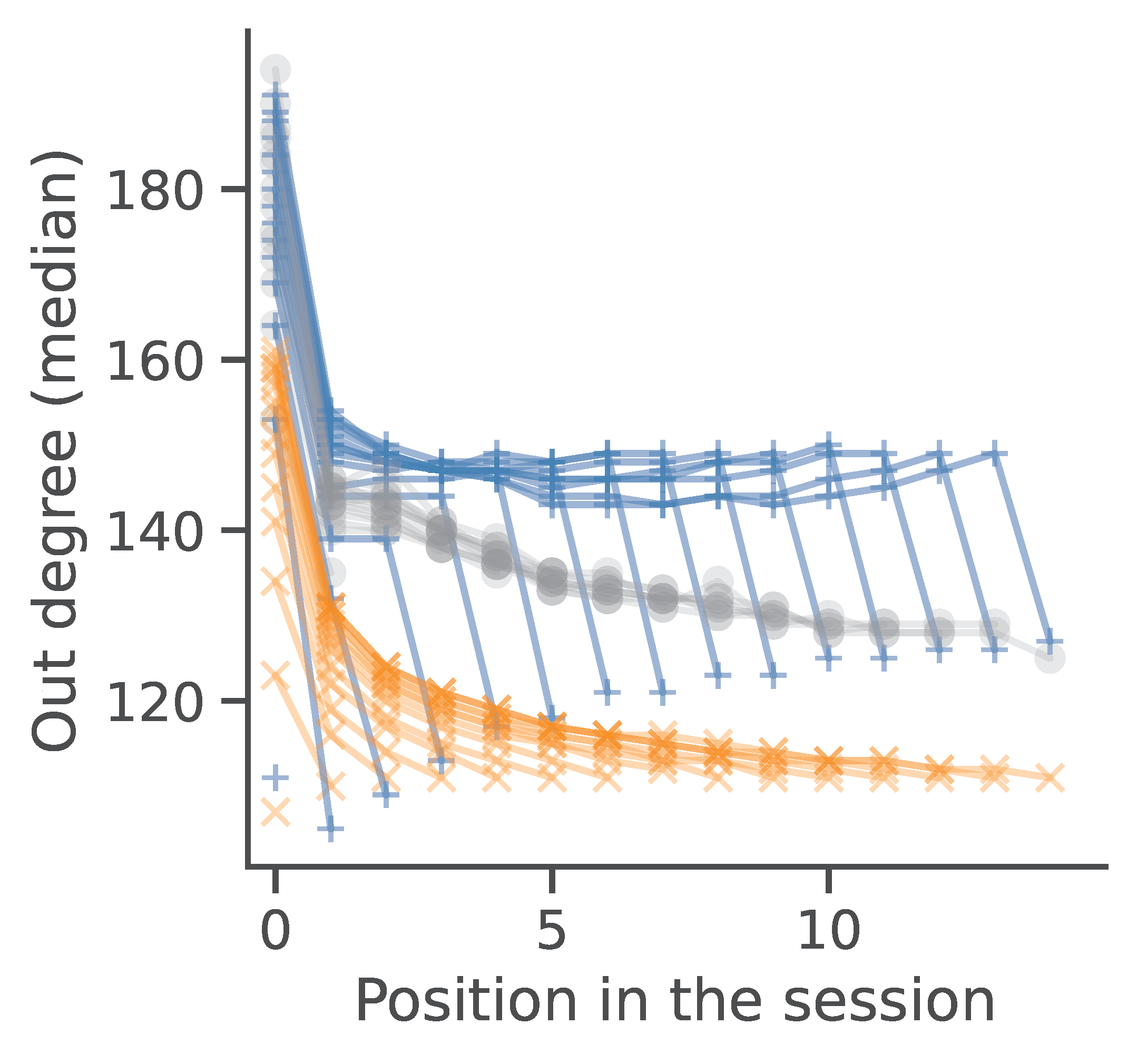}
        \subcaption{Out-degree}\label{fig:out_degree}
    \end{minipage}
    \hfill
    \begin{minipage}[t]{0.19\textwidth}
        \centering
        \includegraphics[height=2.7cm]{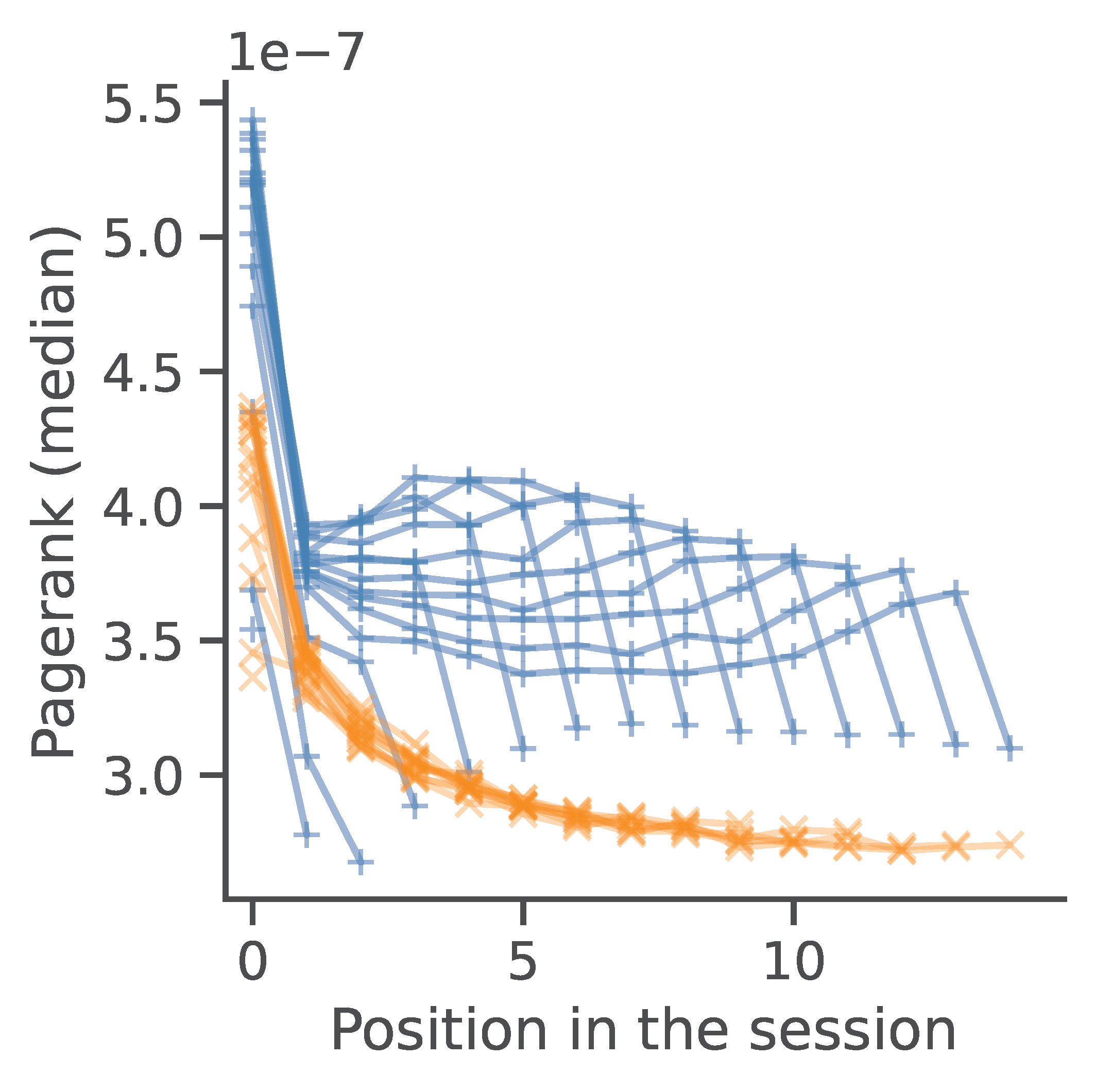}
        \subcaption{Pagerank}\label{fig:pagerank}
    \end{minipage}
    \hfill
% \vspace{-2mm}
\caption{
Within-session evolution of five article properties. Each curve represents sessions of different lengths.
%Comparison of the evolution of five different properties by using the Navigation Trees and Reading Sequences. Grey curves represent the unbiased random walker. Each curve represents sessions of different lengths
}
\label{fig:evolution}
% \vspace{-2mm}
\end{figure*}

\begin{figure*}[t]
% \hfill
    \begin{minipage}[t]{0.19\textwidth}
        \centering
        \includegraphics[height=2.7cm]{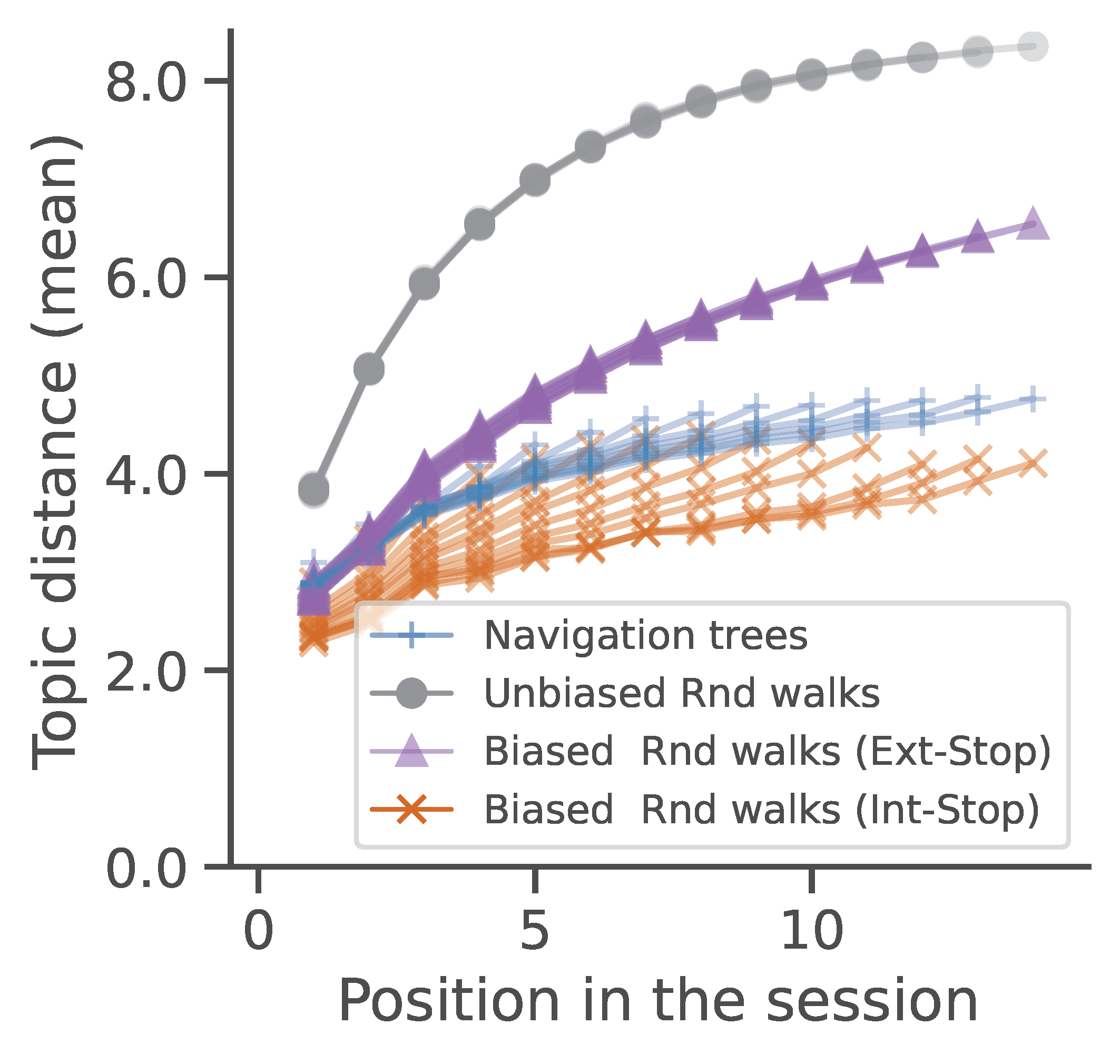}
       \subcaption{From first article}\label{fig:topic_distance_first_rw}
    \end{minipage}
    \hfill
    \begin{minipage}[t]{0.19\textwidth}
        \centering
        \includegraphics[height=2.7cm]{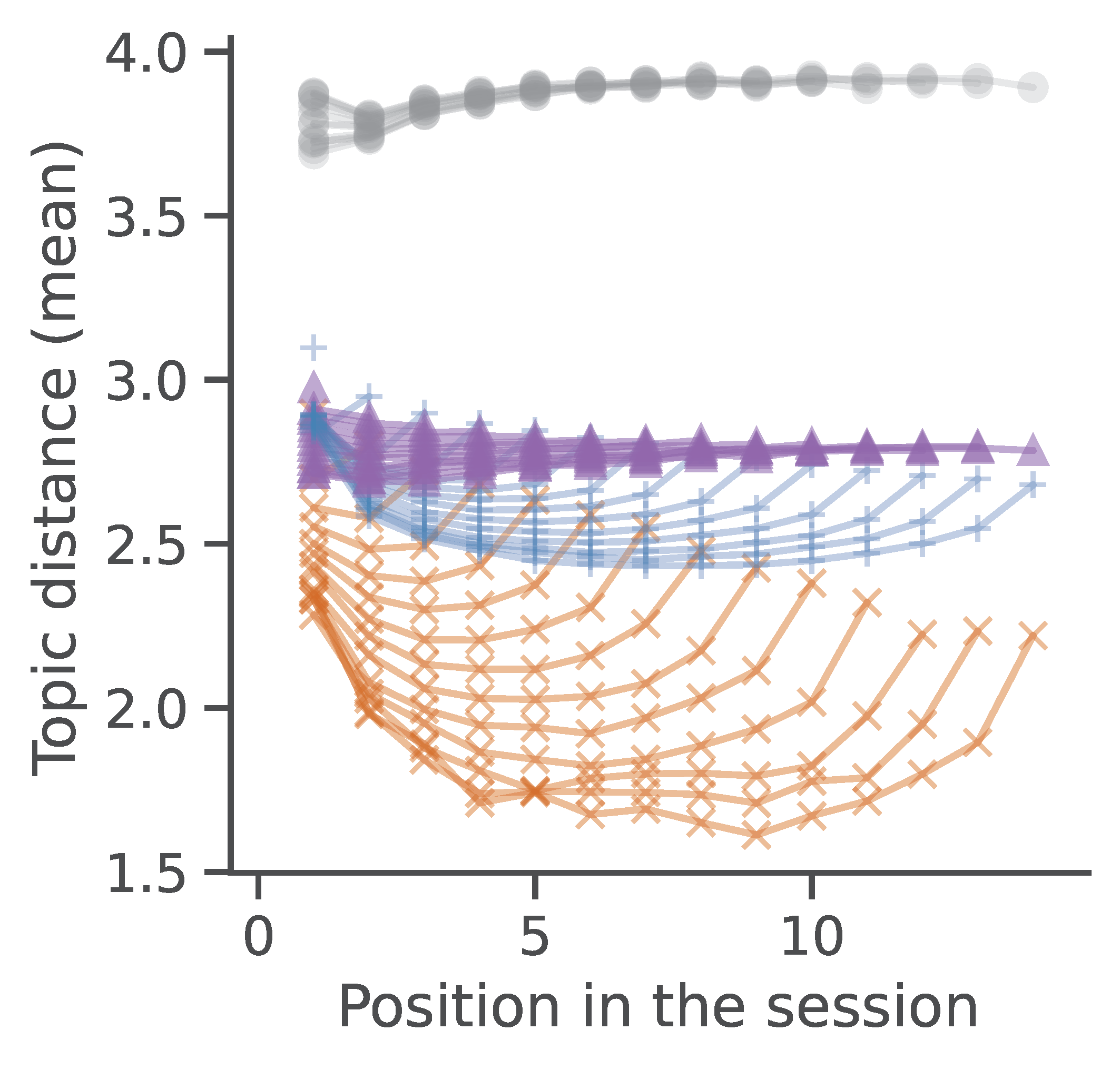}
        \subcaption{From previous article}\label{fig:topic_distance_prev_rw}
    \end{minipage}
    \hfill
    \begin{minipage}[t]{0.19\textwidth}
        \centering
        \includegraphics[height=2.7cm]{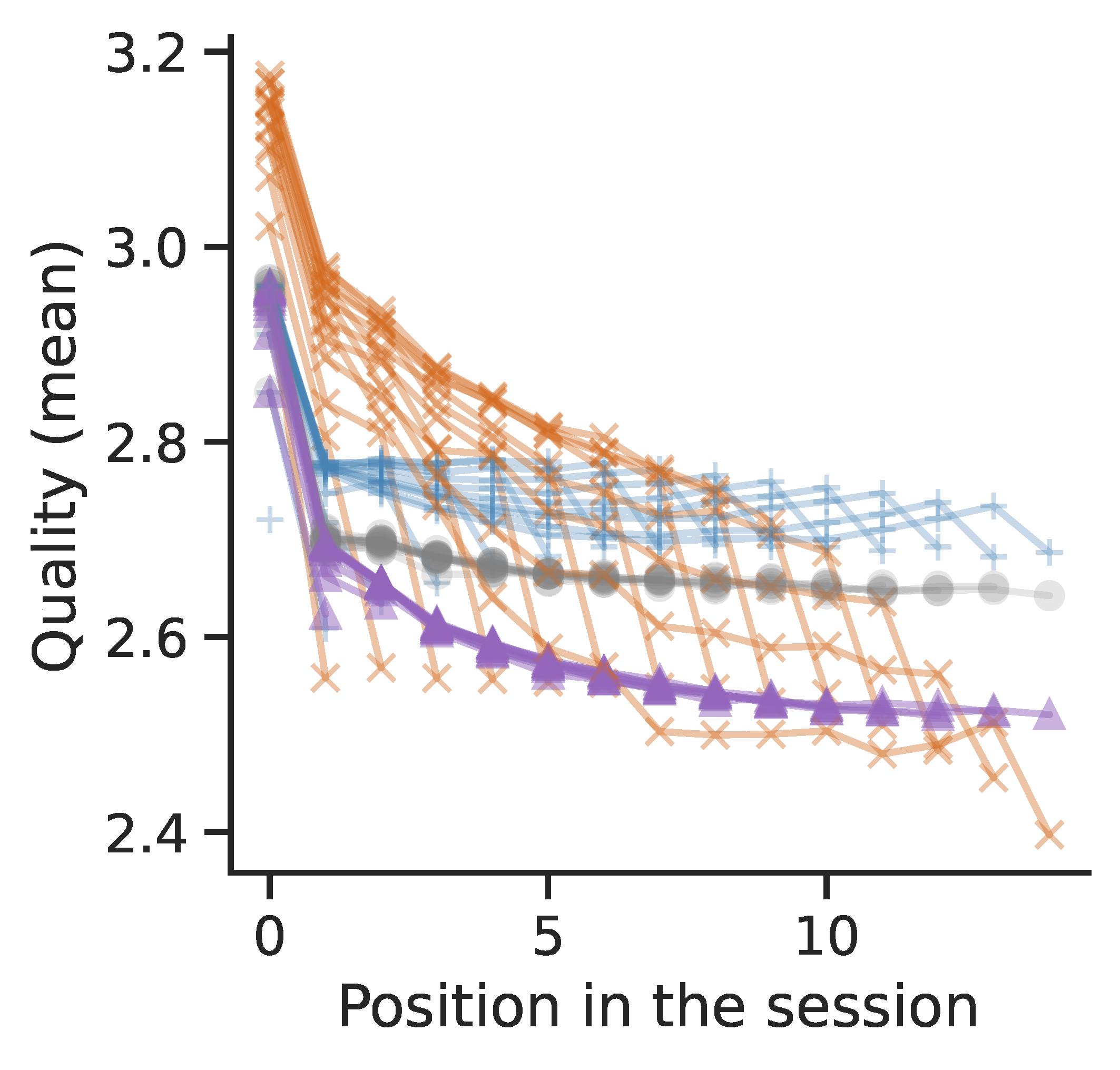}
        \subcaption{Quality}\label{fig:quality_rw}
    \end{minipage}
    \hfill
    \begin{minipage}[t]{0.19\textwidth}
        \centering
        \includegraphics[height=2.7cm]{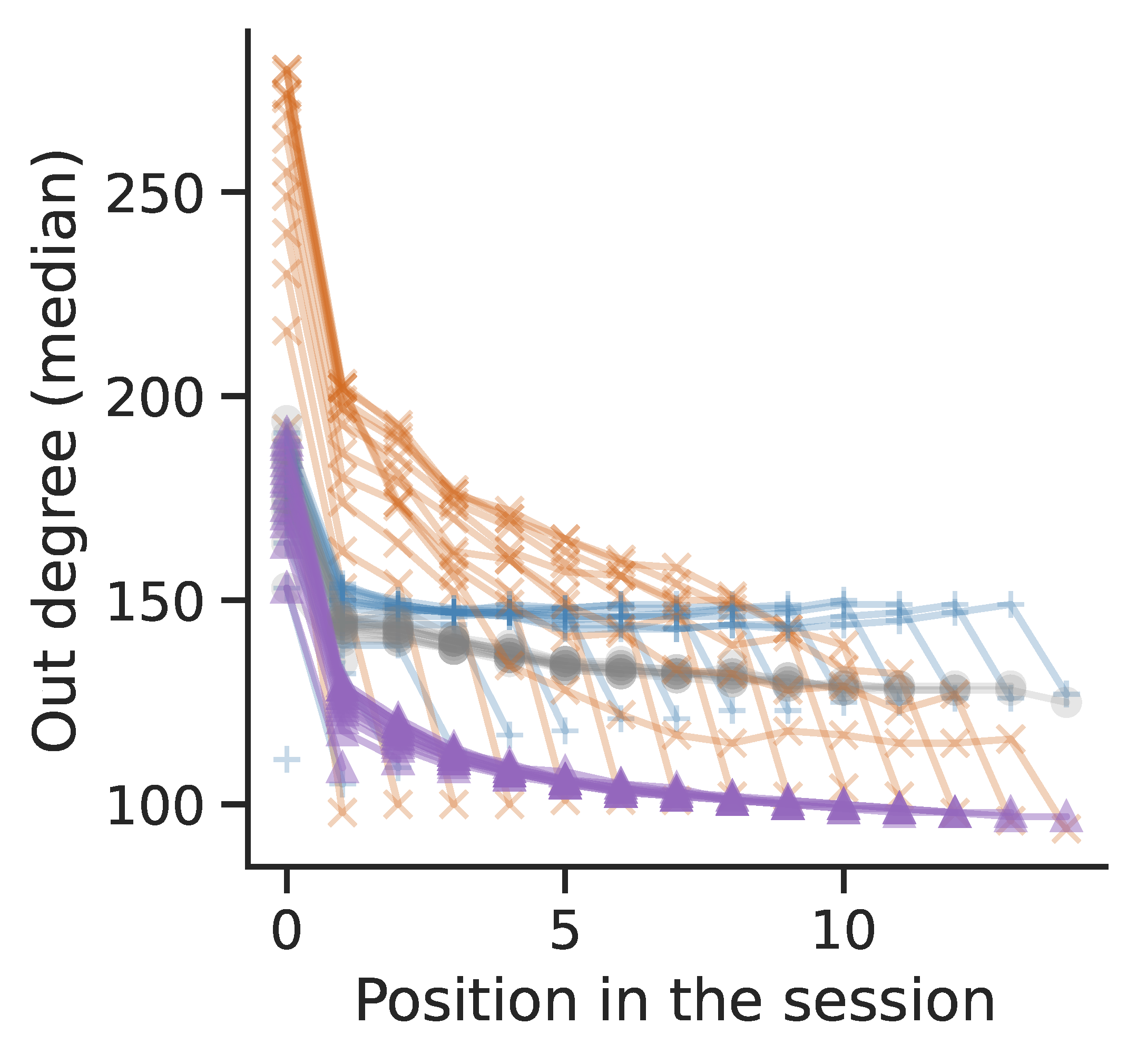}
        \subcaption{Out-degree}\label{fig:out_degree_rw}
    \end{minipage}
    \hfill
    \begin{minipage}[t]{0.19\textwidth}
        \centering
        \includegraphics[height=2.7cm]{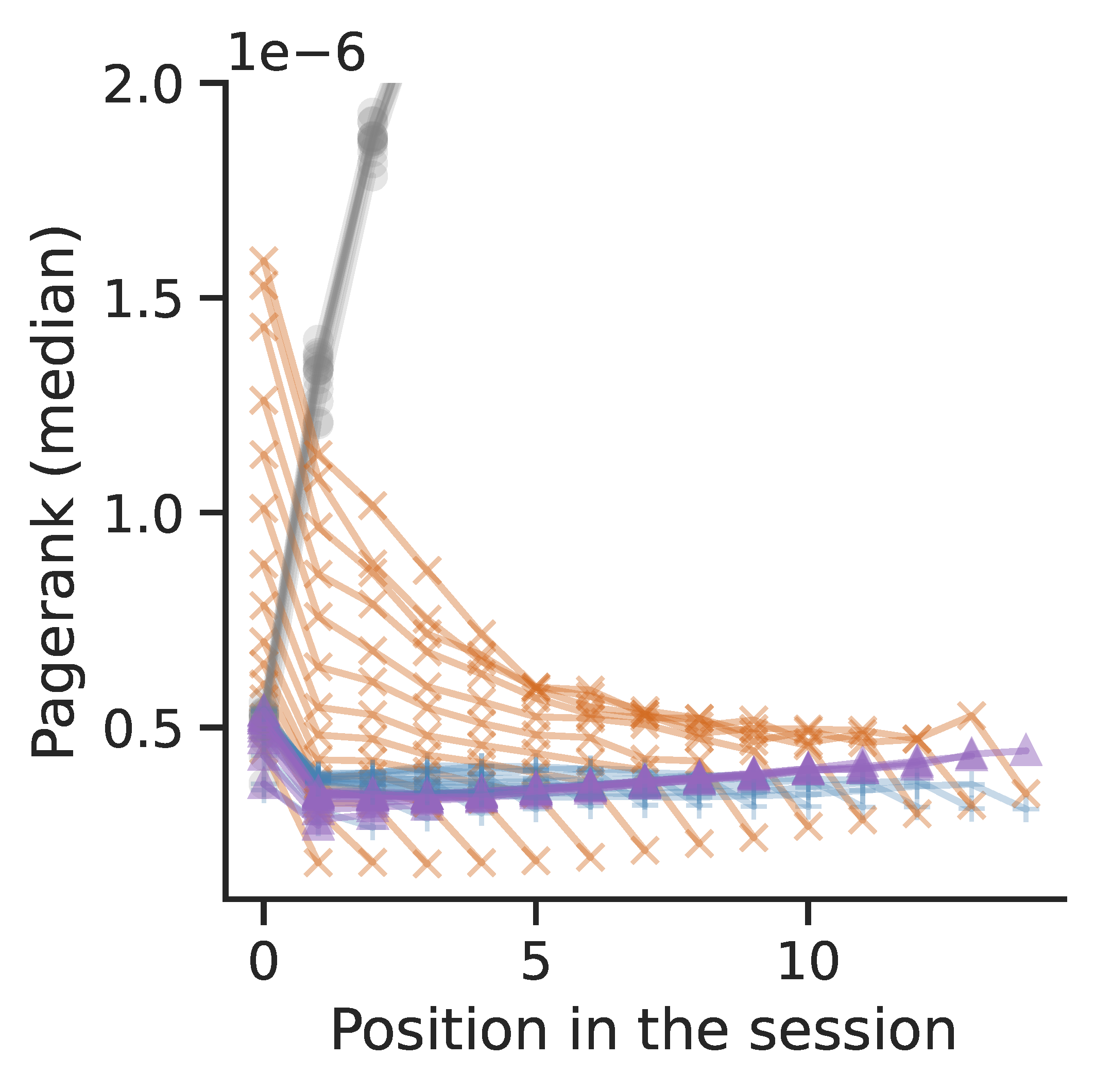}
        \subcaption{Pagerank}\label{fig:pagerank_rw}
    \end{minipage}
    \hfill
    
\caption{Property evolution of the trajectories generated by the three random walk models, compared with natural navigation as captured by navigation trees. Each curve represents sessions of different lengths.}
\label{fig:evolution_rw}
% \vspace{-5mm}
\end{figure*}

\begin{figure*}[t]

    \begin{minipage}[t]{0.19\textwidth}
        \centering
        \includegraphics[height=2.7cm]{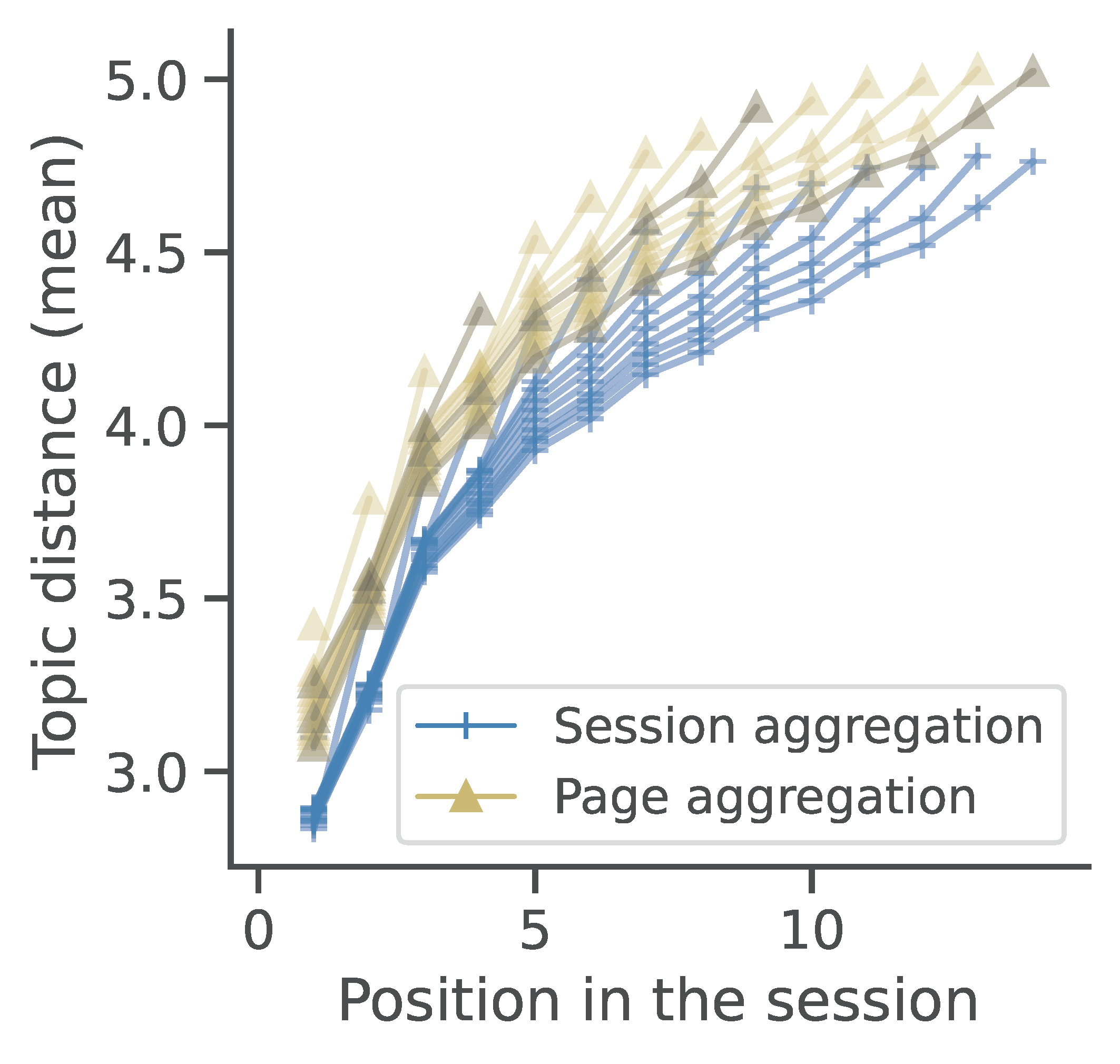}
        \subcaption{From the first article}\label{fig:topic_distance_first:agg}
    \end{minipage}
    \hfill
    \begin{minipage}[t]{0.19\textwidth}
        \centering
        \includegraphics[height=2.7cm]{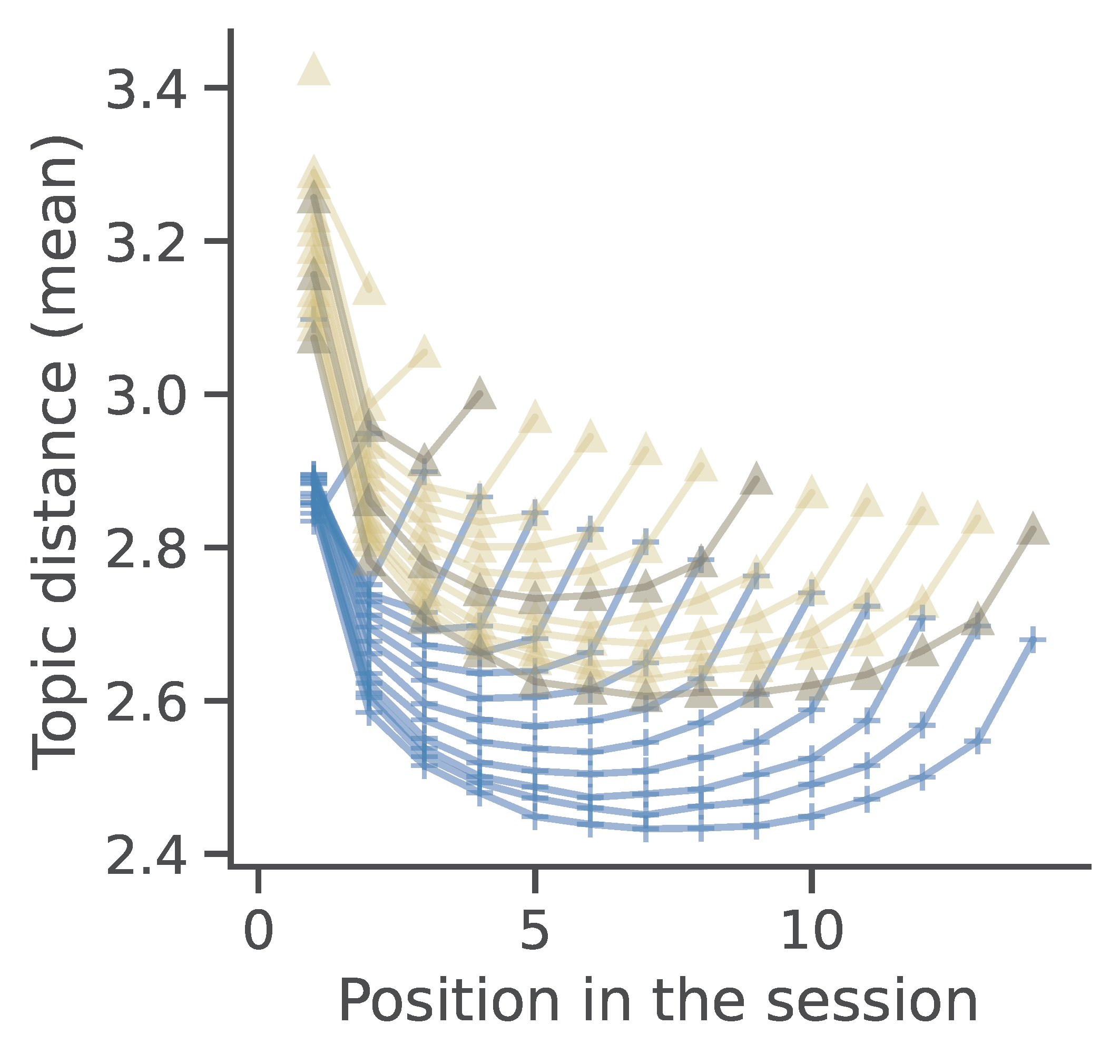}
        \subcaption{From the prev. article}\label{fig:topic_distance_prev:agg}
    \end{minipage}
    \hfill
    \begin{minipage}[t]{0.19\textwidth}
        \centering
        \includegraphics[height=2.7cm]{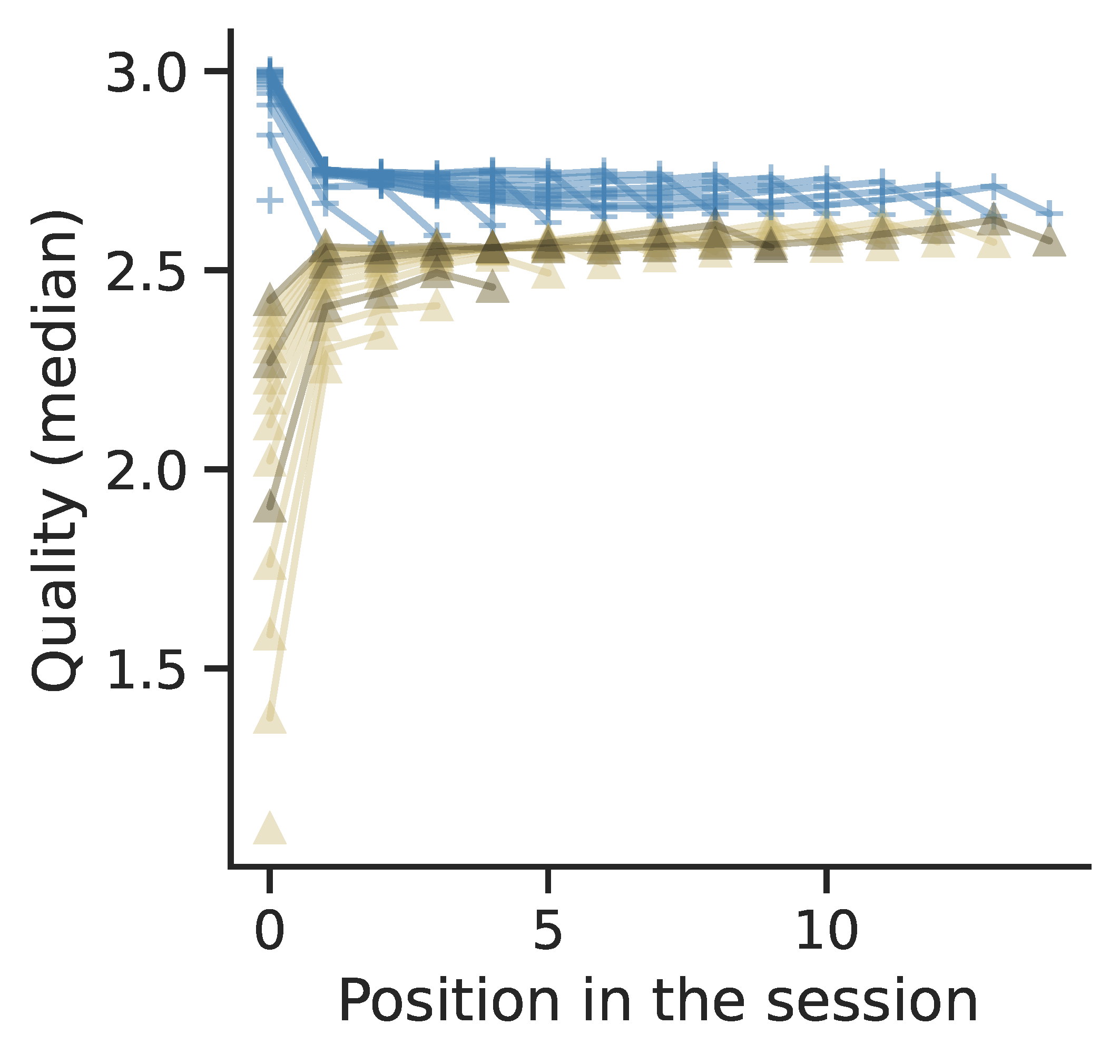}
        \subcaption{Quality}\label{fig:quality:agg}
    \end{minipage}
    \hfill
    \begin{minipage}[t]{0.19\textwidth}
        \centering
        \includegraphics[height=2.7cm]{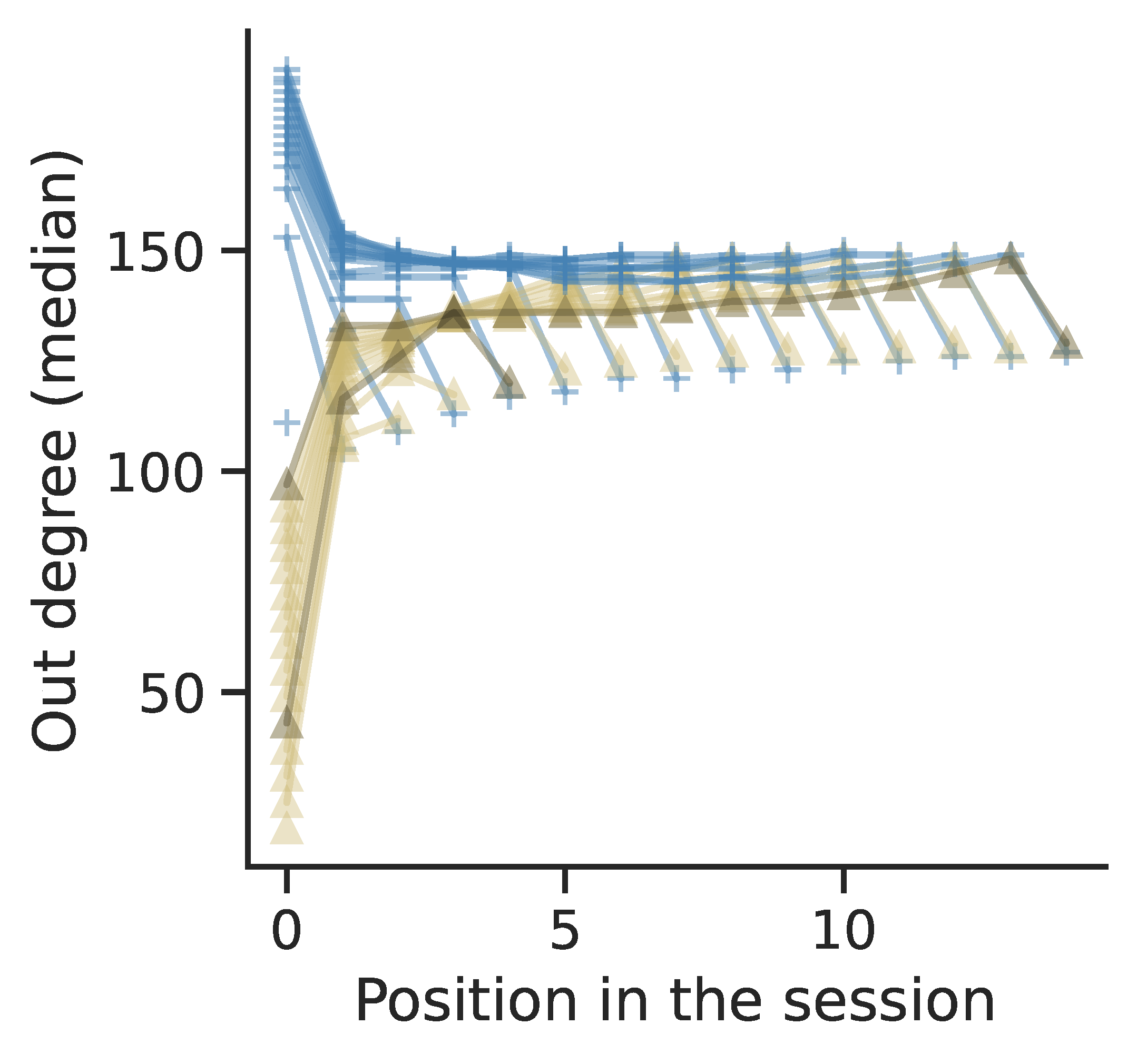}
        \subcaption{Out-degree}\label{fig:out_degree:agg}
    \end{minipage}
    \hfill
    \begin{minipage}[t]{0.19\textwidth}
        \centering
        \includegraphics[height=2.7cm]{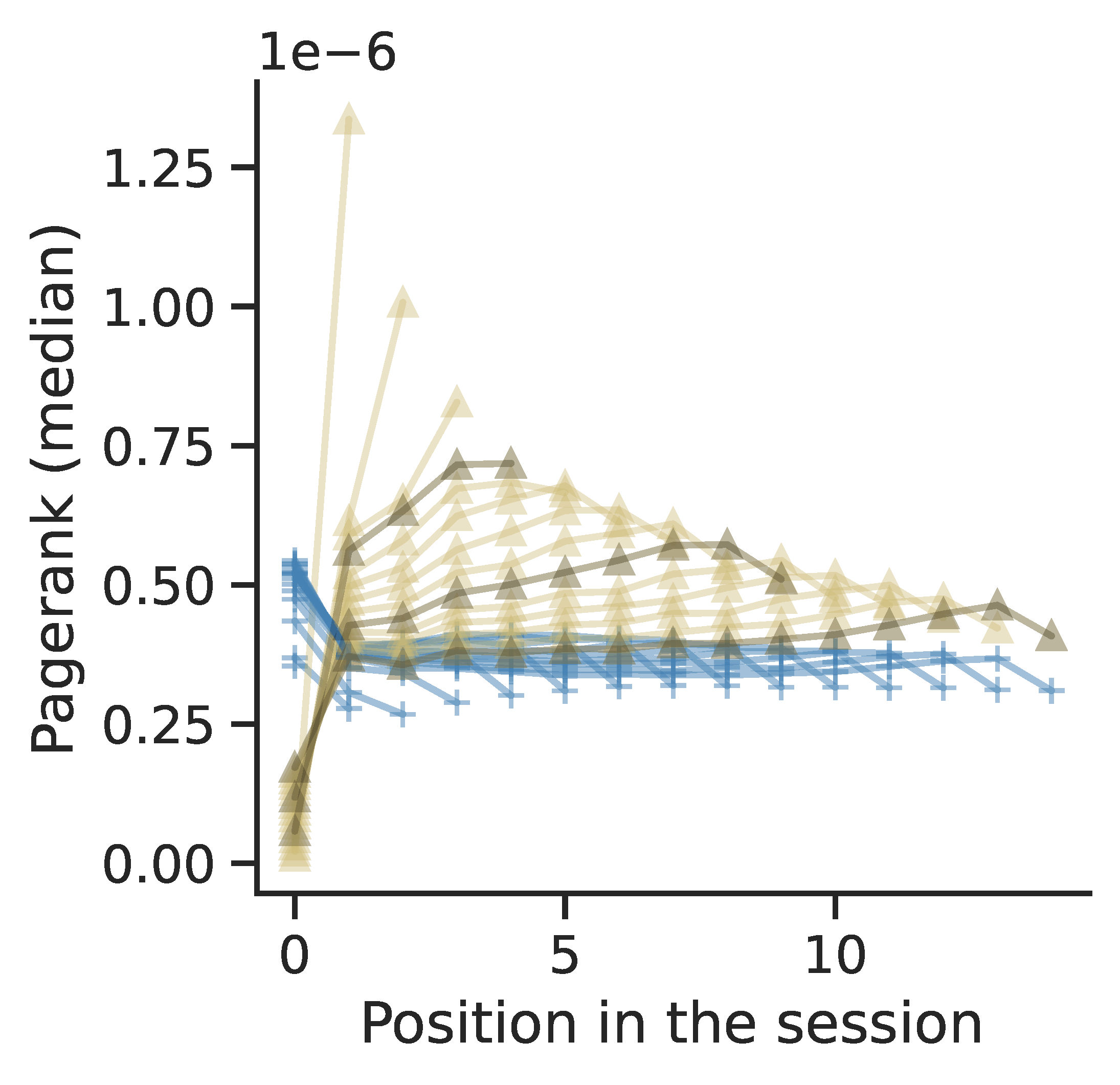}
        \subcaption{Pagerank}\label{fig:pagerank:agg}
    \end{minipage}
    \hfill
    % \vspace{-3mm}
\caption{
Comparison of the evolution of five different properties when aggregating navigation trees by session (micro-average, blue) and by starting page (macro-average, gold). Gray trajectories highlighted for readability. Each curve represents sessions of different lengths.}
\label{fig:evolution_agg}

\end{figure*}

In order to better interpret our observations, we compare them 
%In our analysis, the evolution of the different properties is compared
with three null models corresponding to different random walkers. 
The null models serve as a comparison to assess to which degree the observed properties of the navigation dynamics are due to chance.
We randomly sample 120M paths from the navigation trees, and run (from the tree's starting article)
(1)~an \textit{unbiased random walker} that selects the next step with uniform probability from the available links and generates a sequence of the same length as the original path;
(2)~an \textit{extrinsic-stop biased random walker} that selects the next step based on the pairwise transition probabilities obtained from the public clickstream and generates a sequence of the same length as the original path;
(3)~an \textit{intrinsic-stop biased random walker} that selects the next step---or stops---based on the pairwise transition probabilities from the public clickstream \cite{SearchStrategies}.
We consider sessions up to length 15, stratifying by session length.

\xhdr{Topic space}
% We generate a topic space from the embedding vectors obtained with WikiPDA (\Secref{sec:features}) and
We measure the topical distance between articles via the Kullback--Leibler (KL) divergence of their respective WikiPDA topic distribution vectors (\Secref{sec:features}). 
For robustness, we tried different topic models (WikiPDA and ORES) and different distance metrics (KL divergence, Euclidean, cosine, and Wasserstein), obtaining qualitatively similar results.
%We checked the robustness of our findings by modeling the topic space with vectors obtained both from WikiPDA and ORES and with four different distanced metrics: KL divergence, Euclidean, Cosine, and Wasserstein. All combination gives us qualitatively the same results, and we report the result based on WikiPDA topic vectors using KL divergence.
%\xhdr{Diffusion from first} 
First, we study how readers diffuse in topic space starting from the first article, which plays a special role, as it represents the entry point to Wikipedia.
On average, readers diffuse in topic space, moving further from the first article with every step (\Figref{fig:topic_distance_first}).
Reading sequences and navigation trees exhibit the same trend, with a shift due to the tendency of reading sequences to ignore external navigation.
All the random walkers show similar increasing trajectories (\Figref{fig:topic_distance_first_rw}), diffusing faster than natural navigation when the random walker is unbiased, or biased but extrinsically stopped.

%\xhdr{Divergence from previous}
Second, we measure the semantic step size in topic space by tracking how the topical distance to the previous article evolves. Both navigation trees and reading sequences exhibit a U-shape, suggesting that readers tend to first reduce their semantic step size, before diverging and finally abandoning (\Figref{fig:topic_distance_prev}). The discrepancy between navigation trees and reading sequences is consistent with the previous observation on diffusion from the first article.
Interestingly, this U-shape is similar to the trajectories generated by the intrinsic-stop biased random walker (\Figref{fig:topic_distance_prev_rw}), as also reported in previous work \cite{SearchStrategies}.
%\tp{Interpretation?}
In contrast, the other two random walk models show that by selecting a random link or stopping at predefined lengths, the average distance from the previous article tends to stabilize to an equilibrium value.

\xhdr{Quality}
The evolution of article quality shows a sharp drop at the beginning, for both reading sequences and navigation trees (\Figref{fig:quality}). This behavior can be interpreted as a form of regression to the mean, since many sessions start from popular pages with high quality, 
% (Pearson corr. popularity/quality=0.16 -- Log(popularity)/quality=0.63)
which thus contribute more to the distribution. By moving one step in the link network, readers naturally reach a page that is, on average, of lower quality.
% such that the next step naturally leads to a page with lower page.
The intuition is confirmed by the behavior of the unbiased random walker, which shows the same drop with the first step (\Figref{fig:quality_rw}). 

In contrast to reading sequences, navigation trees show a sharp drop in quality with the \emph{last} pageload. 
This indicates that readers have a higher chance to stop Wikipedia\hyp internal navigation when reaching a low-quality page, and as a result, continue navigating in a different branch of the tree or via an external transition. 

%Notable is the irregularity visible for the sessions consistent for different lengths: navigation trees show a final drop while reading sequences have a small increase. The behavior observer on navigation trees can be interpreted as the abandonment of the internal navigation when the readers end up on a page of lower quality than the previous one. The small increase in the reading sequences could be explained by readers abandoning the internal navigation and loading a different page of higher quality from an external source.

Compared to the random walkers (\Figref{fig:quality_rw}), readers tend to navigate across pages with less variance in quality. The random walkers' traces support the hypothesis that there are articles with a higher chance of terminating the navigation: while the unbiased and extrinsic-stop biased walkers show no termination pattern, the intrinsic-stop biased walker shows a final drop as in human navigation. The organic stopping of this random walker, mirroring readers' behavior more closely, increases the chances to abandon the navigation on pages of low quality that, according to the clickstream data, relay less traffic.

% The evolution of the quality property shows (\Figref{fig:quality}) for the reading sequences and navigation trees a sharp drop at the beginning. This is motivate by the fact that the popular pages that contribute more in the distribution and typically of higher quality (Pearson corr. popularity/quality=0.16 -- Log(popularity)/quality=0.63). The intuition is confirmed by the behavior of the random walker that shows the same drop in the first step. 

% Notable is the anomaly visible for each end of the sequence. Navigation trees show a final drop, while reading session an increase. 

%\subsubsection{Outdegree}
\xhdr{Network centrality}
Finally, we are interested in how reader sessions evolve in the network with respect to  different centrality measures.
%network-related properties evolve with the session.
%
%\xhdr{Outdegree}
We start with out-degree (the number of outgoing links in article bodies).
% To analyze how the out-degree evolves with the readers' navigation, we obtain the number of outgoing links in the body of the article and compute the median at each step in the sequence. 
Similar to article quality, the out-degree shows a sharp drop 
%for both session definitions and the random walker 
with the first step (\Figref{fig:out_degree}) for navigation trees and reading sequences, likely caused by the presence of many sessions starting from pages with a particularly high out-degree. 
We also find a sharp drop for the last pageload in the sequence of the navigation trees, suggesting that readers have a higher chance of stopping Wikipedia\hyp internal navigation upon reaching a page with low out-degree.
% Consistent with the findings on the evolution of the quality, readers tend to interrupt the internal navigation when they reach a page with a lower out-degree than the previous articles. On the other hand, reading sequences show a monotonically decreasing trend suggesting that users move to articles with fewer links and no clear indications that the reader reached the end of the navigation.

In the case of the random walkers, we draw similar conclusions as for article quality. Whereas unbiased random walks and extrinsic-stop biased random walks show a decrease and stabilization of out-degree, the intrinsic-stop random walker, as humans, terminates on pages of lower degree (\Figref{fig:out_degree_rw}). Compared to random walkers, human navigation is more stable: after the initial drop, they have a higher chance to stay on pages with around 150 links.

%\xhdr{PageRank}
Finally, we characterize how the PageRank of visited articles changes during sessions. We observe that the PageRank  mirrors the evolution of quality and out-degree with regard to the initial drop (\Figref{fig:pagerank}).
Readers tend to enter more frequently on popular pages with high centrality and naturally move to a less central node in one step.
Also for this case, a drop is visible in the last step of the navigation trees, indicating that, when the readers reach an article leading to the network periphery, they have higher chances to stop the Wikipedia\hyp internal navigation.
The random walkers (\Figref{fig:pagerank_rw}) show that unbiased walks naturally converge in a few step to the most central pages with very high PageRank. The extrinsic-stop biased walker, on the contrary, after an initial drop, tends to move to central nodes at a much lower speed. Finally, the intrinsic-stop biased walker, again, shows a final drop from a stable value before abandoning the navigation, similar to human readers.

\xhdr{Aggregation by page}
% \subsubsection{Aggregation by page}
The quantities in \Figref{fig:evolution} correspond to a micro-average over all sessions, where the average behavior can be dominated by sessions starting from the most popular pages since the overall distribution of pageviews is highly skewed.
Therefore, we also calculate a macro-average by aggregating on a starting-page level to make each first article contribute equally.  %\Figref{fig:evolution_agg} summarizes the trend by combining all sessions with the same starting page and the same length (\Figref{fig:evolution_agg}).
The diffusion in topic space is qualitatively similar in both aggregation methods (\Figref{fig:topic_distance_first:agg} and \Figref{fig:topic_distance_prev:agg}). In contrast, for quality, out-degree, and PageRank, the overall trend is inverted, i.e., instead of a sharp drop, we observe a sharp increase in these metrics after the first step (\Figref{fig:quality:agg}, \Figref{fig:out_degree:agg}, and \Figref{fig:pagerank:agg}). 
This discrepancy could be caused by the presence of many low-quality \cite{piccardi2018structuring} and low-degree articles, such that readers at the first step tend to move to better articles in search of information (a sort of regression to the mean).
Interestingly, the drop towards the last pageload in a session appears across both aggregation methods.

\section{Discussion}
\label{sec:discussion}

%With Wikipedia being one of the largest platform, this provides a first step towards better understanding how people seek knowledge more generally. 

\subsection{Summary of findings}
We have provided a systematic characterization of the navigation pathways of Wikipedia readers through a large-scale study of the site's server logs.
Starting from the raw logs, we aggregated the data in navigation trails to quantify how readers reach, and transition between, pages.
First, the most common way to reach a page is through an external search engine, followed in frequency by internal navigation from other Wikipedia articles; other sources, such as external websites (mostly social media sites) and other Wikipedia content (such as categories or special pages), are much less frequent, but still substantial in absolute numbers.
Second, readers frequently transition between pages via external search engines instead of using direct Wikipedia links.
These external transitions are characterized by larger topical jumps and larger inter-event times between pageloads; they must, however, still be considered semantically meaningful, for, in many cases, a link for internal navigation---even if not taken---would still be available. 
Third, by analyzing sequential patterns, we find that consecutive reloads and revisits of previously visited articles are common (10\% or more each).

We continued by characterizing how readers combine the above patterns into extended navigation sequences.  
First, we introduced two approaches to capture paths of readers: \textit{navigation trees} based only on internal navigation, and \textit{reading sequences} based on the time-ordered pageloads including internal and external transitions.
Second, we described how sessions are affected by their context in terms of device type and time of day. We find that topics related to STEM [entertainment] are more associated with working [evening and night] hours.
Third, we measured the size and structure of sessions. While most sessions consist of a single pageload (68--78\% depending on the aggregation method), the size distribution shows a long tail with tens of millions of sessions consisting of 10 or more pageloads. 
%We identified several factors for longer sessions: evening hours, weekend days, topics on entertainment. 
The topic not only affects the size but also the shape of trees: while sessions starting from articles on entertainment generally consist of more pageloads, such trees are also broader (higher branching factor) than sessions starting, e.g., from STEM topics, which are smaller and deeper. 
%\tp{These results are aligned with Why we read wikipedia. On STEM people have long focused reading, while other topics are explored when bored}
Fourth, we investigated the within-session evolution of article properties. 
% \todo[]{FIX:}
In topic space, longer sessions diffuse ever further away from the origin, with semantic step size following a characteristic U-shape pattern suggesting that readers reduce their semantic step size first, before diverging in ever larger steps and finally abandoning the session.
The first and last pageload of a session show special behavior regarding the evolution of article quality and network centrality. More popular (and thus higher-quality and higher-centrality) pages are naturally more common as first articles, thus engendering a form of regression to the mean with the second step. An inverted effect appears when sessions are aggregated at the starting-page level, such that every starting article is represented equally. Either way, articles at the end of the navigation are typically lower-quality pages, suggesting that readers stop following the internal navigation when they reach these pages, which thus act as network sinks.

\subsection{Implications}
%In the following we discuss the implications of our findings in a broader context.
\xhdr{Complexity of navigation behavior}
Our results show that the navigation paths extracted from Wikipedia's server logs constitute a non-trivial dataset requiring extreme care in order to avoid drawing spurious conclusions.
First, in contrast to existing pre-processing pipelines for sequence analysis (e.g., tokenization, stopword removal, stemming, etc., in NLP), we still lack an understanding of universal best practices for navigation paths, and as a result we had to investigate and compare alternative strategies for conceptualizing sessions---namely, reading sequences \vs\ navigation trees. 
% pre-processing and filtering the raw data required development of heuristics using domain-knowledge for, e.g., the identification of users, removing automated requests, or traffic from specials pages or namespaces (such as the Main Page). In contrast to existing pre-processing pipelines for generating sequences for further analysis (e.g. tokenization step in NLP), here, we still lack an understanding of best practices. 
%
Second, operationalizing navigation paths makes strong assumptions: while navigation trees from pure internal navigation are more topically coherent with more complex structure, reading sequences from temporally ordering all of the user's pageloads are less coherent but provide a linear sequence that is not broken by external searching (which is common). 
The latter typically introduces an additional cutoff for sessions if consecutive pageloads are separated by more than one hour~\cite{InterActivityTime}; however, our analysis suggests other potential data-informed choices, such as the time separation of internal and external transitions at approximately four minutes (\Figref{fig:probability_referer_by_time}).
% approaches to capture navigation paths include the generation of knowledge-networks as proposed by Lydon-Staley et al.~\cite{Lydon-Staley2021}.
Naturally, the suitable choice depends on the question of interest.
Third, our analysis shows that the data can exhibit Simpson's paradoxes; \eg, the inversion of the within\hyp session evolution of page properties such as PageRank (\Figref{fig:evolution_agg}) depends on the aggregation level.
Fourth, the prevalence of trivial patterns (e.g., reload or revisit) points to potential caveats when applying prediction models to session-based recommendation~\cite{Wang2021survey}.
%; especially, in view of recent results showing that simple heuristic methods are often preferable over conceptually more complex models ~\cite{Ludewig2020empirical}. 
%Page loads could be pre-aggregate in trees 

\xhdr{Diversity}
There is extraordinary diversity in the ways readers browse Wikipedia, modulated by topic, device, time of day, etc. 
This reflects the diversity found in previous studies on the different motivations and information needs of readers across the globe~\cite{singer_why_2017,lemmerich_why_2019,johnson2020global}. 
This heterogeneity indicates caution against simplistic models aiming to capture a single average behavior.
% For example:
% - most people with short sessions, still millions of users with long sessions

% - not just linear browsing but non-trivial patterns:  trees (e.g. multitab), external links

% - variation of context: device, working hours, country (?)

% - variation depending on content: topic not only varies by length but also the shape of the trees (i.e. the exploration strategy) differs. for example, articles on films with longest sessions, but also more breadth-first navigation. even more directly, the number of available links correlates strongly with the length of the session.

\xhdr{Online ecosystem}
The usage of Wikipedia is embedded in a larger online ecosystem.
Multiple studies have shown the importance of Wikipedia to search engines~\cite{McMahon2017Substantial,Vincent2021Deeper}, as a gateway to the Web~\cite{piccardi2020quantifying,piccardi2021gateway}, and as a main educational resource for online learning more generally~\cite{Kross2021Characterizing}.
Our results show that this interplay between external and internal (with respect to Wikipedia) also plays a crucial role on an intra-session level when navigating encyclopedic information.

% Of course, not completely new, there have been multiple studies showing the importance of Wikipedia:
% more generally its position in the general online learning ecosystem as one of the main educational resources online ~\cite{Kross2021Characterizing}. when entering Wikipedia, the importance of Wikipedia links to search engines . when leaving Wikipedia, the engagement with external references or linkes~\cite{piccardi2020quantifying,piccardi2021gateway} 
%
% Here, we observe, that there is a frequent back and forth within the same reading session, e.g. read a Wikipedia page, visit external pages, and re-enter. Not a rare event but almost on the same order of magnitude as internal navigation. 
% This means that distinction between inside and outside is very blurry. 
%This motivates the question how to study Wikipedia and its evolution in isolation. 

\xhdr{Navigation in the wild}
The navigation of readers on Wikipedia differs from targeted navigation in lab-based settings \cite{HumanWayfinding,Helic,West_Leskovec_2012}. 
% Therefore, it is worth to discuss how these observations match previous studies on navigation in Wikipedia conducted in lab-based settings or using semi-syntethic datasets. 
% First, we find noticeable differences to targeted-navigation. the latter typically from experiments such as Wikispeedia or Wikigame (most notably~\cite{HumanWayfinding}). 
We do not observe typical strategies characterized by, e.g., navigation via hubs (an initial increase, followed by a drop, in out-degree) or gradually decreasing the step size in semantic space towards a target. 
Instead, we find a range of other patterns, such as a U-shape for the step-size in semantic space and an immediate sharp drop followed by largely constant centrality measures (out-degree, PageRank).
%In hindsight, this observation might be expected in lack of a well-defined target and an explicit definition of success of navigation. 
This highlights conceptual limitations of targeted-navigation experiments with respect to generalizing their results to how humans seek knowledge more generally.
% - navigation in the wild shows different properties. drop-out rate much higher in wild-navigation (paths are much shorter). outdegree: sharp drop only after first step; maybe due to centrality/quality of the popular entry-pages (mostly through external-search). semantic similarity show characteristic U-shape pattern (zoom-in and then zoom-out). semantic similarity to target reverse-matches our observations to semantic similarity to first; speculation: first page takes the role of the target?. 
% - to some degree, this can be expected as the two modes of navigation differ conceptually. there might not be a well-defined target when navigating in the wild, e.g. there is no explicit definition of what success means in navigation. it is hard (or impossible) to know if the reader found what they were looking for.
% - in lack of large-scale observational data, many studies rely on targeted navigation to understand and model how humans seek knowledge. Our results caution against over-generalization of these results for applications in the wild.

Furthermore, our results provide a more nuanced picture on the conclusions derived from publicly available data, most notably the Wikipedia clickstream~\cite{Wulczyn2015clickstream},
which provides aggregate data on the number of times a link was clicked.
For example, we can observe that the overall tendency to navigate towards peripheral nodes~\cite{LinkSuccessfulWikipedia} is mainly driven by the first step after reaching Wikipedia, with subsequent steps showing much smaller differences in centrality measures (with the exception of the last step, see below). 
One possible interpretation is a regression-to-the-mean effect as popular pages (the starting points of navigation) are generally skewed towards higher centrality and quality. 

% - single click without taking into account the position in the sequence . they found that links are more popular if target-article has smaller out-degree and conclude that users tend to navigate to peripheral nodes. Here, we find that the drop in out-degree and pagerank comes mostly after the first pageview and then remains largely constant on average. Explanation: most clicks come from that first transition, so overall conclusion true; however, behaviour along the navigation path different from that.

% - paths generated from random walks ~\cite{SearchStrategies}. characteristic curves in semantic similarity (U-shape for current vs next page, increase for current to last/first) was already observed in semi-synthetic sequences. conclusion: confidence in generalizability of these results to navigation in the wild. suggests that clickstream captures large part of the navigation behaviour \mg{how much do we want to talk about our other unpublished paper here?}

\xhdr{Content and navigation}
Our results contribute to describing the relation between content and navigation, expanding the prior understanding of how readership and popularity are influenced by visual position~\cite{LinkSuccessfulWikipedia} or quality~\cite{Zhu2020content}.
%Our results go beyond the population level, suggesting that low-quality pages lead readers to stop navigating along a specific branch in the navigation tree (and continuing along a different branch or stopping altogether). 
Our results go beyond the population level, suggesting that upon encountering low-quality pages readers tend to stop navigating along a specific branch in the navigation tree (and continuing along a different branch or stopping altogether).
This is specifically important in the context of knowledge gaps in Wikipedia~\cite{Redi2020taxonomy}, in order to address the uneven representation of, e.g., articles on women, where a better understanding of the interaction between content, readers, and editors~\cite{Shaw2018pipeline,mtp} is crucial to allow for more informed decision\hyp making in designing interventions.

More generally, this finding is aligned with the definition of information scent used in information foraging theory \cite{chi2001using}. The theory states that, in analogy to animals following the scent of food, when seeking information, we rely on our intuition---or ``built-in'' foraging strategies---to pick the path that maximizes information intake while minimizing the investment of time and energy.
In this view, readers foraging for information follow the scent with higher chances of leading to the desired content; when scent loses intensity, they move to more promising information sources. 
Additionally, our work can have implications for developing theoretical frameworks to describe navigation patterns. Understanding how readers follow specific trails can inform researchers about the distinct properties of the information scent that guide our search for information online. These findings can be instrumental in developing novel theories on how humans move in information networks.
% - Previous studies have gained valuable insights how content affects readership: visual position of links correlated to their clickthrough rate (on the top and on the left) \cite{LinkSuccessfulWikipedia}; improving articles leads to more pageviews \cite{Zhu2020content} 
% - Here: much more nuanced picture how it affects the navigation not just in quantity but also qualitatively. indications from semantic similarity and network-features that readers stop navigation when encountering pages of poor quality. 

% - suggests that low quality pages leads to readers stopping navigation

% - indirectly, we also observe how information need affects navigation matching previous studies using surveys~\cite{singer_why_2017,lemmerich_why_2019}. for example, longer sessions predicted in-depth information need. we found many short sessions, correlated to STEM-topics during work-hours. 

% This is important in the context of knowledge gaps in Wikipedia~\cite{Redi2020taxonomy}, for example, uneven coverage or participation of women in Wikipedia. There are different models for the interaction between content, readers, and contributors (e.g. a pipeline\cite{Shaw2018pipeline} or the flywheel\cite{mtp}). 
% Additional quantitative evidence for the intricate interplay between these components.
% The latter will allow for more informed-decision making in designing interventions.

\xhdr{Best practices for Wikipedia log analysis}
One of the challenges in conducting analyses like those presented here is the lack of standard pipelines to preprocess and aggregate the logs.
Unlike fields such as NLP or computer vision, where preprocessing steps are de facto standardized, modeling behavior from access logs does not have a unique standard procedure yet, and researchers are forced to make many modeling choices. The purpose of the study and limitations of the data, such as privacy concerns, may influence how sessions are defined and, consequently, the results obtained. Our work fills this gap in the case of Wikipedia by providing best practices for processing server logs to study reader sessions. We describe two complementary approaches based on trees and temporal sequences and demonstrate their relative advantages and disadvantages. This operationalization is crucial for developing systematic approaches in future studies to understand reader navigation better, capture their information needs, and improve their experience on Wikipedia and the Web more generally.

\subsection{Limitations and future work}
\xhdr{Limitations}
In terms of limitations, we capture navigation paths only via events in the server logs. Moving forward, how people engage with content could be more accurately observed via client-side instrumentation. 
The aggregation based on IP addresses and user agent information also has limitations; \eg, we had to discard the sessions of large organizations with shared IP addresses. 

The navigation logs suggests that Wikipedia fulfills various information needs and readers exhibit diverse navigation patterns. Using large-scale digital traces offers important advantages over other methods when we are interested in the quantitative measurement of behavioral phenomena \cite{salganik2019bit}. However, purely log-based analysis also has limitations, and it should be considered a complementary, and not a substitutive, approach. Previous work indicates that big-data analyses are not immune to biases introduced by algorithmic dynamics \cite{lazer2014parable,wagner2021measuring}, data collection problems, preprocessing errors, and measurement errors \cite{wagner2021measuring,lazer2021meaningful}.

Finally, we only focused on a single language, English. While this already revealed a rich spectrum of phenomena, additional variation can be expected from a comparison across languages \cite{lemmerich_why_2019}.

\xhdr{Future work}
To overcome these limitations, future work should capture the variation in navigation across Wikipedia's over 300 languages.
% Previous studies on motivations for reading Wikipedia revealed strong differences on the cultural context~. %Therefore, one aim should be to understand how these patterns vary with language or geographic region.
%
Moreover, in order to better serve the different information needs of readers,  a better understanding is needed regarding how patterns in navigation correspond to underlying motivations~\cite{singer_why_2017} and other traits such as curiosity~\cite{Lydon-Staley2021}. By enriching the behavioral patterns with qualitative feedback, we can better understand user objectives and design ways to facilitate more efficient access to the desired information. 
% For example, how do patterns of navigation correspond to motivations~\cite{singer_why_2017} or other traits such as curiosity~\cite{Lydon-Staley2021}? More generally, this would allow to understand how nvaigation affects learning.
%

In line with previous studies \cite{ibanez2022comparison,bilenko2008mining,ibanez2022comparison}, future work should also investigate the relationship between search queries (from external and internal origin) and subsequent navigation behavior on encyclopedic platforms such as Wikipedia.
Finally, in order to capture encyclopedic information seeking more generally, researchers should capture navigation beyond individual platforms to take into account the interdependence of Wikipedia with the rest of the Web.
% Examples include the improvement of the search experience, the effect of previews not only to external references~\cite{wikiarchive} but also embedded in external sites~\cite{previewexternal}, or traffic from social media sites such as Twitter~\cite{Morgan2021social}.
% Investigate the interdependence of Wikipedia and other online learning and social platforms. 
% We only capture so much by looking at these platforms (including Wikipedia) in isolation. We have shown there is a continuous back and forth in navigation. 

% - one aspect of this is to improve search experience on Wikipedia (because we see Wikipedia-external-Wikipedia is very common). Not clear whether this is user-experience or whether external search' results are perceived better.

% - existing initiatives. need to better understand impact: Wikipedia previews on external sites~\cite{previewexternal}, traffic from social media sites such as youtube, facebook, twitter ~\cite{Morgan2021social}, previews to external references \cite{wikiarchive}.

% Investigate client-side information to get ground-truth from a sample (e.g. closing tabs)

% Prediction of trajectory.

\subsection{Conclusion}

Seventy-seven years ago, in 1945, Vannevar Bush sketched his vision of an information management
device---the ``memex''---that would allow users to not only retrieve documents quickly, but to also easily interlink documents \cite{as_we_may_think}.
With the advent of the Web, the hyperlink structure envisioned by Bush has since become a reality---but Bush's vision went further: he saw the trails taken by users as first-class citizens of the hypertext environment, as important as the text content itself:
``Wholly new forms of encyclopedias will appear, ready made with a mesh of associative trails running through them, ready to be dropped into the memex and there amplified'' \cite{as_we_may_think}.
% advent of “a new profession of trail blazers, those who find delight in the task of establishing useful trails through the enormous mass of the common record. The inheritance from the master becomes, not only his additions to the world’s record, but for his disciples the entire scaffolding by which they were erected.”
In this regard, our technological reality has not caught up with Bush's vision yet, and the present work should be seen as a small step toward achieving it:
we have started by describing the ``associative trails running through'' Wikipedia,
and we hope that its future versions will build on these insights to incorporate tools and features that will allow readers to continually benefit from each other's encyclopedic trail blazing.

\section*{Availability of data and code}
The underlying data from Wikipedia’s server logs are not publicly available due to privacy reasons. Code is available at \url{https://github.com/epfl-dlab/how_readers_browse_wikipedia}.

\begin{acks}
We thank Leila Zia for insightful discussions.
West's lab is partly supported by grants from
Swiss National Science Foundation (200021\_185043),
Swiss Data Science Center (P22\_08),
H2020 (952215),
Microsoft Swiss Joint Research Center,
and Google,
and by generous gifts from Facebook, Google, and Microsoft.
\end{acks}

\bibliographystyle{ACM-Reference-Format}
\bibliography{references}

\end{document}
\endinput

%% file: dlab_macros.tex
\usepackage{hyphenat}
\usepackage{xspace}
\usepackage{amsmath}
\usepackage{amsfonts}
\usepackage{url}
\usepackage{xcolor}
\usepackage{marvosym}
\usepackage{ifthen}
\newcommand{\chatoDisplayMode}[1]{#1}

\definecolor{MyRed}{rgb}{0.6,0.0,0.0} 
\definecolor{MyBlack}{rgb}{0.1,0.1,0.1} 
\newcommand{\inred}[1]{{\color{MyRed}\sf\textbf{\textsc{#1}}}}
\newcommand{\frameit}[2]{
  \begin{center}
  {\color{MyRed}
  \framebox[.9\columnwidth][l]{
    \begin{minipage}{.85\columnwidth}
    \inred{#1}: {\sf\color{MyBlack}#2}
    \end{minipage}
  }\\
  }
  \end{center}
}

\newcommand{\note}[2][]{\chatoDisplayMode{\def\@tmpsig{#1}\frameit{{\Pointinghand} Note}{#2\ifx \@tmpsig \@empty \else \mbox{ --\em #1}\fi}}}
\newcommand{\todo}[2][]{\chatoDisplayMode{\def\@tmpsig{#1}\frameit{{\Writinghand} To-do}{#2\ifx \@tmpsig \@empty \else \mbox{ --\em #1}\fi}}}

\newcommand{\abbrevStyle}[1]{#1}
\newcommand{\ie}{\abbrevStyle{i.e.}\xspace}
\newcommand{\eg}{\abbrevStyle{e.g.}\xspace}
\newcommand{\cf}{\abbrevStyle{cf.}\xspace}

\newcommand{\vs}{\abbrevStyle{vs.}\xspace}

\newcommand{\Secref}[1]{Sec.~\ref{#1}}

\newcommand{\Tabref}[1]{Table~\ref{#1}}
\newcommand{\Figref}[1]{Fig.~\ref{#1}}

\newcommand{\xhdr}[1]{\vspace{1.7mm}\noindent{{\bf #1.}}}

\newcommand{\xhdrNoPeriod}[1]{\vspace{1.7mm}\noindent{{\bf #1}}}

\newcommand{\textcite}[1]{\citeauthor{#1} \shortcite{#1}}

\newcommand{\hide}[1]{}

\makeatletter
\newcommand{\iffont}[2]{\ifthenelse{\equal{\f@family}{#1}}{#2}{}}
\makeatother

\iffont{ptm}{
  \usepackage{mathptmx}

  \DeclareSymbolFont{greek}{OML}{cmm}{m}{n}
  \DeclareMathSymbol{\alpha}{\mathalpha}{greek}{"0B}
  \DeclareMathSymbol{\beta}{\mathalpha}{greek}{"0C}
  \DeclareMathSymbol{\gamma}{\mathalpha}{greek}{"0D}
  \DeclareMathSymbol{\delta}{\mathalpha}{greek}{"0E}
  \DeclareMathSymbol{\epsilon}{\mathalpha}{greek}{"0F}
  \DeclareMathSymbol{\zeta}{\mathalpha}{greek}{"10}
  \DeclareMathSymbol{\eta}{\mathalpha}{greek}{"11}
  \DeclareMathSymbol{\theta}{\mathalpha}{greek}{"12}
  \DeclareMathSymbol{\iota}{\mathalpha}{greek}{"13}
  \DeclareMathSymbol{\kappa}{\mathalpha}{greek}{"14}
  \DeclareMathSymbol{\lambda}{\mathalpha}{greek}{"15}
  \DeclareMathSymbol{\mu}{\mathalpha}{greek}{"16}
  \DeclareMathSymbol{\nu}{\mathalpha}{greek}{"17}
  \DeclareMathSymbol{\xi}{\mathalpha}{greek}{"18}
  \DeclareMathSymbol{\pi}{\mathalpha}{greek}{"19}
  \DeclareMathSymbol{\rho}{\mathalpha}{greek}{"1A}
  \DeclareMathSymbol{\sigma}{\mathalpha}{greek}{"1B}
  \DeclareMathSymbol{\tau}{\mathalpha}{greek}{"1C}
  \DeclareMathSymbol{\upsilon}{\mathalpha}{greek}{"1D}
  \DeclareMathSymbol{\phi}{\mathalpha}{greek}{"1E}
  \DeclareMathSymbol{\chi}{\mathalpha}{greek}{"1F}
  \DeclareMathSymbol{\psi}{\mathalpha}{greek}{"20}
  \DeclareMathSymbol{\omega}{\mathalpha}{greek}{"21}
  \DeclareMathSymbol{\varepsilon}{\mathalpha}{greek}{"22}
  \DeclareMathSymbol{\vartheta}{\mathalpha}{greek}{"23}
  \DeclareMathSymbol{\varpi}{\mathalpha}{greek}{"24}
  \DeclareMathSymbol{\varrho}{\mathalpha}{greek}{"25}
  \DeclareMathSymbol{\varsigma}{\mathalpha}{greek}{"26}
  \DeclareMathSymbol{\varphi}{\mathalpha}{greek}{"27}
  \DeclareSymbolFont{otone}{OT1}{cmr}{m}{n}
  \DeclareMathSymbol{\Gamma}{\mathalpha}{otone}{0}
  \DeclareMathSymbol{\Delta}{\mathalpha}{otone}{1}
  \DeclareMathSymbol{\Theta}{\mathalpha}{otone}{2}
  \DeclareMathSymbol{\Lambda}{\mathalpha}{otone}{3}
  \DeclareMathSymbol{\Xi}{\mathalpha}{otone}{4}
  \DeclareMathSymbol{\Pi}{\mathalpha}{otone}{5}
  \DeclareMathSymbol{\Sigma}{\mathalpha}{otone}{6}
  \DeclareMathSymbol{\Upsilon}{\mathalpha}{otone}{7}
  \DeclareMathSymbol{\Phi}{\mathalpha}{otone}{8}
  \DeclareMathSymbol{\Psi}{\mathalpha}{otone}{9}
  \DeclareMathSymbol{\Omega}{\mathalpha}{otone}{10}
  \DeclareSymbolFont{syms}{OML}{cmm}{m}{it}
  \DeclareMathSymbol{\partial}{\mathord}{syms}{"40}
  \DeclareMathAlphabet{\mathbold}{OML}{cmm}{b}{it}
  \DeclareSymbolFont{largesymbols}{OMX}{cmex}{m}{n}

}
\usepackage{soul}